\pdfoutput=1
\documentclass[useAMS,usenatbib,pdftex]{mn2e}
\usepackage{yhmath}
\usepackage{mathrsfs}
\usepackage{amssymb}
\usepackage{amsmath}
\usepackage[ruled,lined,linesnumbered]{algorithm2e}
\usepackage{url}
\usepackage{pifont}
\usepackage{xspace}
\usepackage[pdftex]{graphicx,color}
\usepackage[pass,a4paper]{geometry}

\newcommand{\sep}{\hspace{+.2em}}

\newcommand{\pc}{\mathrm{pc}}
\newcommand{\kpc}{\mathrm{kpc}}

\newcommand{\yr}{\mathrm{yr}}
\newcommand{\kyr}{\mathrm{kyr}}
\newcommand{\myr}{\mathrm{Myr}}

\newcommand{\Msolar}{\mathrm{M_{\odot}}}

\newcommand{\kms}{\mathrm{km\sep s^{-1}}}
\newcommand{\nden}{\mathrm{cm^{-3}}}
\newcommand{\mden}{\mathrm{g\sep cm^{-3}}}

\newcommand{\eV}{\mathrm{eV}}

\newcommand{\Hatm}{\mathrm{H_{I}}}
\newcommand{\Hmol}{\mathrm{H_{2}}}

\newcommand{\kBoltz}{k_{\mathrm{B}}}
\newcommand{\MHydro}{m_{\mathrm{H}}}

\newcommand{\mH}{m_{\mathrm{H}}}

\newcommand{\nele}{n_{e}}
\newcommand{\npro}{n_{p}}
\newcommand{\nHI}{n_{\mathrm{H_{I}}}}
\newcommand{\nHtwo}{n_{\mathrm{H_{2}}}}
\newcommand{\nH}{n_{\mathrm{H}}}

\newcommand{\ngr}{n_{\mathrm{gr}}}
\newcommand{\Tgas}{T_{\mathrm{gas}}}

\newcommand{\Tgr}{T_{\mathrm{gr}}}

\newcommand{\agr}{a_{\mathrm{gr}}}
\newcommand{\arad}{\bmath{a}_{\mathrm{rad}}}

\newcommand{\Qabs}{\mathcal{Q}_{\mathrm{abs}}(\nu,a)}
\newcommand{\Qsca}{\mathcal{Q}_{\mathrm{sca}}(\nu,a)}

\newcommand{\sigmagr}{\sigma_{\mathrm{gr}}}
\newcommand{\gammaeff}{\gamma_{\mathrm{eff}}}

\newcommand{\rcl}{r_{\mathrm{cl}}}

\newcommand{\Mcl}{M_{\mathrm{cl}}}
\newcommand{\NHcl}{N_{\mathrm{H,cl}}}
\newcommand{\Lbol}{L_{\mathrm{bol}}}
\newcommand{\Lion}{L_{\mathrm{ion}}}

\newcommand{\Fion}{F_{\mathrm{ion}}}

\newcommand{\vshapp}{v_{\mathrm{sh}}^{\mathrm{app}}}
\newcommand{\vorb}{v_{\mathrm{orb}}}
\newcommand{\tsweep}{t_{\mathrm{sweep}}}
\newcommand{\tsc}{t_{\mathrm{sc}}}
\newcommand{\torb}{t_{\mathrm{orb}}}
\newcommand{\tspin}{t_{\mathrm{spin}}}

\newcommand{\BCion}{\mathrm{BC_{ion}}}
\newcommand{\hnuion}{h\nu_{\mathrm{ion}}}

\newcommand{\nptcl}{n_{\mathrm{ptcl}}}
\newcommand{\MBH}{M_{\mathrm{BH}}}
\newcommand{\bBH}{b_{\mathrm{BH}}}
\newcommand{\MNSC}{M_{\mathrm{NSC}}}
\newcommand{\bNSC}{b_{\mathrm{NSC}}}

\newcommand{\dthydi}{\Delta t_{\mathrm{hyd},i}}
\newcommand{\dtchem}{\Delta t_{\mathrm{chem}}}

\newcommand{\dtengy}{\Delta t_{\mathrm{engy}}}

\newcommand{\dtgrvi}{\Delta t_{\mathrm{grv},i}}
\newcommand{\dtsub}{\Delta t_{\mathrm{sub}}}
\newcommand{\dtglobal}{\Delta t_{\mathrm{g}}}
\newcommand{\dtlimit}{\Delta t_{\mathrm{lim}}}

\newcommand{\SCmodel}{SC00-3D\xspace}
\newcommand{\SCs}{SC00-3D-static\xspace}
\newcommand{\SCff}{SC00-3D-ff\xspace}

\newcommand{\linesegment}[2]{\overline{#1\operatorname{-}#2}}

\title[On the evolution of gas clouds exposed to AGN radiation]{On the evolution of gas clouds exposed to AGN radiation. I. Three-dimensional radiation hydrodynamic simulations}
\author[D. Namekata, M. Umemura, and K. Hasegawa]{
D. Namekata$^{1}$\thanks{E-mail: namekata@ccs.tsukuba.ac.jp},
M. Umemura$^{1}$ \thanks{E-mail: umemura@ccs.tsukuba.ac.jp}, and
K. Hasegawa$^{1}$\thanks{E-mail: hasegawa@ccs.tsukuba.ac.jp}\\
$^{1}$Center for Computational Sciences, University of Tsukuba, 1-1-1 Tennodai, Tsukuba 305-8577 Ibaraki, Japan}

\begin{document}

\date{Accepted; Received; in original form}

\pagerange{\pageref{firstpage}--\pageref{lastpage}} \pubyear{2014}

\maketitle

\label{firstpage}

\begin{abstract}
We perform three-dimensional radiation hydrodynamic simulations of uniform dusty gas clouds irradiated by an active galactic nucleus (AGN) to investigate the dependence of evolution of clouds on the ionization parameter $\mathcal{U}$ and the Str{\"o}mgren number $\mathcal{N}_{S}$. We find that the evolution can be classified into two cases depending on $\mathcal{U}$. In low $\mathcal{U}$ cases ($\mathcal{U}\approx 10^{-2}$), the evolution is mainly driven by photo-evaporation. A approximately spherically-symmetric evaporation flow with velocity of $100\operatorname{-}150\;\kms$ is launched from the irradiated face. The cloud is compressed by a D-type shock with losing its mass due to photo-evaporation and is finally turned into a dense filament by $t\lesssim 1.5\tsc$. In high $\mathcal{U}$ cases ($\mathcal{U}\approx 5\times 10^{-2}$), radiation pressure suppresses photo-evaporation from the central part of the irradiated face, reducing photo-evaporation rate. A evaporation flow from the outskirts of the irradiated face is turned into a high velocity ($\lesssim 500\;\kms$) gas wind because of radiation pressure on dust. The cloud is swept by a radiation pressure-driven shock and becomes a dense gas disk by $t\approx \tsweep$. Star formation is expected in these dense regions for both cases of $\mathcal{U}$. We discuss the influences of the AGN radiation on the clumpy torus. A simple estimate suggests that the clumps are destroyed in timescales shorter than their orbital periods. For the clumpy structure to be maintained over long period, the incident radiation field needs to be sufficiently weaken for most of the clumps, or, some mechanism that creates the clumps continuously is needed.
\end{abstract}


\section{Introduction} \label{sec:intro}
Active galactic nuclei (AGNs) are one of the most brightest objects in the universe
and their bolometric luminosities can be as high as $\approx 10^{47}\;\mathrm{erg\;s^{-1}}$ (e.g., \citealt{croom02:_qso_redsh_survey_ix,dietrich02:_contin}).
Most of their radiation is emitted in the optical/ultraviolet (UV) wavelength as well as in the X-ray wavelength and
is enable to ionize and heat surrounding interstellar medium (ISM) or intragalatic medium (IGM).
Therefore, radiation from the AGNs, along with relativistic jets, are believed to have an great impact on evolution and formation of galaxies. One of indirect evidence supporting this is provided by a comparison between theoretical predictions of galaxy luminosity function and observational data (e.g., \citealt{benson03:_what}).

The activities of AGNs are maintained through mass accretion onto supermassive black holes (SMBHs).
The accreting matter is considered to flow from a dusty molecular torus which surrounds the SMBH. Its existence is suggested in the unified scheme of AGNs (e.g., \citealt{antonucci93:_unified_model_activ_galac_nuclei_quasar,urry95:_unified_schem_radio_loud_activ_galac_nuclei}) and is supported by indirect observational evidence such as polarized light from Seyfert galaxies (e.g., \citealt{young95:_near_ir_ngc,young96:_polar,smith02:_seyfer,smith04:_seyfer,smith05:_equat_seyfer}). A theoretical consideration (\citealt{krolik88:_molec_tori_seyfer_galax}) and spectral energy distribution (SED) modelings of emission from the AGN torus (\citealt{nenkova02:_dust_emiss_activ_galac_nuclei,dullemond05:_clump,onig06:_radiat_agn_ngc,nenkova08a:_agn_dusty_tori,nenkova08b:_agn_dusty_tori}) suggest that the AGN torus consists of a number of dense gas clumps. This is called the clumpy torus model. The physical properties of the gas clumps and the maintenance mechanism of the clumpy structure is currently unknown. The molecular gas in the AGN tori also must come from outer parts of host galaxies such as bulges and galactic disks or from outside the host galaxies. Several scenarios are proposed for mass supply process toward the galactic centers. In large scales ($r \gtrsim 1\;\kpc$), (1) tidal torque driven by major and the minor merger (e.g., \citealt{hernquist89:_tidal_trigg_of_starb_and,barnes91:_fuelin_starb_galax_with_gas_rich_merger,mihos96:_gasdy_and_starb_in_major_merger,taniguchi99:_minor_merger_driven_nuclear_activ,kendall03:_activ_galac_nuclei_and_minor_merger_hypot,saitoh04:_coevol_of_galac_cores_and_spiral_galax,cattaneo05:_activ_galac_nuclei_in_cosmol_simul}), (2) non-axisymmetric perturbations by stellar bars or spiral arms (e.g., \citealt{athanassoula:92}), (3) viscous torque on giant molecular clouds (e.g., \citealt{fukunaga83:_radial_distr_of_giant_molec_cloud,fukunaga84:_radial_distr_of_giant_molec_cloud,fukunaga84:_random_motion_of_giant_molec_cloud}), (4) magnetic stress by the magnetic rotational instability (MRI) (e.g., \citealt{milosavljevic04:_origin_of_nuclear_star_clust}) are suggested. In the outer part of AGN torus (a few pc $\lesssim r \lesssim$ several tens of pc), \citet{wada02:_obscur_mater_aroun_seyfer_nuclei_with_starb} and \citet{wada09:_molec_gas_disk_struc_aroun} demonstrated numerically that supernova-driven turbulence enhances mass inflow towards the center (see also \citealt{schartmann09:_agn,schartmann10:_gas_ngc}). Such process is investigated semi-analytically by \citet{kawakatu08:_coevol_of_super_black_holes,kawakatu09:_format_of_high_redsh_z}. \citet{krolik88:_molec_tori_seyfer_galax} suggested that kinetic viscosity by clump-clump collisions induces angular momentum transfer in the AGN torus and drive inward flow. A possibility of mass supply by gas clumps is also investigated for our Galaxy (e.g., \citealt{sanders98:_circum_mater_in_galac_centr,nayakshin07:_simul_sgr,bonnell08:_star_format_aroud_super_black_holes,wardle08:_format_compac_stell_disks_aroun_sagit,hobbs09:_simul_galac_centr,namekata11:_evolut_nuclear_disk_gas_galac_center}). However, it is not well understood how these gas supply processes (at different scales) are affected by the AGN when it increases its activity. This question must be related to maintenance mechanism of the AGN activity.

In this context, \citet{wada12:_radiat} performed recently 3D RHD simulations of a circumnuclear gas disk (its size is $r\lesssim 30\;\pc$) in a AGN hosting galaxy taking into account X-ray heating and radiation pressure on gas. He showed that a geometrically and optically-thick torus can be naturally formed by hydrodynamic interaction between the back-flow of a biconical gas outflow, which is launched from the inner part of the disk ($r\leq$ a few $\pc$) in a vertical direction by the radiation force, and the disk gas. He also showed that the gas accretion is not stopped completely by the radiation feedback. Because the gas accretion rate to the central parsec is one order magnitude smaller than the gas accretion rate required to maintain the AGN luminosity, he suggested that the AGN activity is intrinsically intermittent or that there are other mechanisms that enhance the mass accretion to the center.

A complementary approach to understand the effects of the AGN radiation on the gas supply processes is to investigate evolution of optically-thick dense clouds in detail, because they are expected to play a main role in gas supply process even under strong radiation field. Especially, survival time and star formation properties (initial mass function and star formation rate) of such clouds are quite important information, since the survival time can be related to the efficiency of angular momentum transfer and star formation and stellar feedback affect not only the gravitational stability of the gas clouds but also that of the galactic gas disk by consuming gas or inputting energy. It is difficult to investigate these important properties by global simulations.

Evolution of an cloud irradiated by an AGN has been studied by several authors. Early studies investigated the effects of radiation pressure on line-emitting clouds in QSOs by ignoring hydrodynamic effects entirely (e.g., \citealt{williams72,mathews74:_radiat_seyfer_galax_nuclei,mckee75:_radiat_press_quasar_cloud,weymann76:_confin,mathews76:_stabil,mathews77:_rayleig_taylor,blumenthal79:_qso,krolik81:_two,mathews82:_quasar,mathews86:_struc}). \citet{pier95:_photoev} investigated hydrodynamic properties of photo-evaporation wind from AGN-irradiated dusty gas clouds by solving the steady-state wind equation assuming spherical symmetry. They showed that radiation pressure force suppresses the photo-evaporation by confining the evaporation flow to near the cloud surface if thermal sputtering of dust is inefficient and that the dusty clouds on eccentric orbits can penetrate well inside the inner edge of the torus because the photo-evaporation time is longer than the orbital period. Recently, \citet{schartmann11:_radiat_seyfer} performed two-dimensional RHD simulations of a dusty gas cloud falling towards the galactic center with ignoring self-gravity and showed that the clouds can be destroyed by hydrodynamic instabilities generated by the complex interplay between the radiation pressure and the ram pressure. \citet{hocuk10:_x,hocuk11:_star_agn_variat} showed numerically the possibility that the X-ray heating from the AGN makes the initial mass function top-heavy. More recently, \citet{proga14:_effec_irrad_cloud_evolut_activ_galac_nuclei} investigated the dependencies of the evolution of an irradiated cloud on the properties of type of opacity (absorption-dominated or scattering-dominated) and the optically-thickness of the cloud by performing two-dimensional RHD simulations taking into account both absorption and scattering, but ignoring photo-ionization and self-gravity.

As already mentioned, a large fraction of the AGN radiation is emitted in the optical/UV wavelength. Most of these photons must be absorbed by neutral atomic hydrogen in an irradiated face of gas clouds and thereby photo-evaporation should occur. Because an ionized gas has temperature of $(1\operatorname{-}3)\times 10^{4}\;\mathrm{K}$, evaporation must be stronger than thermal expansion driven by X-ray heating shown in \citet{hocuk10:_x,hocuk11:_star_agn_variat}. Thus, photo-evaporation due to H{\scriptsize I} ionization can play an important role in evolution of the gas clouds. In addition, radiation pressure also plays a great impact on evolution of the gas clouds as already shown by \citet{schartmann11:_radiat_seyfer}. However, it is not clear how these processes affect evolution of the gas clouds when they operate simultaneously, because no hydrodynamic simulation that includes both processes has been carried out. Therefore, in this paper, we perform 3D RHD simulations of gas clouds exposed to an AGN taking into account non-equilibrium chemistry of $\mathrm{e}^{-}$, $p$, $\Hatm$, $\Hmol$, and dust, in order to study combined effects of the photo-evaporation and the radiation pressure on the evolution of the clouds without enforcing spherical symmetry. Especially, we clarify the dependencies of the survival times of the clouds on the optically thickness of the cloud and the incident radiation field strength, because such information can be useful when we consider the gas supply processes. Effects of star formation and stellar feedbacks are investigated in a subsequent paper.

This paper is organized as follows. In \S\ref{sec:model}, we explain our models and basic assumptions. In \S\ref{sec:numerical_methods}, we describe the detail of the numerical methods. In \S\ref{sec:numerical_results}, we show our numerical results. Evolution of irradiated clouds is explained here and we show the dependence of survival time on the radiation field strength and the optical depth of the clouds. In \S\ref{sec:discussions}, we discuss star formation in the clouds and give some implications for gas clumps in AGN tori. Finally, we summarized this paper in \S\ref{sec:summary}.

\section{Model} \label{sec:model}
\subsection{Basic assumptions and model parameters}\label{subsec:parameters}
\begin{figure}
\centering
\includegraphics[width=\hsize]{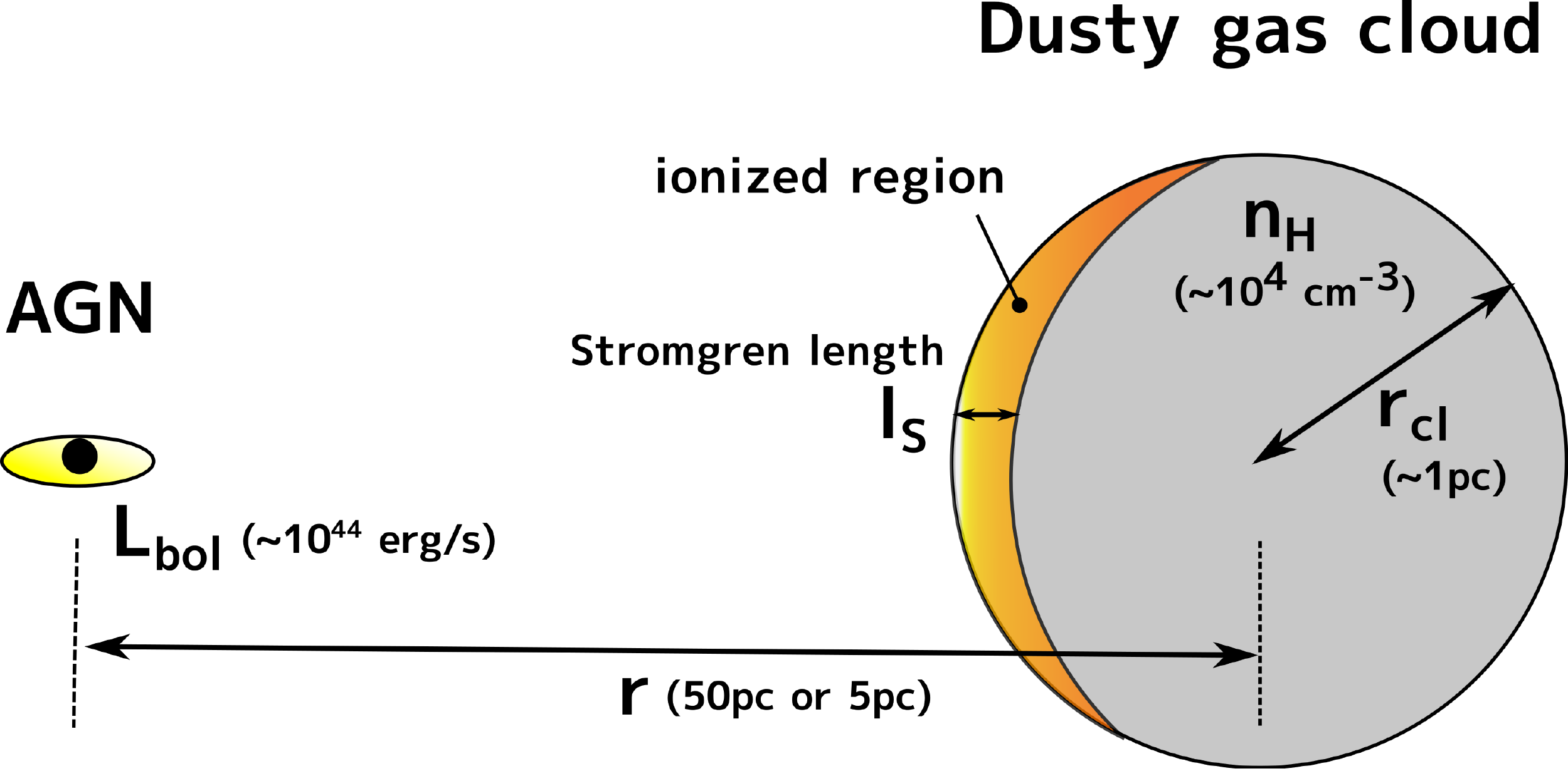}
\caption{A schematic illustration of our model. The system is described by four parameters: the cloud radius $\rcl$, the hydrogen number density of the cloud $\nH$, the distance between the AGN and the center of the cloud $r$, and the bolometric luminosity of the AGN $\Lbol$. Typical values for the parameters assumed in this study are also shown in the parentheses. The surface layer to a depth of the Str{\" o}mgren length, $l_{S}$, at the irradiated side of the cloud will be photo-ionized.}
\label{fig:model_description}
\end{figure}

Figure~\ref{fig:model_description} shows a schematic illustration of our model. To make the analysis easy, we assume that a gas cloud is an initially uniform sphere and consists of neutral molecular hydrogen. The gas cloud is placed at the distance of $r$ from the AGN and $r$ is chosen so that the radiation is almost plane-parallel. The gas cloud is initially at rest. The gravitational potential of the host galaxy and the SMBH is not considered for simplicity except for the model \SCff which is explained later. We do not take into account star formation and stellar feedback processes in the gas cloud. The simulation is started when the AGN radiation is turned on. Such an abrupt increase of luminosity will be occurred when the AGN just becomes active or when the gas cloud moves in the clumpy torus from a shadow region, where most of the AGN radiation is shielded by other gas clouds, to a region near the surface of the torus. We numerically follow the evolution of the cloud taking into account self-gravity of gas and absorption of direct radiation from AGN, until the numerical timestep becomes prohibitively small. Scattering of photons and transfer of diffuse (scattered and re-emitted) photons are not taken into account in this study.

The basic parameters that control the system are the cloud radius $\rcl$, the hydrogen number density of the cloud $\nH$, the distance of the cloud center from the AGN $r$, and the bolometric luminosity of the AGN $\Lbol$ (see Fig.~\ref{fig:model_description}). Thus, the number of free parameters is 4. It is not practical to quest the entire of the four-dimensional parameter space. Instead, we investigate the dependence on two parameters that we consider important, the ionization parameter $\mathcal{U}$ and the Str{\"o}mgren number $\mathcal{N}_{S}$. They are defined by
\begin{eqnarray}
\mathcal{U} & = & \dfrac{\Fion}{c \nH}, \\
\mathcal{N}_{S} & = & \frac{2\rcl}{l_{S}},
\end{eqnarray}
where $\Fion$ is the number flux of ionizing photon and is defined as
\begin{eqnarray}
\Fion & = & \dfrac{1}{4\pi r^{2}}\int^{\infty}_{\nu_{\mathrm{L}}}\dfrac{L_{\nu'}}{h\nu'}\:d\nu',
\end{eqnarray}
where $L_{\nu}$ is the monochromatic luminosity of the AGN, $h$ is the Planck constant, and $\nu_{\mathrm{L}}$ is the frequency at the Lyman limit. $c$ is the speed of light, $l_{S}$ is the Str{\"o}mgren length (see Fig.~\ref{fig:model_description}) defined by
\begin{eqnarray}
l_{S} = \frac{\Fion}{\alpha_{\mathrm{B}}\nH^{2}},
\end{eqnarray}
where $\alpha_{\mathrm{B}}\approx 2.59\times 10^{-13}\;\mathrm{cm^{3}\;s^{-1}}$ is the case B recombination coefficient at the gas temperature of $10^{4}\;\mathrm{K}$ (\citealt{hui97:_equat}). The Str{\"o}mgren number $\mathcal{N}_{S}$ represents what fraction of gas in the cloud is photo-ionized and therefore corresponds to the \textit{effective} optical depth of the gas cloud for the photo-ionization of $\Hatm$, whereas the ionization parameter $\mathcal{U}$ is the ratio of the photon number density to the hydrogen number density and is an indicator of the strength of incident radiation fields. The latter can be understood by considering the ratio of the radiation energy density to the thermal energy density in a fully-ionized region which are formed at an irradiated surface of the gas cloud.
The thermal energy density in the fully-ionized gas is given by
\begin{eqnarray}
e_{\mathrm{th}} = \dfrac{3}{2}(\nele + \npro) \kBoltz \Tgas = 3\nH \kBoltz \Tgas.
\end{eqnarray}
On the other hand, the radiation energy density is given by
\begin{eqnarray}
e_{\mathrm{rad}} = \dfrac{\Lbol}{4\pi r^{2}c}.
\end{eqnarray}
The ratio of both values is written as
\begin{eqnarray}
\dfrac{e_{\mathrm{rad}}}{e_{\mathrm{th}}} & = & \dfrac{\Lbol}{4\pi r^{2}c\nH}\dfrac{1}{3\kBoltz \Tgas} \\
& \approx & \dfrac{1\;\eV}{3\kBoltz \Tgas}\left(\dfrac{\mathcal{U}}{1.5\times 10^{-2}}\right)\left(\dfrac{\hnuion}{44\;\eV}\right)\left(\dfrac{\BCion}{1.5}\right),
\end{eqnarray}
where $\hnuion (\equiv h \int^{\infty}_{\nu_{\mathrm{L}}}\nu' L_{\nu'}\:d\nu'/\int^{\infty}_{\nu_{\mathrm{L}}}L_{\nu'}\:d\nu')$ is the mean-energy of ionizing photon and $\BCion (\equiv \Lbol/\Lion)$ is the bolometric correction factor. The normalizations for $\hnuion$ and $\BCion$ are based on the adopted SED (see \S~\ref{subsec:AGN_SED} and Table.~\ref{tbl:SED_properties}). Thus, the ratio $e_{\mathrm{rad}}/e_{\mathrm{th}}$ is proportional to $\mathcal{U}$.
In the fully-ionized gas, the gas temperature is in the range of $\operatorname{(1-3)} \times 10^{4}\;\mathrm{K}$ and can be considered to be a constant\footnote{Note that this temperature range is true only for the small range of $\mathcal{U}$ investigated in this study ($\mathcal{U}\approx 0.01\operatorname{-}0.2$; see Table.~\ref{tbl:simulation_runs}). The maximal temperature that fully-ionized gas can reach is the Compton temperature ($\sim 10^{7}\;\mathrm{K}$).}. Therefore, higher $\mathcal{U}$ directly implies stronger radiation fields. Similar conclusion can be obtained by an order estimation of the force balance in the fully-ionized region. The pressure-gradient acceleration is of the order of 
\begin{eqnarray}
|\bmath{a}_{\mathrm{p}}| & = & \dfrac{\nabla p}{\rho} \sim \frac{p}{\rho l_{S}} \\
& \approx & \dfrac{2\nH \kBoltz \Tgas}{\MHydro \nH l_{S}}.
\end{eqnarray}
The order of the radiative acceleration can be estimated as
\begin{eqnarray}
|\bmath{a}_{\mathrm{rad}}| \sim \dfrac{\Lbol}{4\pi r^{2}c}\dfrac{1}{\mH \nH l_{S}}.
\end{eqnarray}
The ratio becomes 
\begin{eqnarray}
\dfrac{|\bmath{a}_{\mathrm{rad}}|}{|\bmath{a}_{\mathrm{p}}|} & = & \dfrac{\Lbol}{4\pi r^{2}c\nH} \dfrac{1}{2 \kBoltz \Tgas} \\
& \approx & \dfrac{1\;\eV}{2\kBoltz \Tgas}\left(\dfrac{\mathcal{U}}{1.5\times 10^{-2}}\right)\left(\dfrac{\hnuion}{44\;\eV}\right)\left(\dfrac{\BCion}{1.5}\right).
\end{eqnarray}
Again, the ratio is proportional to $\mathcal{U}$.
According to these order estimation, it is expected that the radiation pressure becomes important when $\mathcal{U}$ is larger than $\approx 0.05$.

In this paper, we investigate two cases of $\mathcal{U}$ ($5.2\times 10^{-2}$ and $1.3\times 10^{-2}$). This choice is motivated by the discussion above, that is, we expect that the radiation pressure plays an more important role in the evolution of the cloud for the case of $\mathcal{U}=5.2\times 10^{-2}$ than $\mathcal{U}=1.3\times 10^{-2}$. Thus, this range of $\mathcal{U}$ is sufficient to examine relative importance of the photo-evaporation and the radiation pressure (but, a larger $\mathcal{U}$ case is also investigated in this study; see next paragraph). For each of $\mathcal{U}$, three different values (5, 10, and 20) of $\mathcal{N}_{S}$ are examined. In reality, there are more optically thick clouds than $\mathcal{N}_{S}=20$. As we will discuss in \S\ref{subsec:AGN_gas_clumps}, gas clumps in AGN tori can have $\mathcal{N_{S}}>10^{4}$. Unfortunately, it is impossible to numerically resolve such extremely optically thick clouds in the current computational power. Therefore, we use the results of these runs to investigate the dependency of the cloud evolution on the optical depth and attempt to predict the fates of larger $\mathcal{N}_{S}$ clouds based on the derived dependency. For the given $\mathcal{U}$ and $\mathcal{N}_{S}$, an arbitrary realization is possible. In this study, we assume that $\nH=10^{4}\;\mathrm{cm^{-3}}$ and $r=50\;\pc$ for all the models except the model \SCmodel. For $\mathcal{U}=1.3\times 10^{-2}$ and $5.2\times 10^{-2}$, we assume $\Lbol=1.25\times 10^{44}\;\mathrm{erg\;s^{-1}}$ and $5\times 10^{44}\;\mathrm{erg\;s^{-1}}$, respectively. Then, $\rcl$ is uniquely determined if $\mathcal{N}_{S}$ is given. The choice of $\nH$ is intended to mimic physical conditions similar to molecular cores in the Galaxy. Table~\ref{tbl:simulation_runs} summarizes the simulation runs. As shown in the table, the Jeans ratio $r_{\mathrm{J}}$ of the models L05, L10, L20, and H05 are larger than unity and therefore they are initially stable for their self-gravity. We note that $\nH$ can affect the evolution of the clouds independently of $\Mcl$, since the thermodynamic nature of the gas depends on $\nH$.

The model \SCmodel in Table~\ref{tbl:simulation_runs} is the run with almost the same calculation conditions as the run SC00 in \citet{schartmann11:_radiat_seyfer} (see Table 2 in their paper) and has a more higher $\mathcal{U}$ than High-$\mathcal{U}$ models. The major differences between ours and theirs are as follows. First, ours is a 3D simulation, while their simulation is 2D. Therefore, the dynamical response to the tidal force by the external gravities is expected to be different between ours and theirs. Second, the self-gravity of gas and the photo-ionization are taken into account in our simulation. On the other hand, these effects are not considered in their study. This also may affect the evolution of the cloud. The comparison between \SCmodel and SC00 will be useful to investigate effects of these differences on the evolution of cloud. The detail of the simulation conditions is explained in \S~\ref{subsec:SC_models}.

 In order to compare different runs, it may be useful to use time normalized by a characteristic timescale. Here, we introduce two characteristic timescales. Since the irradiated gaseous layer of the gas cloud is expected to expand in the direction of the AGN, one of characteristic timescale can be the sound (rarefaction wave) crossing time $\tsc$ defined as
\begin{eqnarray}
\tsc = \dfrac{2\rcl}{c^{\mathrm{irr}}_{s}} = 7\times 10^{4}\;\yr \left(\dfrac{\rcl}{1\;\pc}\right)\left(\dfrac{c^{\mathrm{ion}}_{s}}{28.75\;\kms}\right)^{-1}, \label{eq:tsc}
\end{eqnarray}
where $c^{\mathrm{irr}}_{s}$ is the sound speed of the irradiated gas layer. Here, we normalize it by the sound speed of fully-ionized pure hydrogen gas of $\Tgas=30000\;\mathrm{K}$. Another can be the sweeping time of the shocked layer which is formed at the irradiated side of the cloud due to the back reaction of the photo-evaporation flow or the radiation pressure (see \S~\ref{sec:numerical_results}). If we assume that (i) only the radiation pressure force is important, and (ii) all the radiation is absorbed in this shocked layer, the motion of the shocked layer can be modeled as $d(\rho SRv)/dt=|\bmath{F}_{\mathrm{rad}}|S$, where $S$ is surface area, $R$ is the position of the layer, $v$ is the velocity of the layer, $|\bmath{F}_{\mathrm{rad}}|=\Lbol/(4\pi r^{2}c)$ is the radiation pressure force acting on the unit area. In this razor-thin approximation, the velocity of the shocked layer is
\begin{eqnarray}
\vshapp & = & \sqrt{\frac{|\bmath{F}_{\mathrm{rad}}|}{\rho}} \nonumber \\
& = & 9.4\;\kms\left(\dfrac{\Lbol}{1.25\times 10^{44}\;\mathrm{erg\;s^{-1}}}\right)^{0.5} \nonumber \\ 
&   & \quad \times \left(\dfrac{r}{50\;\pc}\right)^{-1}\left(\dfrac{\nH}{10^{4}\;\mathrm{cm^{-3}}}\right)^{-0.5}. \label{eq:vshapp}
\end{eqnarray}
Using the approximate relation $\Lbol/(4\pi r^{2})\approx \BCion \hnuion \mathcal{U}$, we can rewrite Eq.(\ref{eq:vshapp}) into
\begin{eqnarray}
\vshapp & = & \sqrt{\dfrac{\BCion\hnuion\mathcal{U}}{m_{\mathrm{H}}}}. \label{eq:vshapp2}
\end{eqnarray}
Thus, $\vshapp$ is a function of $\mathcal{U}$ only for a fixed SED.
The sweeping time is given by
\begin{eqnarray}
\tsweep & = & \frac{2\rcl}{\vshapp} \nonumber \\
& = & 2.13\times 10^{5}\;\yr \left(\dfrac{\rcl}{1\;\pc}\right)\left(\dfrac{\Lbol}{1.25\times 10^{44}\;\mathrm{erg\;s^{-1}}}\right)^{-0.5} \nonumber \\
&   & \quad \times \left(\dfrac{r}{50\;\pc}\right)\left(\dfrac{\nH}{10^{4}\;\mathrm{cm^{-3}}}\right)^{0.5} \label{eq:tsweep}.
\end{eqnarray}

\begin{table*}
\centering
\begin{minipage}{\hsize}
\caption{Simulation runs.}
\label{tbl:simulation_runs}
\begin{tabular}{@{}lllllllllllll@{}}
\hline
Model Family & Model Name & $\mathcal{U}$ & \multicolumn{3}{c}{$\mathcal{N}_{S}$$^{\sharp}$} & $\Lbol$ & $r$ & $\nH$ & $\rcl$ & $\Mcl$ & $\Omega$ & $r_{\mathrm{J}}^{\dagger}$ \\ \cline{4-6}
 & & &  near & center & far & $[\mathrm{erg\;s^{-1}}]$ & $[\pc]$ & $[\mathrm{cm^{-3}}]$ & $[\pc]$ & $[\Msolar]$ & $[\mathrm{ster.}]$  & \\ 
\hline
Low-$\mathcal{U}$   & L05      & $1.3\times 10^{-2}$ & $5.085$ & $5.111$ & $5.136$ & $1.25\times 10^{44}$ & $50$    & $10^{4}$            & $0.125$ & $2.022$   & $1.96\times 10^{-5}$ & $23.9$ \\
                    & L10      & $1.3\times 10^{-2}$ & $10.12$ & $10.22$ & $10.32$ & $1.25\times 10^{44}$ & $50$    & $10^{4}$            & $0.250$ & $16.18$   & $7.85\times 10^{-5}$ & $5.98$ \\
                    & L20      & $1.3\times 10^{-2}$ & $20.04$ & $20.44$ & $20.85$ & $1.25\times 10^{44}$ & $50$    & $10^{4}$            & $0.500$ & $129.4$   & $3.14\times 10^{-4}$ & $1.50$ \\
                                                                      
High-$\mathcal{U}$  & H05      & $5.2\times 10^{-2}$ & $5.009$ & $5.111$ & $5.213$ & $5.0\times 10^{44}$  & $50$    & $10^{4}$            & $0.5$   & $129.4$   & $3.14\times 10^{-4}$ & $1.50$ \\
                    & H10      & $5.2\times 10^{-2}$ & $9.816$ & $10.22$ & $10.63$ & $5.0\times 10^{44}$  & $50$    & $10^{4}$            & $1.0$   & $1035$    & $1.26\times 10^{-3}$ & $0.374$ \\
                    & H20      & $5.2\times 10^{-2}$ & $18.83$ & $20.44$ & $22.11$ & $5.0\times 10^{44}$  & $50$    & $10^{4}$            & $2.0$   & $8284$    & $5.03\times 10^{-3}$ & $0.093$ \\
                                                                      
SC                  & \SCmodel & $0.17$ & $12.10$ & $18.92$ & $27.24$  & $1.04\times 10^{44}$ & $5$     & $6\times 10^{4}$    & $1.0$   & $6213$    & $0.127$            & $0.062$\\  
\hline
\end{tabular}
\end{minipage}
\begin{flushleft}
$^{\sharp}$We show the Str{\"o}mgren number $\mathcal{N}_{S}$ at three different positions in a cloud. The sub-entries ``near'',''center'', and ``far'' correspond to the positions at distances of $r-\rcl$, $r$, and $r+\rcl$ from the AGN, respectively. A small difference between $\mathcal{N}_{S}$ at three positions means that the incident radiation field is effectively plane-parallel. \\
$^{\dagger}$The ratio of total thermal energy to gravitational energy ($r_{\mathrm{J}}\equiv E_{\mathrm{th}}/|E_{\mathrm{grv}}|$, where $E_{\mathrm{th}}=\frac{1}{\gammaeff-1}\frac{k_{\mathrm{B}}\Tgas}{\mu m_{\mathrm{H}}}\Mcl$ and $E_{\mathrm{grv}}=-\frac{3}{5}\frac{G\Mcl^{2}}{\rcl}$). For a critically stable cloud, $r_{\mathrm{J}}$ is written as $\frac{5}{\pi^{2}\gammaeff(\gammaeff-1)}$ and $r_{\mathrm{J}}=0.91$ if the cloud consists of a molecular hydrogen only ($\gammaeff=7/5$). If $r_{\mathrm{J}}$ of a cloud is much smaller than unity, the cloud must be bounded by its self-gravity. As a reference, we show the Jeans length for typical parameter values: $\lambda_{\mathrm{J}}\approx 1.34\;\pc\:(\gammaeff/\frac{7}{5})^{\frac{1}{2}}(\mu/2)^{-\frac{1}{2}}(\Tgas/10^{2}\;\mathrm{K})^{\frac{1}{2}}(\nH/10^{4}\;\mathrm{cm^{-3}})^{-\frac{1}{2}}$.
\end{flushleft}
\end{table*}

\subsection{AGN spectral energy distribution} \label{subsec:AGN_SED}
\begin{figure}
\centering
\includegraphics[width=\hsize]{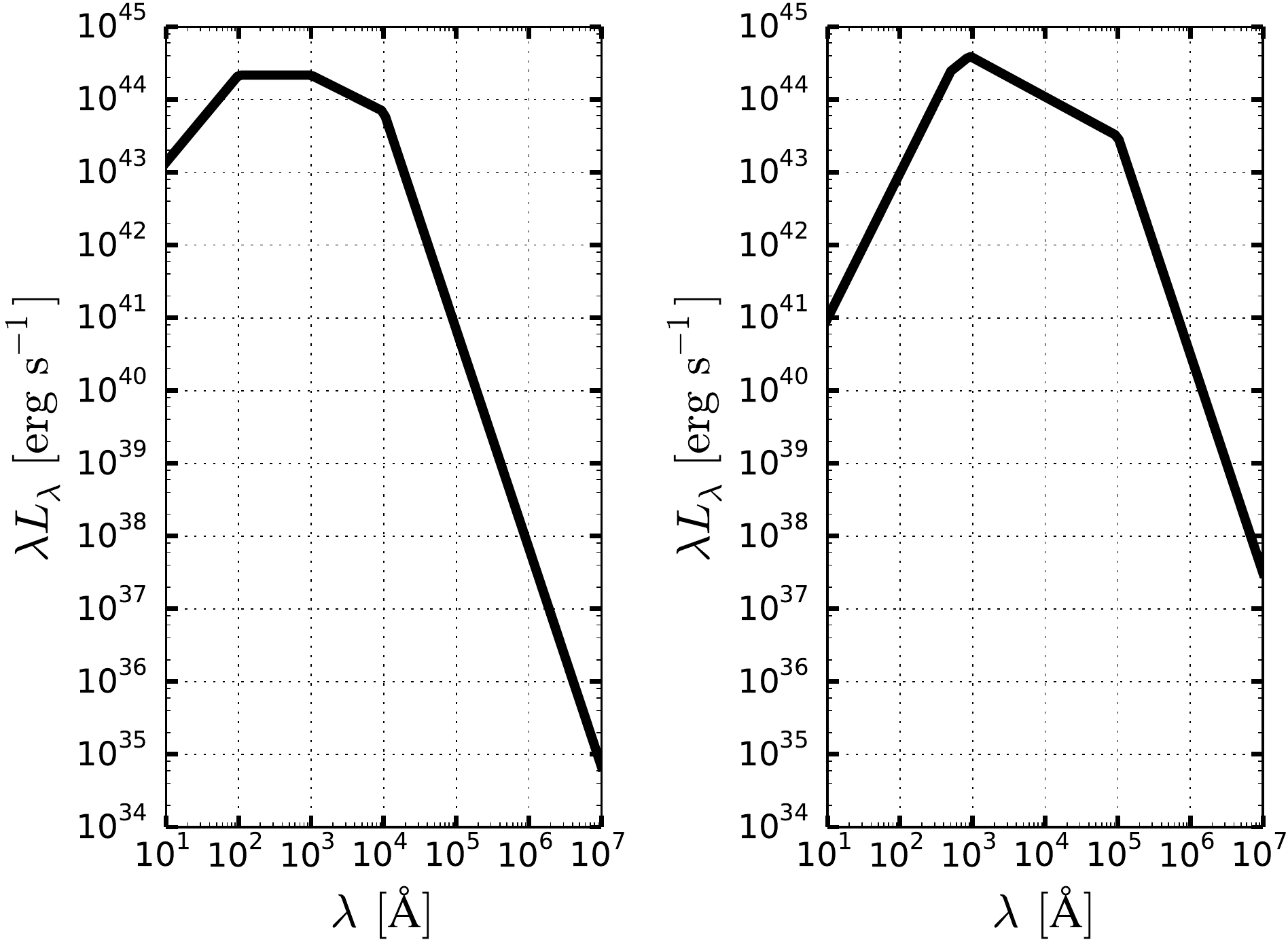}
\caption{The model SEDs of AGN for $\Lbol=10^{45}\;\mathrm{erg\;s^{-1}}$. The left panel shows the model SED by \citet{nenkova08a:_agn_dusty_tori} and the right panel shows the one by \citet{schartmann11:_radiat_seyfer}.}
\label{fig:AGN_model_SEDs}
\end{figure}

\begin{figure}
\centering
\includegraphics[width=\hsize]{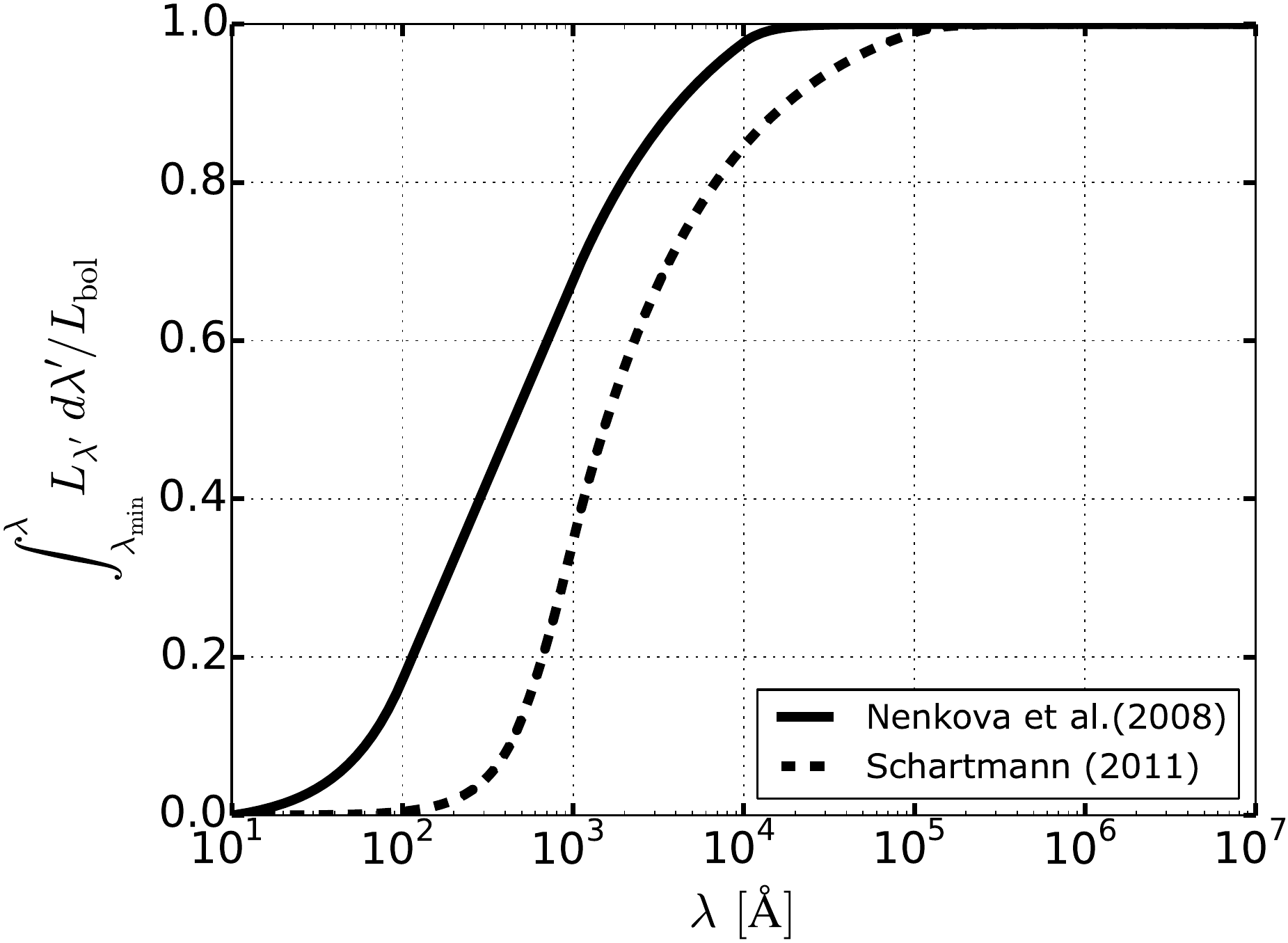}
\caption{$\int^{\lambda}_{\lambda_{\min}}L_{\lambda'}d\lambda'/\Lbol$ for the model SEDs (\textit{solid}: \citet{nenkova08a:_agn_dusty_tori}, \textit{dashed}: \citet{schartmann11:_radiat_seyfer}), where $\lambda_{\min}=10\;[\mathrm{\AA}]$.}
\label{fig:AGN_fractional_luminosity}
\end{figure}

We assume the spectral energy distribution (SED) given by \citet{nenkova08a:_agn_dusty_tori} except for \SCmodel models,
\begin{eqnarray}
\lambda F_{\lambda} \propto \left \{
\begin{array}{ll}
(\lambda/\lambda_{h})^{1.2}, & \lambda < \lambda_{h} \\
1, & \lambda_{h}\leq \lambda \leq \lambda_{u} \\
(\lambda/\lambda_{u})^{-p}, & \lambda_{u}\leq \lambda \leq \lambda_{\mathrm{RJ}} \\
(\lambda_{u}/\lambda_{\mathrm{RJ}})^{p}(\lambda/\lambda_{\mathrm{RJ}})^{-3}, & \lambda_{\mathrm{RJ}} \leq \lambda
\end{array}
\right., \label{eq:AGN_SED_Nenkova08}
\end{eqnarray}
where $\lambda_{h}=0.01\;\micron$, $\lambda_{u}=0.1\;\micron$, $\lambda_{\mathrm{RJ}}=1\;\micron$, and $p=0.5$.
For SC00 model, we use the same SED used in \citet{schartmann11:_radiat_seyfer} (see \citet{schartmann05:_towar_activ_galac_nuclei} for details):
\begin{eqnarray}
\lambda F_{\lambda} \propto \left \{
\begin{array}{ll}
\lambda^{2},  & \lambda < 500\;\mathrm{\AA} \\
\lambda^{0.8}  & 500\;\mathrm{\AA} \leq \lambda \leq 912\;\mathrm{\AA} \\
\lambda^{-0.54} & 912\;\mathrm{\AA} \leq \lambda \leq 10\;\micron \\
\lambda^{-3} & 10\;\micron \leq \lambda
\end{array}
\right.. \label{eq:AGN_SED_Schartmann05}
\end{eqnarray}
The profiles and  $\int^{\lambda}_{\lambda_{\min}}L_{\lambda'}\:d\lambda'/\Lbol$ for the both SEDs are shown in Fig. \ref{fig:AGN_model_SEDs} and \ref{fig:AGN_fractional_luminosity}, respectively, and the properties of the SEDs are summarized in Table.~\ref{tbl:SED_properties}. In this study, we take into account the wavelength range of $[10\mathrm{\AA},10^{7}\mathrm{\AA}]$ and assume isotropic radiation.

\begin{table}
\centering
\begin{minipage}{\hsize}
\caption{Properties of the model SEDs.}
\label{tbl:SED_properties}
\begin{tabular}{@{}lll@{}}
\hline
SED & $\BCion^{\sharp}$ & $\hnuion\;[\eV]^{\dagger}$ \\
\hline
\citet{nenkova08a:_agn_dusty_tori}  & 1.514 & 44.03 \\
\citet{schartmann11:_radiat_seyfer} & 3.201 & 22.27 \\
\hline
\end{tabular}
\begin{flushleft}
$^{\sharp}$ Bolometric correction factor, $\BCion$, which is defined as $\Lbol/\Lion$, where $\Lion \equiv \int^{\infty}_{h\nu_{\mathrm{L}}}L_{\nu'}\:d\nu'$ and $\nu_{\mathrm{L}}$ is the frequency at the Lyman limit.  \\
$^{\dagger}$ Mean-energy of ionizing photon, $\hnuion$, which is defined as $h \int^{\infty}_{\nu_{\mathrm{L}}}\nu' L_{\nu'}\:d\nu'/\int^{\infty}_{\nu_{\mathrm{L}}}L_{\nu'}\:d\nu'$.
\end{flushleft}
\end{minipage}
\end{table}

\subsection{ISM and dust model} \label{subsec:ISM_dust_model}
We assume that an interstellar medium consists of $e^{-}$, $p^{+}$, $\Hatm$, $\Hmol$, and dust and assume that gas and dust are dynamically coupled. For simplicity, we ignore helium and metals. The effects of helium and metals are briefly discussed in \S~\ref{sec:discussions}. The chemical reactions and radiative processes considered in this study are summarized in Table \ref{tbl:chemical_reactions} and \ref{tbl:radiative_processes}, respectively. In this paper, we denote the number of hydrogen nuclei as $\nH$ and it is calculated by $\nH=\npro + \nHI + 2\nHtwo$.

\begingroup
\renewcommand{\arraystretch}{1.0}
\begin{table*}
\centering
\begin{minipage}{\hsize}
\caption{Chemical reactions. For the definition of the parameters below, refer to the original papers.}
\label{tbl:chemical_reactions}
\begin{tabular}{@{}llll@{}}
\hline
Number & Reaction & Rate Coefficient or Cross Section$^{\sharp}$ & Reference \\
\hline
R1  & $\Hatm + e^{-} \rightarrow p^{+} + 2e^{-}$  &
$k_{\rm R1}=\exp(-32.71396786375$ & 1 \\
&& \hspace{4.9em}$+13.53655609057\{\ln[\Tgas(\eV)]\}$ & \\ 
&& \hspace{4.9em}$-5.739328757388\{\ln[\Tgas(\eV)]\}^{2}$ & \\ 
&& \hspace{4.9em}$+1.563154982022\{\ln[\Tgas(\eV)]\}^{3}$ & \\
&& \hspace{4.9em}$-2.877056004391 \times 10^{-1} \{\ln[\Tgas(\eV)]\}^{4}$ & \\
&& \hspace{4.9em}$+3.482559773736999 \times 10^{-2} \{\ln[\Tgas(\eV)]\}^{5}$ & \\
&& \hspace{4.9em}$-2.63197617559 \times 10^{-3} \{\ln[\Tgas(\eV)]\}^{6}$ & \\
&& \hspace{4.9em}$+1.119543953861 \times 10^{-4} \{\ln[\Tgas(\eV)]\}^{7}$ & \\
&& \hspace{4.9em}$-2.039149852002 \times 10^{-6} \{\ln[\Tgas(\eV)]\}^{8})$, & \\
&& \hspace{1.0em}if $\Tgas(\eV)>0.8$; otherwise $k_{\rm R1}=0$. & \smallskip\\

R2  & $p^{+}  + e^{-} \rightarrow \Hatm + \gamma$ &
$\displaystyle k_{\rm R2}=2.753\times 10^{-14}\dfrac{\lambda^{1.500}_{\Hatm}}{[1+(\lambda_{\Hatm}/2.740)^{0.407}]^{2.242}}$, & 2 \\
&& \hspace{2.0em} $\lambda _{\Hatm}=2(157807/\Tgas)$. & \smallskip\\

R3  & $\Hmol + e^{-} \rightarrow 2\Hatm + e^{-}$  &
$k_{\rm R3}=4.38\times 10^{-10}\Tgas^{0.35}\exp(-102000/\Tgas)$ & 1 \smallskip\\

R4  & $\Hatm + \gamma \rightarrow p^{+} + e^{-}$  &
$\displaystyle \sigma_{\Hatm}(\nu)=6.3\times 10^{-18}\left(\dfrac{\nu}{\nu_{1}}\right)^{-4}\frac{\exp[4-4\tan^{-1}(\varepsilon)/\varepsilon]}{1-\exp(-2\pi/\varepsilon)}$, & 3 \\
&& \hspace{2.0em} $\displaystyle \varepsilon=\sqrt{\dfrac{\nu}{\nu_{1}}-1}$, & \\
&& \hspace{2.0em} $h\nu_{1}=13.6\;\eV$. & \smallskip\\

R5  & $\Hmol + \gamma \rightarrow 2\Hatm$ & see text & 4  \smallskip\\

R6  & $\Hmol + \Hatm \rightarrow 3\Hatm$         &
$\displaystyle k_{\rm R6}=d\left(\frac{8E}{\pi \mu}\right)^{0.5}\dfrac{aE^{b-1}\Gamma(b+1)\exp(-E_{0}/E)}{(1+cE)^{b+1}}$, & 5 \\
&& \hspace{2.0em} $\displaystyle E=\dfrac{k_{\mathrm{B}}\Tgas}{27.21\;\eV}$, $E_{0}=0.168$, $\mu=2\mH/3$, & \\
&& \hspace{2.0em} $a=54.1263$, $b=2.5726$, $c=3.4500$, $d=1.849\times 10^{-22}$. & \smallskip\\

R7  & $\Hmol + \Hmol \rightarrow \Hmol + 2\Hatm$ &
$k_{\rm R7}$ is obtained by the same formula used in $k_{\rm R6}$, & 5 \\
&& but with the following parameters: & \\
&& \hspace{2.0em} $E_{0}=0.1731$, $\mu=\mH$, & \\
&& \hspace{2.0em} $a=40.1008$, $b=4.6881$, $c=2.1347$.  & \smallskip\\

R8  & $2\Hatm + \mathrm{dust} \rightarrow \Hmol + \mathrm{dust}$ &
$\displaystyle k_{\rm R8}= \sqrt{\dfrac{8k_{\mathrm{B}}\Tgas}{\pi \mH}}S_{\mathrm{H}}f_{a}\sigmagr$, & 6,7,8,9,10 \\
&& \hspace{2.0em} $\displaystyle S_{\mathrm{H}}=\dfrac{1}{[1+0.04(\Tgas+\Tgr)^{0.5}+0.002\Tgas + 8\times 10^{-6}\Tgas^{2}]}$, & \\
&& \hspace{2.0em}
$
f_{a}=\left\{
\begin{array}{ll}
1   & 5\;\mathrm{K} \leq \Tgr \leq 20\;\mathrm{K} \\
0.2 & 20\;\mathrm{K}< \Tgr \leq 500\;\mathrm{K} \\
0   & \mathrm{otherwise}
\end{array}
\right.
$.
& \smallskip\\

R9  & $3\Hatm \rightarrow \Hmol + \Hatm$ &
$k_{\rm R9}=5.5\times 10^{-29}/\Tgas$ & 11 \smallskip\\

R10 & $2\Hatm + \Hmol \rightarrow 2\Hmol$ &
$k_{\rm R10}=6.875\times 10^{-30}/\Tgas$  & 11 \\ 
\hline
\end{tabular}
\begin{flushleft}
{\footnotesize REFERENCES.}---
(1) \citet{abel97:_model}$^{\dagger}$;
(2) \citet{hui97:_equat};
(3) \citet{osterbrock06:_astrop_gaseous_nebul_activ_galac_nuclei};
(4) \citet{draine96:_struc_of_station_photod_front};
(5) \citet{martin98:_collis_induc_dissoc_of_molec};
(6) \citet{hollenbach79:_molec_format_and_infrar_emiss};
(7) \citet{cazaux04:_format_on_grain_surfac};
(8) \citet{cazaux04:_molec_hydrog_format_on_dust};
(9) \citet{cazaux10:_errat};
(10) \citet{hirashita02:_effec};
(11) \citet{palla83:_primor_star_format}. \\
$^{\sharp}$ The rate coefficient is in $\mathrm{cm^{3}\;s^{-1}}$ except for the reaction R9 and R10. They are in $\mathrm{cm^{6}\;s^{-1}}$. The temperatures are in $\mathrm{K}$ unless otherwise stated. The cross section is in $\mathrm{cm}^{2}$. The definition of the formation efficiency $f_{a}$ in the reaction R8 is based on the results of \citet{cazaux04:_molec_hydrog_format_on_dust}. \\
$^{\dagger}$ We actually use the rate coefficients adopted in the T0D code, which is published in \url{http://www.slac.stanford.edu/~tabel/PGas/codes.html}.\end{flushleft}
\end{minipage}
\end{table*}
\endgroup

\begin{table*}
\centering
\begin{minipage}{\hsize}
\caption{Rates of the radiative processes and thermal processes.
We shorten the name of the processes into symbols and their meanings are as follows:
$\mathrm{RC}$ \ding{212} case B recombination cooling;
$\mathrm{BC}$ \ding{212} Bremsstrahlung cooling of $\Hatm$;
$\mathrm{CIC}_{\Hatm}$ \ding{212} collisional ionization cooling of $\Hatm$;
$\mathrm{CEC}_{\Hatm}$ \ding{212} collisional excitation cooling of $\Hatm$;
$\mathrm{CEC}_{\Hmol}$ \ding{212} collisional excitation cooling of $\Hmol$ rovibrational lines;
$\mathrm{Chem}_{\Hmol}$ \ding{212} chemical heating and cooling of $\Hmol$;
$\mathrm{GG}$ \ding{212} collisional gas-grain energy transfer.
 }
\label{tbl:radiative_processes}
\begin{tabular}{@{}lccccccc@{}}
\hline
Process & $\mathrm{RC}$ & $\mathrm{BC}$ & $\mathrm{CIC}_{\Hatm}$ & $\mathrm{CEC}_{\Hatm}$ & $\mathrm{CEC}_{\Hmol}$ & $\mathrm{Chem}_{\Hmol}$ & $\mathrm{GG}$ \\
\hline
Reference & 1 & 7 & 4 & 4 & 2,3 & 2 & 6 \\
\hline
\end{tabular}
\end{minipage}
\begin{flushleft}
{\small REFERENCES.}---
(1) \citet{hui97:_equat};
(2) \citet{hollenbach79:_molec_format_and_infrar_emiss};
(3) \citet{galli98:_univer};
(4) \citet{cen92:_hydrod_approac_to_cosmol};
(5) \citet{fukugita94:_reion};
(6) \citet{burke83:_the_gas_grain_inter_in};
(7) \citet{kellogg75:_studies_clust_x_sourc}.
\end{flushleft}
\end{table*}

The effective specific heat ratio $\gammaeff$ of gas is computed from
\begin{eqnarray}
\frac{1}{\gammaeff -1}=\sum_{i}\frac{X_{i}}{\gamma_{i}-1},
\end{eqnarray}
where $\gamma_{i}$ is the specific heat ratio of species $i$,
$X_{i}$ is the number fraction of species $i$ defines as
\begin{eqnarray}
X_{i} & = & \frac{n_{i}}{n_{\mathrm{tot}}}, \\ 
n_{\mathrm{tot}} & = & \frac{\rho}{\mu \MHydro},
\end{eqnarray}
where $\rho$ is the mass density of gas and $\mu$ is the mean molecular weight relative to the mass of hydrogen atom $\mH$.
We assume $\gamma_{i}=5/3$ except for $\gamma_{\Hmol}$, for which we use
\begin{eqnarray}
\frac{1}{\gamma_{\Hmol}-1}=\frac{1}{2}\left[5+2x^{2}\frac{e^{x}}{(e^{x}-1)^{2}}\right],
\end{eqnarray}
where $x=6100\;\mathrm{K}/\Tgas$ (\citealt{landau80:_statis_physic}; see also \citealt{yoshida06:_format_of_primor_stars_in}).

We assume that dust consists of amorphous silicon whose composition is $\mathrm{Mg Fe Si O_{4}}$ and whose density is $3.36\;\mden$ (\citealt{laor93:_spect_const_on_the_proper}). We do not consider a size distribution of dust for simplicity. We control the mass abundance of dust by the parameter $f_{\mathrm{gr}}$, which is the dust mass per hydrogen nuclei mass. Using $f_{\mathrm{gr}}$, the number density of dust is expressed as
\begin{eqnarray}
\ngr = \frac{f_{\mathrm{gr}}\mH}{\frac{4}{3}\pi \rho_{\mathrm{gr}}a_{\mathrm{gr}}^{3}} \nH \equiv \mathcal{A}_{\mathrm{gr}} \nH,
\end{eqnarray}
where $a_{\mathrm{gr}}$ is the radius of a dust particle.
In all the simulations, we assume $a_{\mathrm{gr}}=0.05\;\micron$ and $f_{\mathrm{gr}}=0.01$, leading to $\mathcal{A}_{\mathrm{gr}}\approx 9.5\times 10^{-12}$.

\begin{figure}
\includegraphics[width=\hsize]{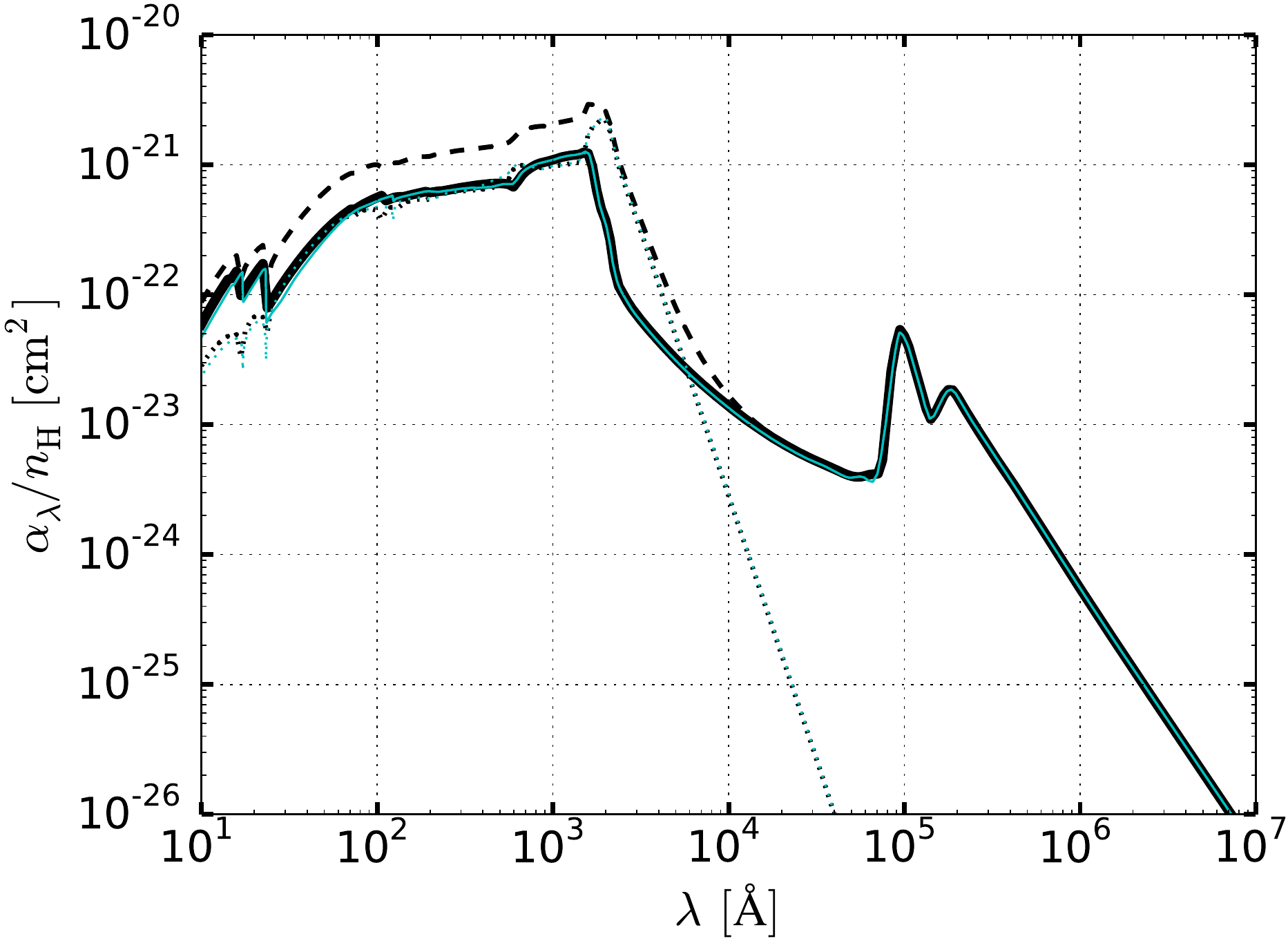}
\caption{The wavelength dependency of the absorption coefficient per hydrogen nuclei for $a_{\mathrm{gr}}=0.05\;\micron$ (\textit{thick black solid line}). For reference, we also show the scattering coefficient per hydrogen nuclei (\textit{thin black dotted line}) and the extinction coefficient per hydrogen nuclei (\textit{thin black dashed line}). We express the extinction coefficient as the sum of the absorption coefficient and the scattering coefficient for simplicity. The absorption and scattering coefficients per hydrogen nuclei for $a_{\mathrm{gr}}=0.05\;\micron$ used in the \textsc{Cloudy} are shown \textit{thin cyan solid line} and \textit{thin cyan dotted line}, respectively (see \S~\ref{subsubsec:chemical_abundance_diffuse_photons}).}
\label{fig:alpha_dust}
\end{figure}

The optical constants of dust, such as the absorption efficiency $\Qabs$ and the scattering efficiency $\Qsca$ are taken from \citet{laor93:_spect_const_on_the_proper}\footnote{Prof. Draine kindly publishes the optical constants of dust calculated in \citet{draine84:_optic_proper_of_inter_graph} and \citet{laor93:_spect_const_on_the_proper} in his site \url{http://www.astro.princeton.edu/~draine/dust/dust.diel.html}}.
We show the wavelength dependency of the absorption coefficient per hydrogen nuclei $\alpha_{\mathrm{abs/H}}(\lambda)$ in Fig.\ref{fig:alpha_dust}.

Gas and dust exchange their thermal energy each other through collisions.
The rate of change of thermal energy is written as
\begin{equation}
\Lambda_{\operatorname{g-gr}}=\ngr \nH \sigmagr \left(\frac{8\kBoltz \Tgas}{\pi \MHydro}\right)^{1/2}\overline{\alpha}_{T}2\kBoltz (\Tgr-\Tgas),
\end{equation}
where $\sigmagr$ is the geometrical cross section of a dust particle, $\Tgr$ is the dust temperature, $\Tgas$ is the gas temperature, and $\overline{\alpha}_{T}$ is the average accommodation coefficient (\citealt{burke83:_the_gas_grain_inter_in}). As shown in Fig.4 in \citet{burke83:_the_gas_grain_inter_in}, $\overline{\alpha}_{T}$ depends on $\Tgr$, $\Tgas$, dust composition, and gas composition. Because it is difficult to take into account all of these dependency, we assume $\overline{\alpha}_{T}=0.4$ in this study.

We assume that dust is instantaneously settled in the thermal equilibrium.
In this case, $\Tgr$ is determined by the following equation,
\begin{eqnarray}
&\int^{\infty}_{0} \frac{L_{\nu}}{4\pi r^{2}} \exp(-\tau_{\nu}(r)) \Qabs  \pi a^{2} \ngr d\nu & \nonumber \\
& - \int^{\infty}_{0} 4\pi a^{2} \pi B_{\nu}(\Tgr) \Qabs \ngr d\nu - \Lambda_{\operatorname{g-gr}} & = 0. \label{eq:Tgr_eq}
\end{eqnarray}

\begin{figure*}
\begin{tabular}{cc}
\begin{minipage}{0.5\linewidth}
\includegraphics[clip,width=\linewidth]{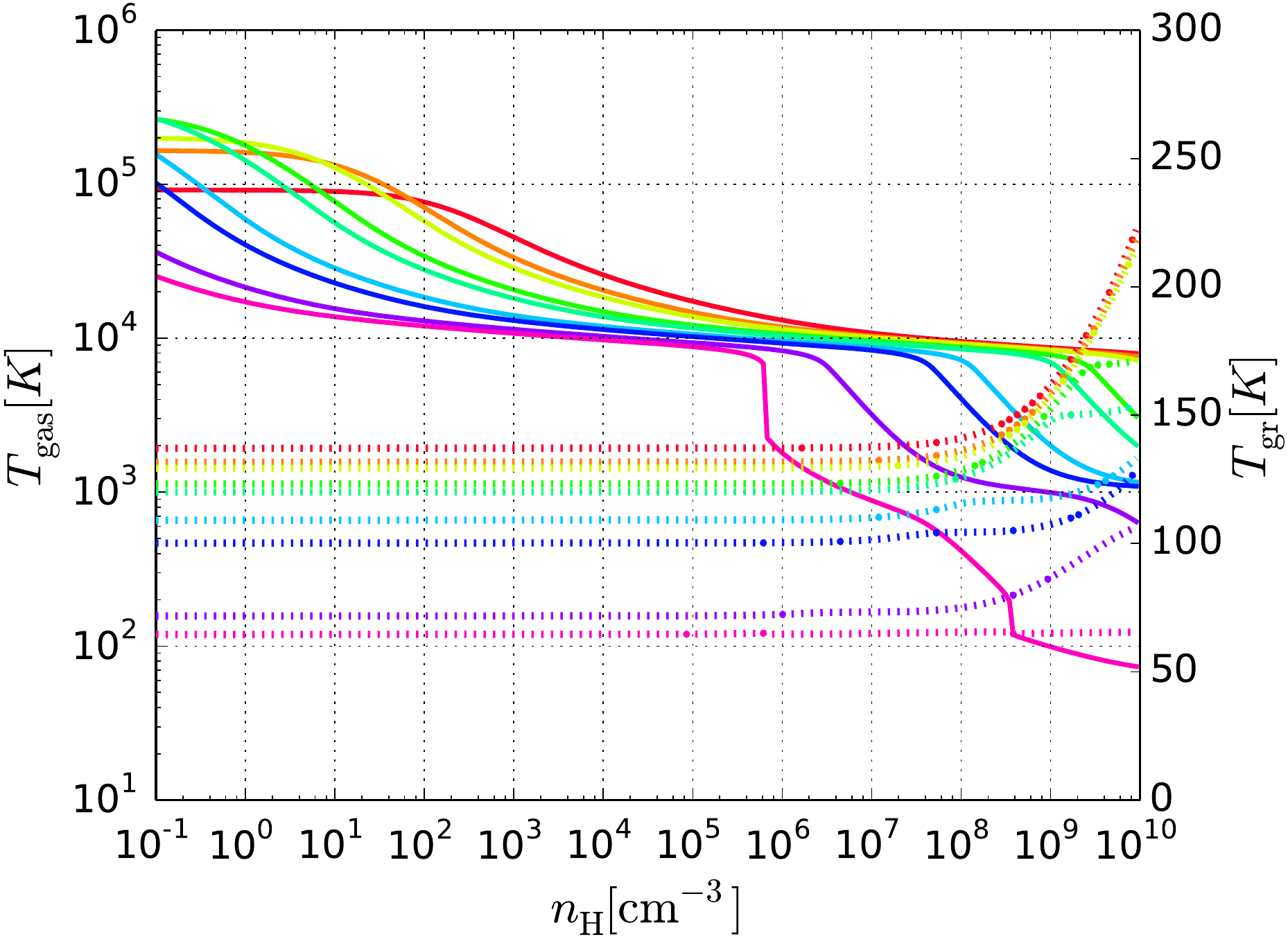}
\end{minipage}
\begin{minipage}{0.5\linewidth}
\includegraphics[clip,width=\linewidth]{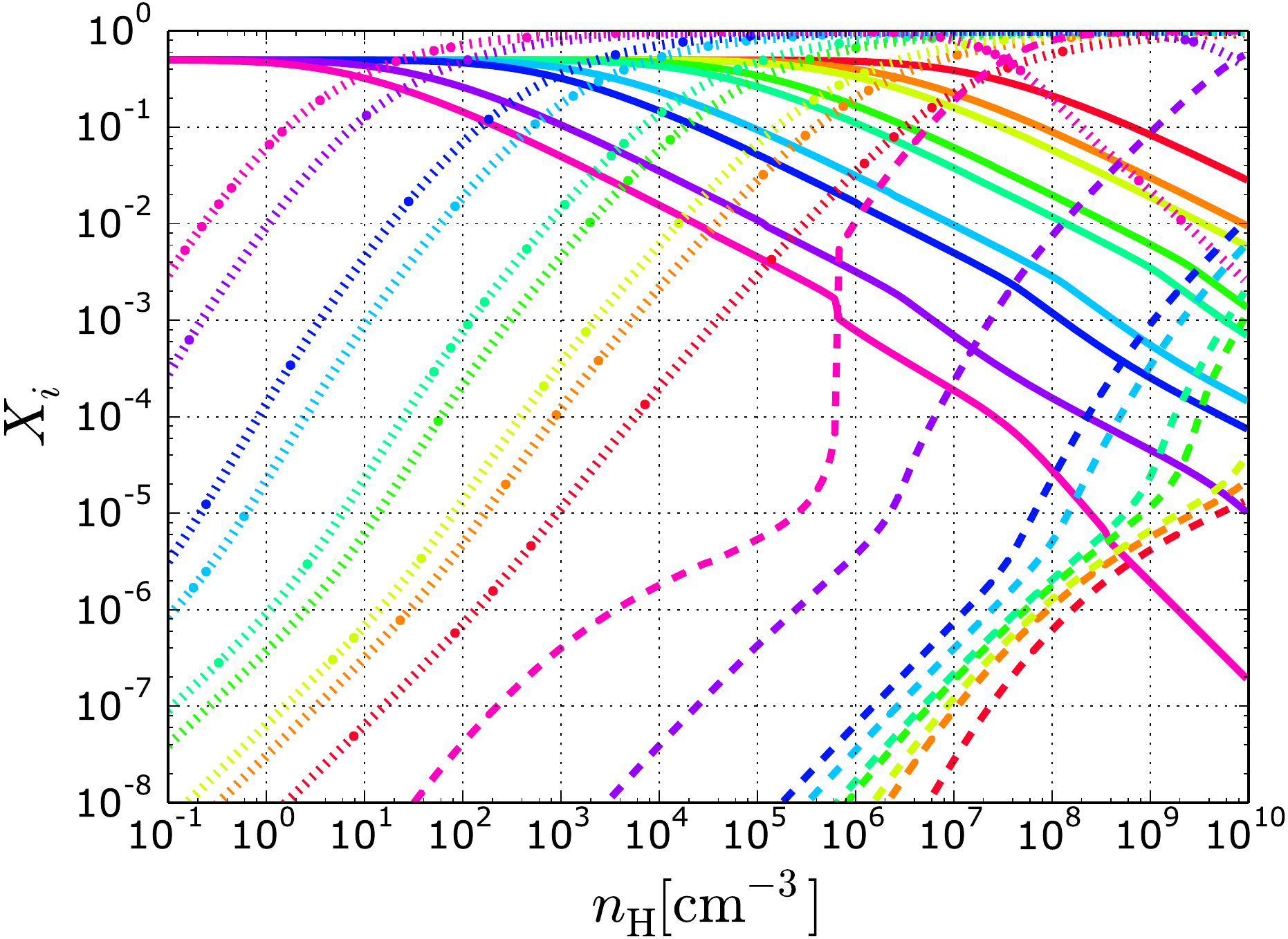}
\end{minipage}
\end{tabular}
\caption{Equilibrium temperatures and number fractions in the adopted ISM model are plotted as a function of $\nH$ for various $N^{\mathrm{obs}}_{\Hatm}$ assuming a fixed radiation field ($\Lbol = 5\times 10^{44}\;\mathrm{erg\;s^{-1}}$, $r=50\;\pc$), where $N^{\mathrm{obs}}_{\Hatm}$ is the obscuring $\Hatm$ column density. Obscuring gas contains dust with $f_{\mathrm{gr}}=0.01$ and its column density is simply used to attenuate an incident radiation field. The ISM is assumed to be at the rest and therefore hydrodynamic effects such as thermal expansion cooling is not taken into account. In both panels, the line colors, in the order of red to magenta, correspond to $N^{\mathrm{obs}}_{\Hatm}=10^{18}$, $5\times 10^{18}$, $10^{19}$, $5\times 10^{19}$, $10^{20}$, $5\times 10^{20}$, $10^{21}$, $5\times 10^{21}$, and $10^{22}\;\mathrm{cm^{-2}}$. \textit{Left}: the gas temperature (\textit{solid}; left $y$ axis) in the logarithmic scale and dust temperatures (\textit{dotted}; right $y$ axis) in the linear scale. Since the obscuring $\Hatm$ makes the incident spectrum hard by preferentially absorbing photons of energy near $13.6\;\mathrm{eV}$ and since the recombination cooling becomes inefficient in low density, $\Tgas$ becomes higher in low $\nH$ regime. $\Tgr$ increases with $\nH$ because the dust-gas coupling becomes tight. \textit{Right}: the number fractions of $e$ (\textit{solid}), $\Hatm$ (\textit{dotted}), and $\Hmol$ (\textit{dashed}).}
\label{fig:equilibrium_states_NHI_only}
\end{figure*}

\begin{figure*}
\begin{tabular}{cc}
\begin{minipage}{0.5\linewidth}
\includegraphics[clip,width=\linewidth]{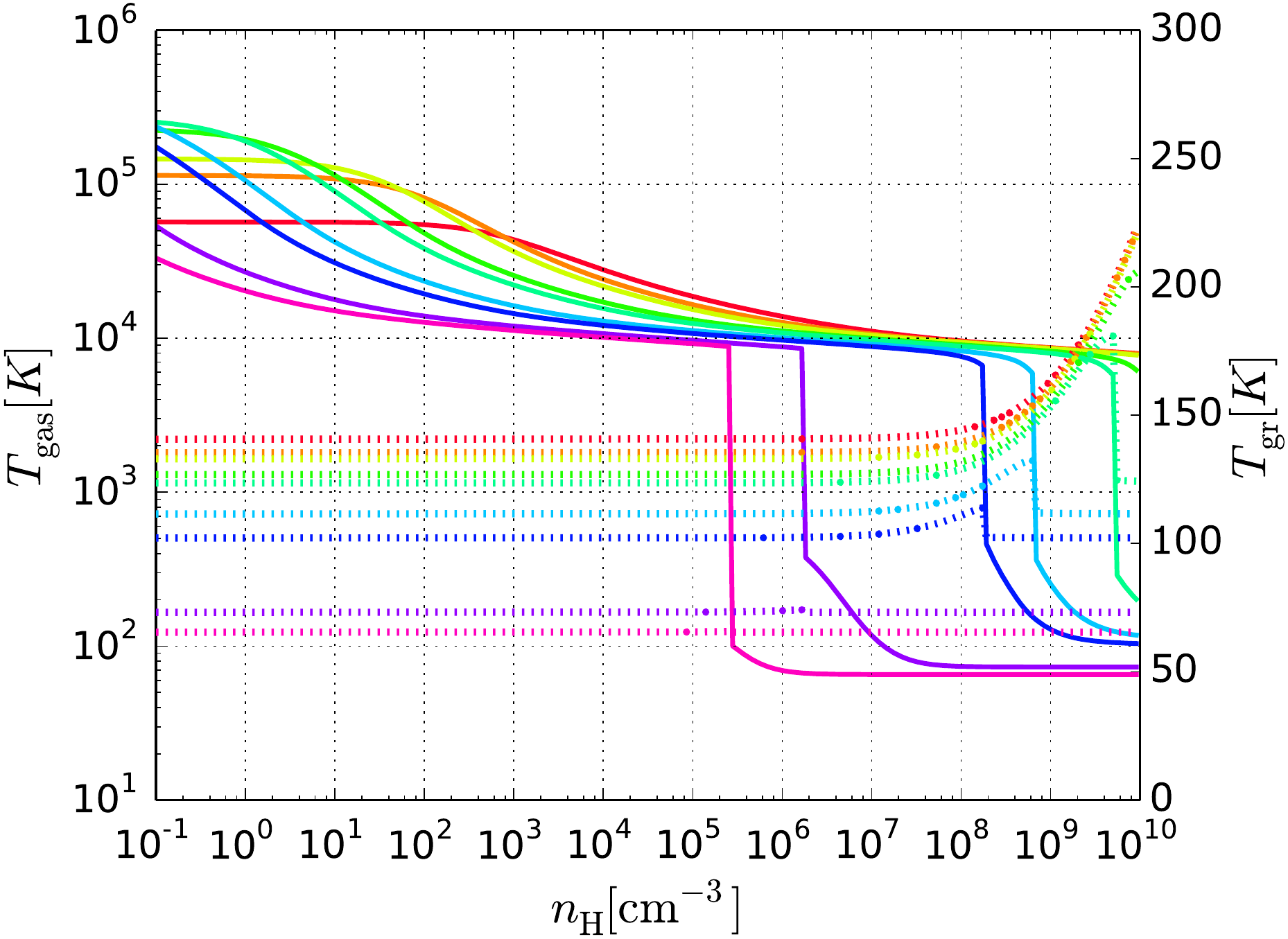}
\end{minipage}
\begin{minipage}{0.5\linewidth}
\includegraphics[clip,width=\linewidth]{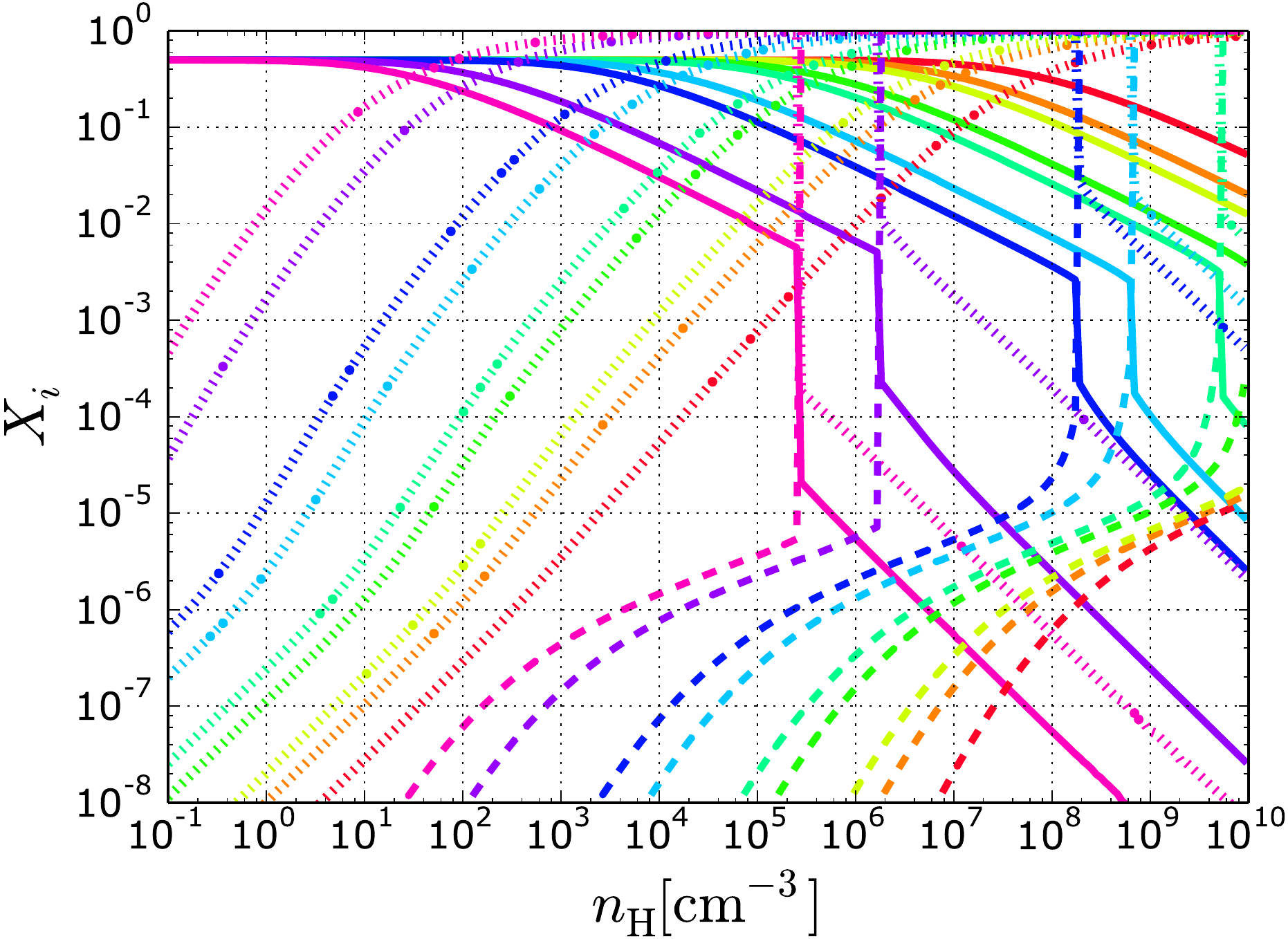}
\end{minipage}
\end{tabular}
\caption{The same as Fig.~\ref{fig:equilibrium_states_NHI_only}, but for the case that the obscuring column consists of $N^{\mathrm{obs}}_{\Hatm}$ and $N^{\mathrm{obs}}_{\mathrm{\Hmol}}$. In both panels, the line colors correspond to $N^{\mathrm{obs}}_{\mathrm{H}}=10^{18}$, $5\times 10^{18}$, $10^{19}$, $5\times 10^{19}$, $10^{20}$, $5\times 10^{20}$, $10^{21}$, $5\times 10^{21}$, and $10^{22}\;\mathrm{cm^{-2}}$. In all cases, the ratio of $N^{\mathrm{obs}}_{\Hatm}$ to $N^{\mathrm{obs}}_{\Hmol}$ is unity. Because of the self-shielding of $\Hmol$, $\Hmol$ formation occurs in lower $\nH$ compared to the cases in Fig.~\ref{fig:equilibrium_states_NHI_only}.}
\label{fig:equilibrium_states_NHI_NH2I_mix}
\end{figure*}

For the adopted ISM model and AGN SED, we calculate the equilibrium temperatures and number fractions for various obscuring column densities, assuming $\Lbol=5\times 10^{44}\;\mathrm{erg\;s^{-1}}$ and $r=50\;\pc$ and the results are shown in Fig.~\ref{fig:equilibrium_states_NHI_only} and \ref{fig:equilibrium_states_NHI_NH2I_mix}.

\section{Numerical Methods} \label{sec:numerical_methods}
In this section, we explain the numerical method in detail.

\subsection{Hydrodynamics and gravity} \label{subsec:hydro_gravity}
We use the pressure-energy SPH formulation\footnote{This formulation is a variant of the density-independent formulation proposed by \citet{saitoh13:_densit_formul_smoot_partic_hydrod} and can treat contact discontinuity more accurately than the standard SPH formulation (e.g., \citealt{lucy77,gingold77:_smoot}).} proposed by \citet{hopkins13:_lagran} to solve hydrodynamics. We employ the $M_{4}$ cubic spline kernel (\citealt{monaghan85}; see also \citealt{springel05:_gadget}) with the kernel gradient modification used in \citet{thomas92:_simul}. The artificial viscosity formulation proposed by \citet{monaghan97:_sph_rieman_solver} is used. To reduce artificial viscosity in smooth regions of the flow, we vary the viscous parameter $\alpha^{\mathrm{SPH}}_{\mathrm{vis}}$ within the range of $\alpha^{\mathrm{SPH}}_{\mathrm{vis}} \in [0.1,1]$ according to \citet{morris97:_switc_reduc_sph_viscos}. We also employ the Balsara switch (\citealt{balsara95:_neuman_stabil_analy_smoot_partic_hydrod}). The density estimates are computed by using the `gather` neighbors\footnote{The meanings of the `gather` and the `scatter` are referred to \citet{hernquist89:_trees}.}, while we calculate the pressure gradient, the artificial viscosity, and the time derivative of the internal energy by using the `gather-and-scatter` neighbors. The smoothing length is adjusted so that the number of the `gather` neighbors $N^{\mathrm{gat}}_{\mathrm{nb}} = 50\pm 2$.

The gravitational force is computed by the tree method with the opening angle of $\theta=0.5$ (\citealt{barnes86:_nlog_n}). We take into count monopole components only. We adopt the optimal binary tree structure described in
\citet{anderson99:_tree_data_struc_for_n_body_simul}.
In the construction of the interaction lists, we use the group interaction list technique (\citealt{barnes90:_modif_tree_code,makino91:_treec_special_purpos_proces}) and use the maximum side lengths of each tree node to evaluate the opening criterion. The calculation of the gravitational force is accelerated by the Phantom-GRAPE libraries which use Streaming SIMD Extensions (SSE) or Advanced Vector Extensions (AVX) instructions\footnote{In Cray XT4 system, we use the limited accuracy version of the Phantom-GRAPE library which is developed by Keigo Nitadori. Unfortunately, the detail of the library is not published as a paper, but we can find the detail description for the full accuracy version of the library in \citet{nitadori06:_perfor_n}. In Cray XC30 system, we use the AVX version of the Phantom-GRAPE library which is developed by \citet{tanikawa13:_phant_grape}. The Phantom-GRAPE library for collisional $N$-body system is also developed and see \citet{tanikawa12:_n_simd_advan_vector} for details if you are interested in it.}.

The code is parallelized with the Message Passing Interface (MPI).
The domain decomposition is done by the sampling method (e.g., \citealt{makino04:_fast_paral_treec_grape}).
The time integration is performed by the standard leapfrog method.

\subsection{Radiative transfer} \label{subsec:radtr}
We explain here our treatment of the radiative transfer (RT) calculation.
First of all, we summarize the approximations used in this study.
(1) We do not consider any scattering processes and effects of the Doppler-shift. In other words, we only take into account absorption processes in the simplest way, (2) we employ the so-called on-the-spot approximation for the ground state $\Hatm$ recombination photons (\citealt{osterbrock06:_astrop_gaseous_nebul_activ_galac_nuclei}), (3) we employ an approximate formula for the photo-dissociation rate of $\Hmol$, and (4) we assume optically-thin for photons arisen from the radiative processes listed in Table.~\ref{tbl:radiative_processes}. In the following, we explain the numerical treatment in more detail.

In the absence of scattering processes and Doppler-shift, the formal solution of the steady radiative transfer equation for a ray can be written as
\begin{eqnarray}
I_{\nu}(r) & = & I_{\nu}(0)\exp\left(-\sum_{i}\sigma^{i}_{\mathrm{abs}}(\nu)N_{i}(r)\right), 
\end{eqnarray}
where $\sigma^{i}_{\mathrm{abs}}(\nu)$ is the absorption cross section of species $i$ at the frequency $\nu$ and $N_{i}(r)$ is the column density of species $i$ at a distance of $r$ from the radiation source, respectively. Thus, the specific intensities at arbitrary positions are written as a function of the column densities. Using the point source approximation, the photo-ionization rate of HI, the photo-heating rates of HI and dust, and the radiative accelerations of HI and dust are given by
\begin{eqnarray}
k_{\mathrm{HI}}(r) & = & \nHI \int^{\infty}_{\nu_{\mathrm{L}}} \dfrac{L_{\nu} e^{-\tau_{\nu}}}{4\pi r^{2} h\nu}\sigma^{\mathrm{HI}}_{\mathrm{abs}}(\nu)\:d\nu, \label{eq:kHI} \\
\Gamma_{\mathrm{HI}}(r) & = & \nHI \int^{\infty}_{\nu_{\mathrm{L}}} \dfrac{L_{\nu} e^{-\tau_{\nu}}}{4\pi r^{2} h\nu}(h\nu-h\nu_{\mathrm{L}})\sigma^{\mathrm{HI}}_{\mathrm{abs}}(\nu) \:d\nu, \label{eq:GamHI} \\
\Gamma_{\mathrm{gr}}(r) & = & \ngr \int^{\infty}_{0} \dfrac{L_{\nu} e^{-\tau_{\nu}}}{4\pi r^{2}}\sigma^{\mathrm{gr}}_{\mathrm{abs}}(\nu)\:d\nu, \label{eq:GamDust} \\
\bmath{a}_{\mathrm{HI}}(r) & = & \nHI \dfrac{1}{c\rho} \int^{\infty}_{\nu_{\mathrm{L}}} \dfrac{L_{\nu} e^{-\tau_{\nu}}}{4\pi r^{2}} \sigma^{\mathrm{HI}}_{\mathrm{abs}}(\nu) \:d\nu, \label{eq:aHI} \\
\bmath{a}_{\mathrm{gr}}(r) & = & \ngr \dfrac{1}{c\rho} \int^{\infty}_{0} \dfrac{L_{\nu} e^{-\tau_{\nu}}}{4\pi r^{2}} \sigma^{\mathrm{gr}}_{\mathrm{abs}}(\nu) \:d\nu, \label{eq:aDust}
\end{eqnarray}
where $\nu_{\mathrm{L}}$ is the frequency at the Lyman limit, $L_{\nu}$ is the monochromatic luminosity of the radiation source, and $\tau_{\nu}=\sum_{i}\sigma_{\mathrm{abs}}^{i}(\nu)N_{i}(r)$.

As in \citet{tajiri98:_criter_photoion_pregal_cloud_expos}, \citet{hasegawa10:_start}, and \citet{okamoto12:_argot}, we make look-up tables for the integrals in Eq.(\ref{eq:kHI})-(\ref{eq:aDust}) except for the geometrical dilution factor $(4\pi r^{2})^{-1}$ as a function of the column densities, before we start the simulations. The frequency dependencies of the incident spectrum and relevant cross sections are properly taken into account when making the look-up tables. In the simulations, we evaluate the photo-ionization rate and the rest by interpolating from these pre-computed look-up tables to specified column densities and applying the geometrical dilution. The column densities of each species for each particle are computed by the tree-accelerated long characteristics method described in Appendix \ref{appendix:sec:TreeLONG}. The photo-ionization rate of HI and others are evaluated by the method described in Appendix \ref{appendix:sec:RT_optically_thick}. In this study, the look-up tables are made for ($N_{\Hatm}$, $N_{\mathrm{H}}$)-grid with $128$ grid points in each dimension.

In order to evaluate the photo-dissociation rate of $\Hmol$ accurately, we should solve the frequency-dependent radiative transfer equation of the Lyman-Werner band photons coupled with the $\Hmol$ level population equations. However, such computation is too costly to be coupled with hydrodynamics except for one-dimensional calculation. Therefore, we use an approximation formula given by \citet{draine96:_struc_of_station_photod_front}. They performed detailed radiative transfer calculations of a plane-parallel stationary photo-dissociation region and derived simple analytic approximate formulae for the photo-dissociation rate of $\Hmol$ as a function of $\Hmol$ column density $N_{\Hmol}$ and dust extinction. The approximate formula is written as
\begin{eqnarray}
k_{\mathrm{diss}}(N_{\Hmol})=k_{\mathrm{diss,0}}\nHtwo\exp(-\tau^{\mathrm{LW}}_{d})f_{\mathrm{sh}}(N_{\Hmol}), \label{eq:kH2I}
\end{eqnarray}
where $k_{\mathrm{diss,0}}$ is unshielded photo-dissociation rate, $\tau^{\mathrm{LW}}_{d}$ is the dust optical depth at the LW band, $f_{\mathrm{sh}}$ is the self-shielding function of $\Hmol$ and is defined as 
\begin{eqnarray}
f_{\mathrm{sh}}(N_{\Hmol})=\min\left[1,\left(\frac{N_{\Hmol}}{10^{14}\;\mathrm{cm^{-2}}}\right)^{-0.75} \right].
\end{eqnarray}
Following \citet{abel97:_model}, we adopt
\begin{eqnarray}
k_{\mathrm{diss,0}}=1.1\times 10^{8}F^{\mathrm{LW}}_{\nu},
\end{eqnarray}
where $F^{\mathrm{LW}}_{\nu}$ is the radiation flux at $h\nu_{\mathrm{LW}} = 12.87\;\eV$.
\citet{draine96:_struc_of_station_photod_front} gave an another more accurate approximate formula. Also, \citet{wolcott-green11:_photod} propose recently an improved version of this second formula. In this study, we do not use these formulae, because they contain a Doppler broadening parameter and it is difficult to adjust the parameter to optimal value in imhomogeneous medium.

The radiative acceleration by absorption of the $\Hmol$ photo-dissociating photon is calculated by
\begin{eqnarray}
|\bmath{a}_{\mathrm{rad},\Hmol}| = \frac{k_{\mathrm{diss}} h\nu_{\mathrm{LW}}}{c \rho}. \label{eq:aH2I}
\end{eqnarray}

Finally, we note that our approximate treatment of dust and $\Hmol$ somewhat overestimates energy and momentum transfer from radiation to matter in the Lyman-Werner band, since the change of the incident spectrum shape by absorption of the Lyman-Werner photons is not taken into account when we make the look-up tables.

\subsection{Non-equilibrium chemistry and radiative processes} \label{subsec:chemistry}
We solve time-dependent chemical reaction equations for $e^{-}$, $p^{+}$, $\Hatm$, $\Hmol$ coupled with the internal energy equation of gas implicitly. 

In the implicit method, a physical variable $A$ is integrated by solving iteratively the following equation,
\begin{eqnarray}
A^{t+\dtglobal} = A^{t} + \left(\frac{dA}{dt}\right)^{t+\dtglobal}\dtglobal,
\end{eqnarray}
where $\dtglobal$ is the global timestep with which the hydrodynamics is solved (defined later).
However, the straightforward implicit integration of the coupled equations is numerical unstable, because timescales of the chemical reactions and the internal energy are largely different from $\dtglobal$ in general. In order to solving the coupled equations efficiently and stably, we follow a similar approach adopted in \citet{whalen06:_multis_algor_for_the_radiat} and \citet{okamoto12:_argot}. The our approach is shown in Algorithm \ref{alg:chemistry}.
In the method, we divide $\dtglobal$ into subcycle. The number of subcycle is determined by dividing $\dtglobal$ by $\dtsub \equiv \min(\dtchem,\dtengy)$, where $\dtchem$ and $\dtengy$ are the chemical timestep and the internal energy timestep, respectively (these are also defined later). If $\dtsub$ are longer than $\dtglobal$, the number of the subcycle is taken to be 1. In the subcycle, we solve the time-dependent chemical reaction equations and the internal energy equation alternately for a fixed radiation field. After updating the chemical abundance and the internal energy, we perform the radiative transfer calculation using the previously updated physical variables. Thus, we update the matter field and the radiation field alternatively. This procedure is repeated until $\bmath{n}$, $\Tgas$, $\Tgr$, and $\bmath{a}_{\mathrm{rad}}$ are converged, where $\bmath{n}$ is the number density vector and $\bmath{a}_{\mathrm{rad}}$ is the radiative acceleration. In this study, we decide that the convergence is achieved, if the maximum relative difference of these physical variables over all the particles is less than $10^{-2}$. We do not find a significant difference between numerical results such as density distribution with this convergence criterion ($\epsilon^{\mathrm{tol}}_{\mathrm{rel}}=10^{-2}$) and with a more strict one (e.g., $\epsilon^{\mathrm{tol}}_{\mathrm{rel}}=10^{-3}$). 

\begin{algorithm}
\caption{Non-equilibrium chemistry}
\label{alg:chemistry}
\KwData{$k$ is the iteration number, \\
\hspace{2.5em} $\mathcal{P}$ is the set of the particles, \\
\hspace{2.5em} $\mathcal{M}^{(k)} = \{ p.\bmath{n}^{(k)}, p.\Tgas^{(k)}, p.\Tgr^{(k)}| p \in \mathcal{P} \} $ \\
\hspace{2.5em} is the matter variables, \\
\hspace{2.5em} $\mathcal{R}^{(k)} = \{ p.k^{(k)}_{\mathrm{ion}}, p.\Gamma^{(k)}_{\mathrm{ph}}, p.\arad^{(k)} | p \in \mathcal{P} \} $ \\
\hspace{2.5em} is the radiation fields, and \\
\hspace{2.5em} $\mathcal{C}^{(k)} = \{ p.\bmath{n}^{(k)}, p.\Tgas^{(k)}, p.\Tgr^{(k)}, p.\arad^{(k)} | p \in \mathcal{P} \}$ \\
\hspace{2.5em} is the variables that we want to converge.}
\BlankLine
RT calculation and obtain $\mathcal{R}^{(0)}$\;
save $\mathcal{M}^{(0)}$  \;
$k = 0$ \;
\Repeat{convergence achieved}{
   $k = k + 1$ \;
   \ForEach{$p \in \mathcal{P}$}{
      calculate $\Delta t_{\mathrm{sub}}$\;
      $t = t_{\mathrm{start}}$ \;
      \While{$t<t_{\mathrm{end}}$}{
         update chemical abundance \;
         update gas and dust temperatures \;
         $t = t+\Delta t_{\mathrm{sub}}$ \;
         update $\Delta t_{\mathrm{sub}}$ \;
      }
   }
   RT calculation and obtain $\mathcal{R}^{(k)}$\;
   compare $\mathcal{C}^{(k)}$ with $\mathcal{C}^{(k-1)}$ and calculate the maximum relative difference $\epsilon^{\max}_{\mathrm{rel}}$\;
   \uIf{$\epsilon^{\max}_{\mathrm{rel}} \leq \epsilon^{\mathrm{tol}}_{\mathrm{rel}}$}{
      exit the infinite loop\;
   }
   \Else{
      save $\mathcal{C}^{(k)}$ \;
      reset $\mathcal{P}$ using $\mathcal{M}^{(0)}$ and $\mathcal{R}^{(k)}$\;
   }
}
\end{algorithm}

We use the $\alpha$-QSS method (\citealt{mott00:_quasi_stead_state_solver_stiff}) to solve the chemical reaction equations and the internal energy equation. The $\alpha$-QSS method is a predictor-corrector type integrator and is originally developed to solve stiff chemical reaction equations. In actual calculations, we solve time evolution of the gas temperature rather than the internal energy. The conversion of the equation is performed assuming that a change of $\gammaeff$ is small. In the following, we describe the adopted method in terms of the number density and the gas temperature. In order to apply the $\alpha$-QSS method, we split source terms into a increasing rate and a decreasing rate and rewrite the equations in the form of
\begin{eqnarray}
\frac{dn_{i}}{dt} & = & C_{i} - D_{i}n_{i}, \label{eq:2} \\
\frac{d\Tgas}{dt} & = & C-D\Tgas, 
\end{eqnarray}
where $n_{i}=\rho Y_{i}/m_{i}$, $\rho$ is the mass density of gas, $Y_{i}$ is the mass fraction of species $i$, $m_{i}$ is the mass of species $i$. The index $i$ is either of $e^{-}$, $p^{+}$, $\Hatm$, $\Hmol$ in our study. The $C_{i}$ are the collective source terms responsible for the creation of species $i$. The second terms $D_{i}n_{i}$ in the equation (\ref{eq:2}) represent the destruction for species $i$. $C$ and $D$ are defined as such that $C$ and $D \Tgas$ is the increasing and decreasing rate of the gas temperature, respectively. $C_{i}$, $D_{i}$, $C$, and $D$ are the functions of $\Tgas$, $\Tgr$, $\bmath{n}$. But, hereafter, we omit dependency for brevity.

The integration scheme for the gas temperature is written as
\begin{eqnarray}
\Tgas^{p} & = & \Tgas^{0} + \frac{\Delta t(C^{0}-D^{0}\Tgas^{0})}{1+\alpha \Delta t D^{0}}, \quad (\mathrm{Predictor}) \label{eq:aQSS_predictor} \\
\Tgas^{c} & = & \Tgas^{0} + \frac{\Delta t(\widetilde{C}-\overline{D}\Tgas^{0})}{1+\overline{\alpha}\Delta t \overline{D}}, \quad (\mathrm{Corrector}) \label{eq:aQSS_corrector}
\end{eqnarray}
where $\Tgas^{0}$ is the gas temperature at the present time, $C^{0}\equiv C(\Tgas^{0},\Tgr^{0})$, $D^{0}\equiv D(\Tgas^{0},\Tgr^{0})$, and $\alpha$ is defined as
\begin{eqnarray}
\alpha(D\Delta t) = \frac{180r^{3} + 60r^{2} + 11r + 1}{360r^{3} + 60r^{2} + 12r + 1}, 
\end{eqnarray}
where $r=1/(D\Delta t)$.
$\overline{D}$ and $\widetilde{C}$ in the corrector are defined as
\begin{eqnarray}
\overline{D}  & = & \frac{1}{2}(D^{0}+D^{p}), \\
\widetilde{C} & = & \overline{\alpha} C^{p} + (1-\overline{\alpha})C^{0},
\end{eqnarray}
where $C^{p}\equiv C(\Tgas^{p},\Tgr^{p})$, $D^{p}\equiv D(\Tgas^{p},\Tgr^{p})$, and $\overline{\alpha}=\alpha(\overline{D}\Delta t)$. $\Tgr^{p}$ and $\Tgr^{c}$ are computed by equation (\ref{eq:Tgr_eq}) using $\Tgas^{p}$ and $\Tgas^{c}$, respectively. The accuracy of the solution can be improved by the multiple corrections. In this study, we continue to correct until the relative difference becomes less than $10^{-5}$. We note that equations (\ref{eq:aQSS_predictor}) and (\ref{eq:aQSS_corrector}) are also definitely positive. Finally, we note that we need to carefully classify the dust-gas energy transfer term into $C$ or $D$, because its sign changes depending on $\Tgas$ and $\Tgr$.

The integration scheme of the number density used in this study is basically the same as that for the gas temperature. One difference is the existence of the normalization step: the predicted and corrected number densities, $n^{p}_{i}$ and $n^{c}_{i}$, are scaled to satisfy $\rho=\sum_{i}m_{i}n^{p}_{i}=\sum_{i}m_{i}n^{c}_{i}$.

We determine the chemical timestep and the internal energy timestep by
\begin{eqnarray}
\dtchem & = & 0.01\min_{i}\frac{(n_{i}+n_{\min})}{\dot{n}_{i}}, \\
\dtengy & = & 0.01\frac{e_{\mathrm{th}}}{\dot{e_{\mathrm{th}}}}, \label{eq:energy_timestep}
\end{eqnarray}
where $e_{\mathrm{th}}$ is the internal energy per unit volume and $n_{\min}=0.001n_{\mathrm{H}}$.
By introducing $n_{\min}$, we avoid a very small timestep, which occurs when the number density of a species is nearly zero.

\subsection{Global timestep} \label{subsec:timestep}

Following \citet{monaghan97:_sph_rieman_solver}, we determine the hydrodynamic timestep of particle $i$ by 
\begin{eqnarray}
\dthydi = \frac{C_{\mathrm{hyd}}h_{i}}{v^{\mathrm{sig}}_{i}},
\end{eqnarray}
where $h_{i}$ is the smoothing length, $C_{\mathrm{hyd}}=0.25$, and $v^{\mathrm{sig}}_{i}$ is the local maximum signal velocity, which is defined as
\begin{eqnarray}
v^{\mathrm{sig}}_{i}=\max_{j} (c_{s,i}+c_{s,j}-3w_{ij}),
\end{eqnarray}
where $j$ denotes the indices of neighbor particles, $c_{s,i}$ and $c_{s,j}$ are the adiabatic sound speeds of particles $i$ and $j$, and $w_{ij}=\min (0,\bmath{v}_{ij}\cdot\bmath{r}_{ij}/|\bmath{r}_{ij}|)$ is the relative velocity projected onto the separation vector $\bmath{r}_{ij}\equiv \bmath{r}_{i}-\bmath{r}_{j}$, where $\bmath{v}_{ij}=\bmath{v}_{i}-\bmath{v}_{j}$.

We determine the gravitational timestep of particle $i$ by
\begin{eqnarray}
\dtgrvi = \frac{C_{\mathrm{grv}}\max(|\bmath{v}_{i}|,c_{s,i})}{|\bmath{a}_{i}|},
\end{eqnarray}
where $\bmath{a}_{i}$ is the total acceleration and $C_{\mathrm{grv}}=0.1$. The introduction of the sound speed in the numerator is intended to avoid $\dtgrvi=0$.

The global timestep $\dtglobal$ is calculated by $\min_{i}(\dthydi,\dtgrvi)$. However, if we use $\dtglobal$ directly, it often happens that the iteration in the calculation of the non-equilibrium chemistry is not converged, because $\dtglobal$ is usually much larger than the timescales of chemical reactions and change of the internal energy. Therefore, we restrict $\dtglobal$ by $\dtlimit$, which is calculated as
\begin{eqnarray}
\dtlimit = \dtglobal^{\mathrm{prev}}\times \left\{
\begin{array}{ll}
2^{1/8}  &, N_{\mathrm{iter}} \leq 4, \\
2^{-1/8} &  N_{\mathrm{iter}} \geq 6,
\end{array}
\right.
\end{eqnarray}
where $\dtglobal^{\mathrm{prev}}$ is the global timestep at the previous step, $N_{\mathrm{iter}}$ is the iteration number at the previous step. This device adjusts the global timestep automatically so that the iteration number is close to $4\sim 6$.

\subsection{Temperature floor} \label{subsec:temperature_floor}
In the simulations, a dense shocked layer is formed in the gas cloud by the counteraction of the thermal expansion of the gas at the irradiation surface or the radiation pressure force acting on the irradiation surface. Insufficient mass resolution induces artificial self-gravitational fragmentation in the shocked layer. In order to avoid it, we impose a lower limit for the gas temperature on each particle according to
\begin{eqnarray}
\Tgas^{\min}(\rho) = 6^{2/3}\frac{\mu m_{\mathrm{H}}}{k_{\mathrm{B}}}\left(\frac{G^{3}\rho(2N^{\mathrm{gat}}_{\mathrm{nb}}\times m_{\mathrm{SPH}})^{2}}{\gammaeff^{3}\pi^{5}}\right)^{1/3}\;\mathrm{K}, \label{eq:temperature_floor}
\end{eqnarray}
where $m_{\mathrm{SPH}}$ is the mass of the SPH particle.
The almost same temperature floor is used in \citet{R.saitoh:06}.
We note that $\gammaeff$ in the right-hand side depends on the gas temperature and therefore we need an iteration to determine the minimum temperature and also note that we compute $\dtsub$ by $\dtchem$ only for the SPH particle at which the temperature floor is active.

\subsection{Initial Condition}
Initial uniform SPH particle distributions are obtained by cutting a sphere of $\approx 2^{18}$ particles from hydrodynamically-relaxed periodic cube, and scaling the mass and the position of each particle to fit the specified model parameters. Then, the cloud is shifted so that the cloud center is located at $(x,y,z)=(50\;\pc,0,0)$.

In all the simulation runs, we assume (i) that the initial gas and dust temperatures are $\Tgas =100\;\mathrm{K}$ and $\Tgr =20\;\mathrm{K}$, respectively, and (ii) that the length of the gravitational softening is $50\;\mathrm{AU}$. In this resolution, the Str{\"o}mgren length corresponds to $3\ell_{\mathrm{mid}}$ in the case of $\mathcal{N}_{S}=20$, where $\ell_{\mathrm{mid}}$ is the mean inter-particle distance. The numerical results are not sensitive to the initial temperatures.

\section{Numerical Results} \label{sec:numerical_results}
Here, we first explain in \S~\ref{subsec:Time_Evolution_No_U_models} the time evolution of the clouds without the AGN radiation to make clear the effects of the AGN radiation. Next, we show the results of Low-$\mathcal{U}$ models in \S~\ref{subsec:Time_Evolution_Low_U_models} and High-$\mathcal{U}$ models in \S~\ref{subsec:Time_Evolution_High_U_models}, respectively. Then, the results of \SCmodel are shown in \S~\ref{subsec:SC_models}. Finally, in \S~\ref{subsec:cloud_evaporation}, we compare the time evolution of dense gas fraction in each model.

\subsection{No-$\mathcal{U}$ models} \label{subsec:Time_Evolution_No_U_models}
\begin{figure}
\centering
\includegraphics[clip,width=\linewidth]{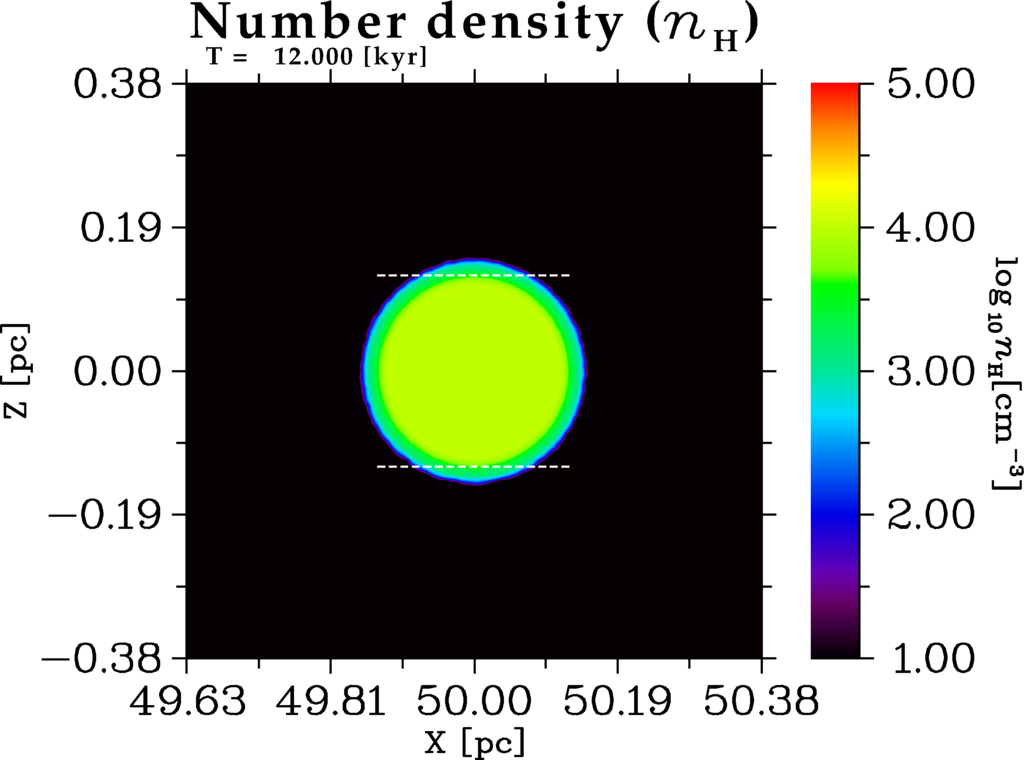}
\caption{The number density distribution of the model L05 at $t=12\;\kyr$ \textit{without} AGN radiation. The white dotted lines show the initial cloud size.}
\label{fig:L05_noU}
\end{figure}

Since their Jeans ratios are larger than unity (see Table.~\ref{tbl:simulation_runs}), the clouds in the models L05, L10, L20, and H05 expand by their thermal pressure without the AGN radiation. This will make it difficult to estimate the contribution of the photo-heating in cloud expansion. In order to make clear the effects of the AGN radiation, we first perform a numerical simulation \textit{without} the AGN radiation for the cloud model L05, which has the largest $r_{\mathrm{J}}$. The simulation is performed to $t=12\;\kyr$, which are the same as the final calculation time shown in \S~\ref{subsec:Time_Evolution_Low_U_models}. Figure \ref{fig:L05_noU} shows the number density slice of the model L05 at $t=12\;\kyr$. The initial cloud size is shown by the white dotted lines in the figure. We can see from this figure that the cloud radius is virtually constant until $t=12\;\kyr$. Thus, the thermal expansion of the cloud is negligible.

\subsection{Low-$\mathcal{U}$ models} \label{subsec:Time_Evolution_Low_U_models}
\begin{figure*}
\centering
\includegraphics[clip,width=\linewidth]{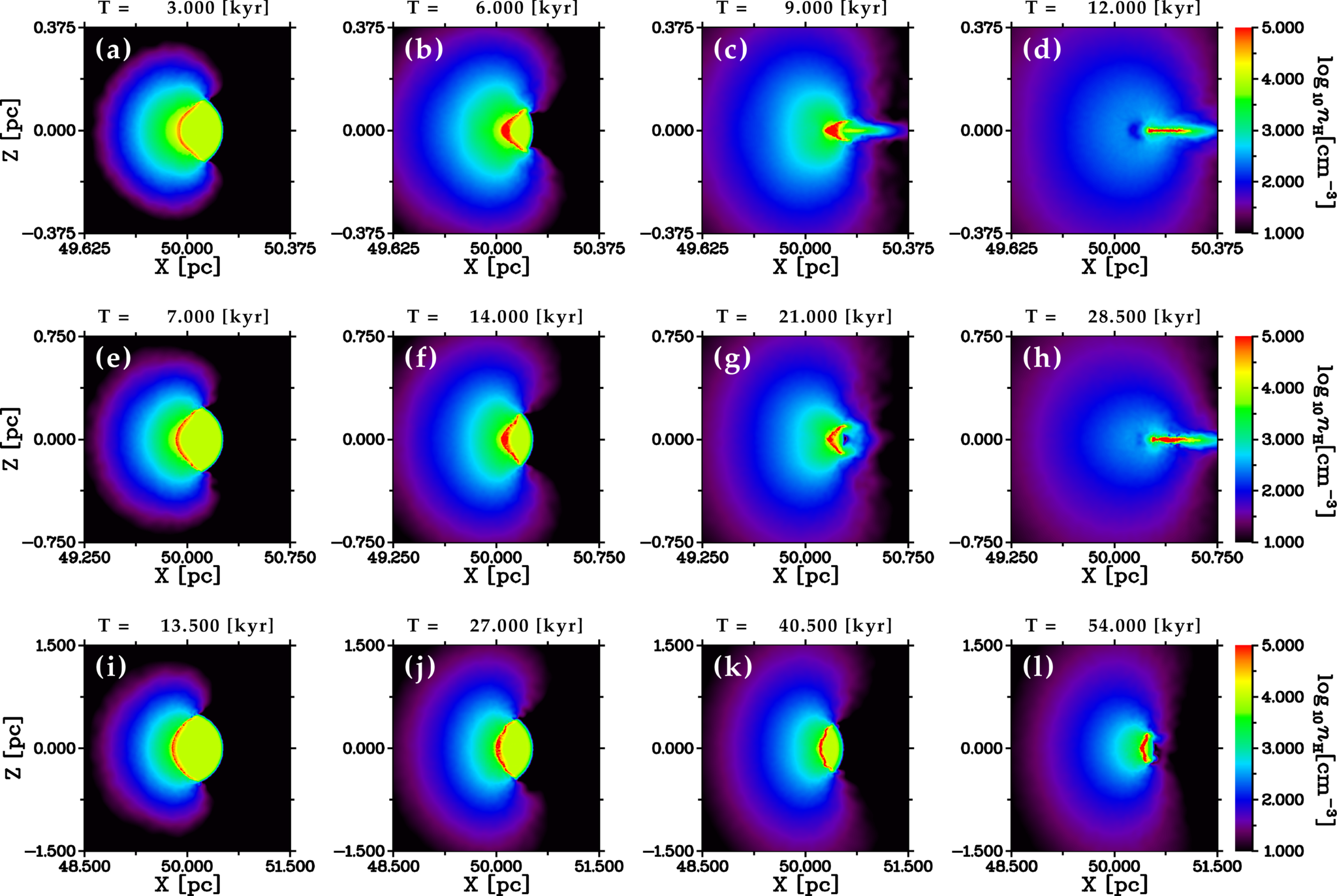}
\caption{Time evolution of the number density distribution in Low-$\mathcal{U}$ models (From the top to the bottom, L05, L10, and L20 are plotted). The AGN, the radiation source, is located at $(x,y,z)=(0,0,0)$. The each panel shows the number density distribution at $y=0$ slice. We show the calculation time on the top of each panel. For the spherical very low density regions found in the models L10 and L05 at a later calculation time, see the text.}
\label{fig:Time_Evolution_Low_U_models}
\end{figure*}

Figure \ref{fig:Time_Evolution_Low_U_models} shows the time evolution of the number density distribution of Low-$\mathcal{U}$ models at the $y=0$ plane. Immediately after the simulation starts, most of the molecular hydrogen from the irradiated face to a depth of $N_{\mathrm{H}}\approx 10^{22}\;\mathrm{cm^{-2}}$ is photo-dissociated and is turned into the atomic hydrogen. This atomic hydrogen is also rapidly ionized and the photo-ionization region forms at the irradiated face of the cloud. During this initial evolution, a distinct dense gas layer is formed at the depth of $N_{\mathrm{H}} \approx 5\times 10^{21}\;\mathrm{cm^{-2}}$, which corresponds to the Str{\"o}mgren length $l_{S}$,
\begin{eqnarray}
l_{S} & = & 0.15\;\pc \left(\dfrac{\Lion}{1.25\times 10^{44}\;\mathrm{erg\;s^{-1}}}\right) \nonumber\\
&& \quad \times \left(\dfrac{\nH}{10^{4}\;\mathrm{cm^{-3}}}\right)^{-2}\left(\dfrac{r}{50\;\pc}\right)^{-2}.
\end{eqnarray}
Most of the $\Hatm$-photo-ionizing photons are absorbed by this layer. Because the photo-heating rates are large in this layer, the photo-evaporation flow occurs at the layer. The counteraction of the photo-evaporation flow pushes the layer and convert it into the D-type shock (e.g., Fig.~\ref{fig:Time_Evolution_Low_U_models}a). For the model L20, the time-averaged velocity of the shocked layer, $\overline{v}_{\mathrm{sh}}$, is $\approx 18\;\kms$ and is larger than the estimated value by the razor-thin approximation $\vshapp(\Lbol=1.25\times 10^{44}) = 9.4\;\kms$. This enhancement is due to the so-called rocket effect (\citealt{oort55:_accel_inter_cloud_o_stars}). On the other hand, $\overline{v}_{\mathrm{sh}}\approx 20.8\;\kms$ for the model L05. Thus, the rocket effect is more effective for smaller $\mathcal{N}_{S}$ model.

\begin{figure*}
\centering
\includegraphics[clip,width=\linewidth]{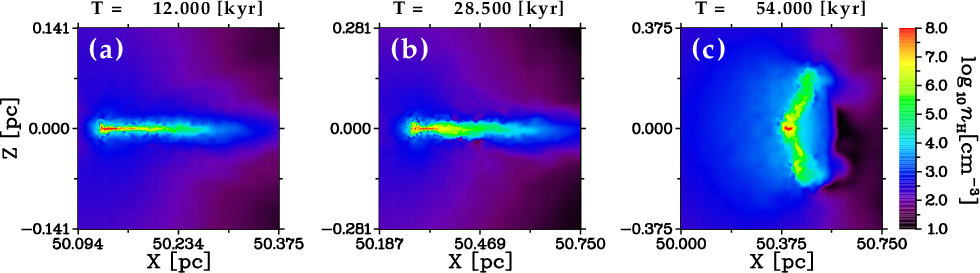}
\caption{Enlarged views of Fig.~\ref{fig:Time_Evolution_Low_U_models}d,h,l. Note that the upper limit of the color bar is different from Fig.~\ref{fig:Time_Evolution_Low_U_models}.}
\label{fig:Low_U_models_Density_enlarged}
\end{figure*}

The direction of the photo-evaporation flow is roughly spherically-outward and its back reaction, by necessity, points spherically inwards. Therefore, the shocked layer gradually bends and takes the shape of an circular cone (e.g., Fig.~\ref{fig:Time_Evolution_Low_U_models}f). As the calculation advances, it continues to collapse and finally forms a very dense molecular filament (Fig.~\ref{fig:Time_Evolution_Low_U_models}d, \ref{fig:Time_Evolution_Low_U_models}h). In the model L10, its number density is $\nH \approx 10^{5}\operatorname{-}10^{8}\;\mathrm{cm^{-3}}$ at $T=28.5\;\kyr$ (Fig.~\ref{fig:Time_Evolution_Low_U_models}h, Fig.~\ref{fig:Low_U_models_Density_enlarged}). A part of the photo-evaporation flow comes around behind the cloud and collides with itself. As a consequence, it takes part in the formation of the dense filament. Since self-shielding is effective in the filament, it promptly becomes neutral. After the dense filament is formed, the rate of the photo-evaporation from the irradiated face decreases rapidly. It is mainly because the surface area of the filament is very small in this stage. The high density of the filament may be another factor, since the $\Hatm$ recombination cooling and the $\Hmol$ formation becomes efficient and consequently $\Tgas$ and $X_{\Hatm}$ becomes small. By this stage, a considerable fraction of the initial gas is evaporated. Thus, the evolution of the gas clouds are mainly determined by the photo-evaporation in Low-$\mathcal{U}$ models.

\begin{figure*}
\centering
\includegraphics[clip,width=\linewidth]{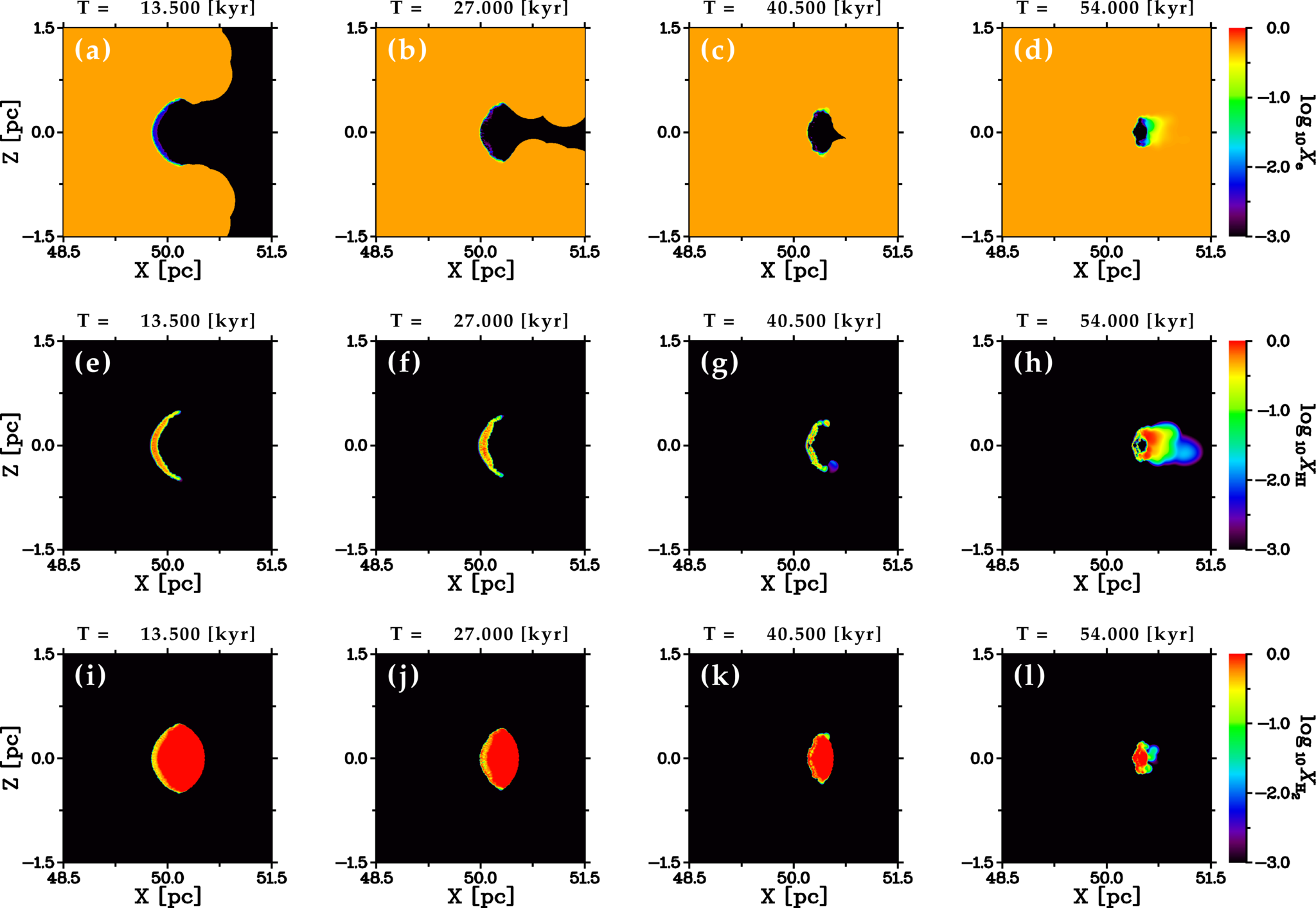}
\caption{Time evolution of the number fraction $X_{i}$ of the model L20 at the $y=0$ slice.
From the tow row to the bottom row, $X_{e}$, $X_{\Hatm}$, and $X_{\Hmol}$ are shown.}
\label{fig:Time_Evolution_L20_Xi}
\end{figure*}

In order to show the chemical structure in the photo-evaporation flow, we show the time evolution of the number fraction $X_{i}$ distribution of the model L20 in Figure~\ref{fig:Time_Evolution_L20_Xi}. From the comparison Fig.~\ref{fig:Time_Evolution_L20_Xi}a-d with Fig.~\ref{fig:Time_Evolution_Low_U_models}i-l, the low density ($\nH \lesssim 10^{2}\;\mathrm{cm^{-3}}$) gas that encloses the dense filament is ionized completely. The irradiated side of the shocked gas layer predominantly consists of $\Hatm$ (Fig.~\ref{fig:Time_Evolution_L20_Xi}e) and the opposite side of the shocked layer consists of $\Hatm$-$\Hmol$ mixtures. 

\begin{figure*}
\centering
\includegraphics[clip,width=\linewidth]{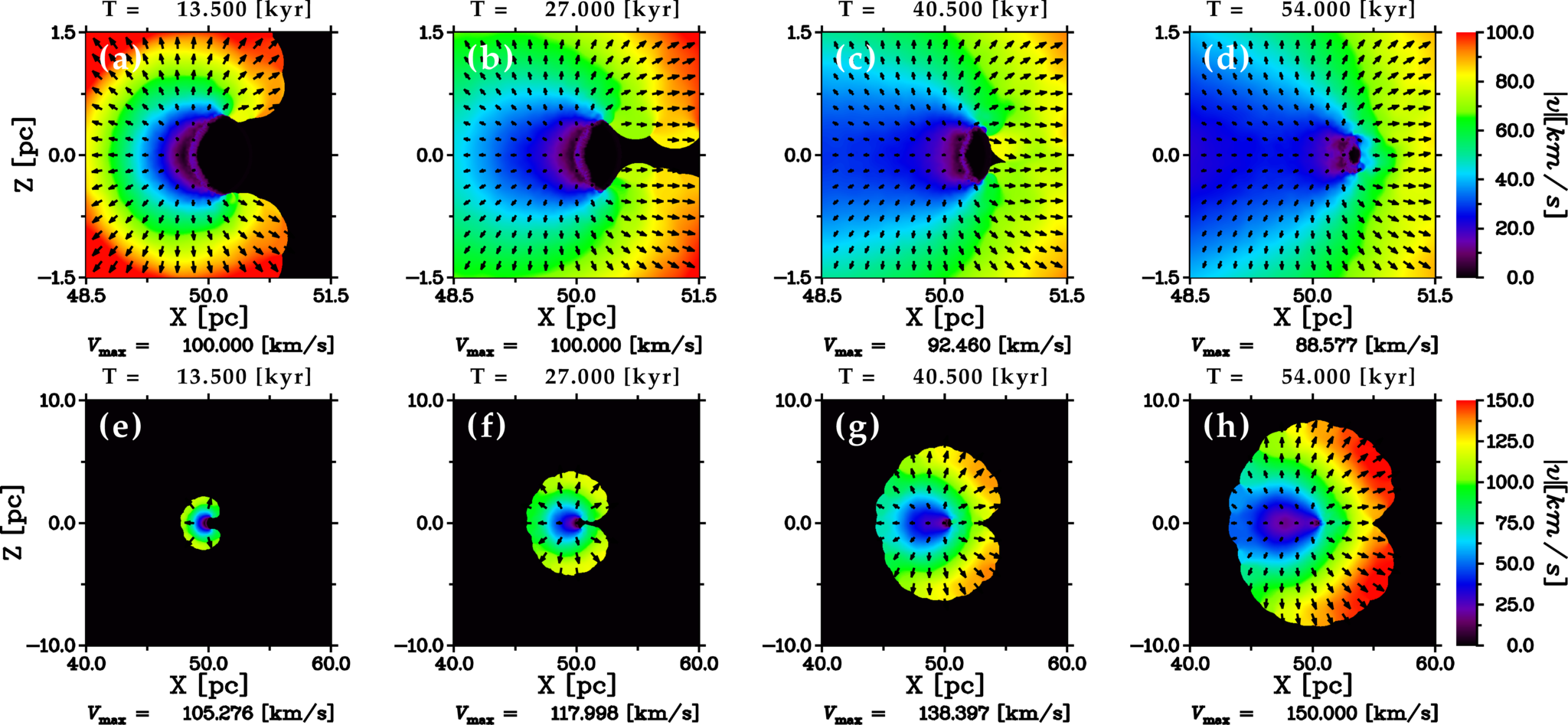}
\caption{Time evolution of the velocity fields of the model L20 at the $y=0$ slice. The lower panels are extensive versions of the upper panels. The black arrow show the velocity vector projected onto the $xz$ plane and the maximum arrow length corresponds to $v_{\max}$ which is described in the bottom of each panel.}
\label{fig:Time_Evolution_L20_Velocity}
\end{figure*}

Figure~\ref{fig:Time_Evolution_L20_Velocity} shows the time evolution of the velocity field at different plotting scales. The velocity field is almost spherical at a early time ($T<27\;\kyr$). At later times, the evaporated gas drifts slightly toward the opposite direction to the AGN because of the radiation pressure. The velocity of the front of the photo-evaporation flow is more than $100\;\kms$ and is supersonic.

\subsection{High-$\mathcal{U}$ models} \label{subsec:Time_Evolution_High_U_models}
\begin{figure*}
\centering
\includegraphics[clip,width=\linewidth]{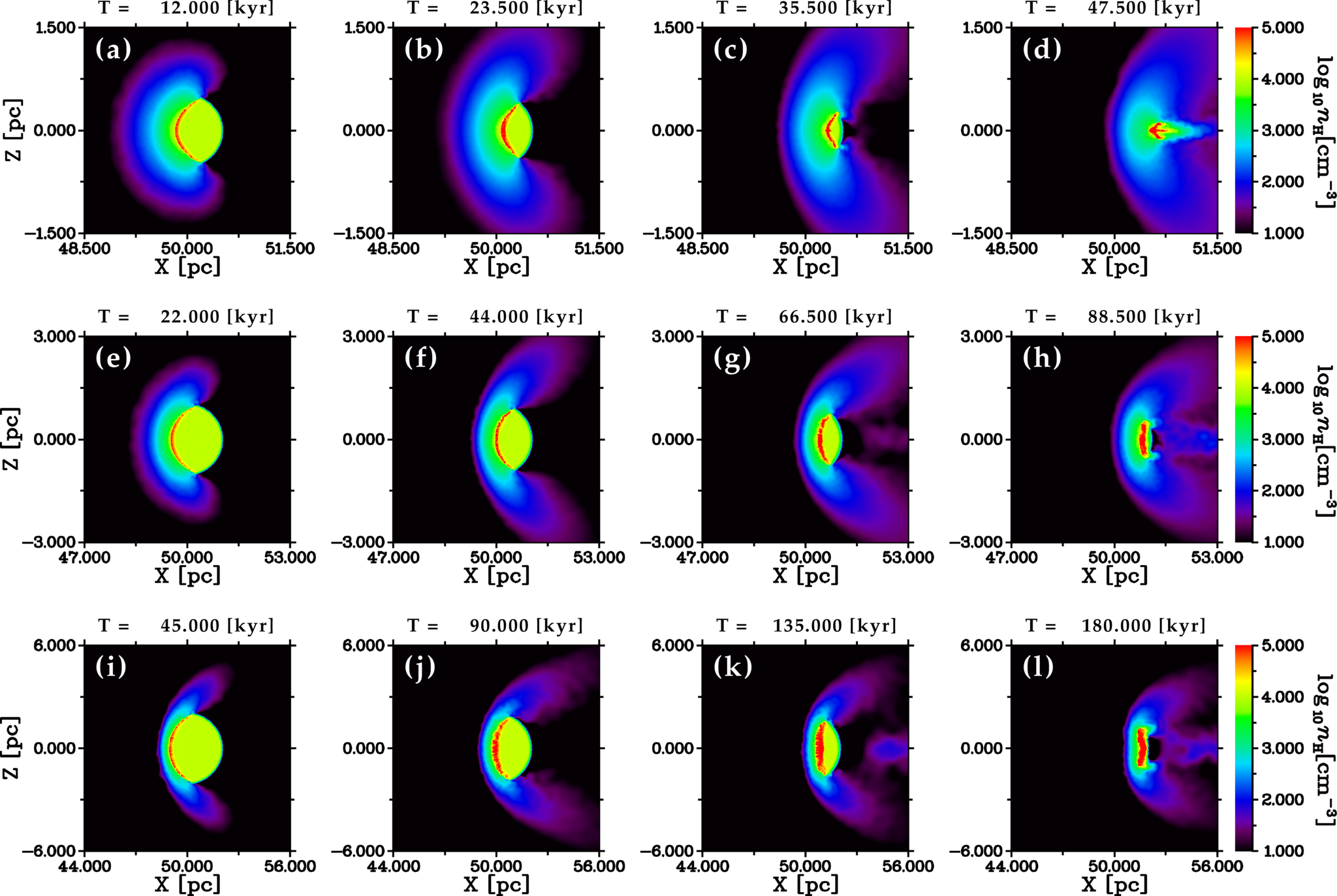} 
\caption{The same as Fig.~\ref{fig:Time_Evolution_Low_U_models},
but for High $\mathcal{U}$ models (\textit{top}:H05, \textit{middle}:H10, \textit{bottom}:H20).}
\label{fig:Time_Evolution_High_U_models}
\end{figure*}

Figure \ref{fig:Time_Evolution_High_U_models} shows the time evolution of the number density distribution of High-$\mathcal{U}$ models at the $y=0$ slice. As in the Low-$\mathcal{U}$ models, High-$\mathcal{U}$ models also show the formation of the shocked layer at a very early time. The photo-evaporation flow launched from the shocked layer interacts with the incident photons. Since the total force, which is mainly sum of the pressure-gradient force and the radiation force (the self-gravity is much smaller than these), is parallel to the $x$-axis near the central part of the irradiated face, the photo-evaporation flow cannot expand in the direction of the AGN. On the other hand, at the outskirts of the irradiated face (at the distance of $\approx \rcl$ from the $x$-axis), the radiation force bends the photo-evaporation flow in the opposite direction of the AGN (e.g., Fig.~\ref{fig:Time_Evolution_High_U_models}a). Thus, the radiation pressure stripping occurs. We can verify these things in the left panel of Fig.~\ref{fig:Acceleration_H20} in which the acceleration fields of the model H20 at $t=90\;\kyr$ is shown.

The shocked layer moves in the direction opposite to the AGN, keeping its shape almost flat except for $\mathcal{N}_{S}=5$ case, which shows an evolution similar to those of Low-$\mathcal{U}$ models. For the model H20, $\overline{v}_{\mathrm{sh}}$ is almost constants until it crosses the cloud and is $\approx 22\;\kms$, which is roughly consistent with $\vshapp(\Lbol=5\times 10^{44}) \approx 18.8\;\kms$. For the model H05, $\overline{v}_{\mathrm{sh}}\approx 21\;\kms$. Thus, $\overline{v}_{\mathrm{sh}}$ is nearly independent of $\mathcal{N}_{S}$ in High-$\mathcal{U}$ models and the rocket effect seems to not be effective compared to Low-$\mathcal{U}$ models. The mass and thickness of the shocked layer gradually increases with time. At a later time, the unirradiated part of the shocked layer becomes self-gravitationally unstable and some part of gas start to collapse. This is seen in Fig. \ref{fig:Time_Evolution_H20_rhoT} which shows the time evolution of $\nH\operatorname{-}\Tgas$ plane of the model H20. At $t=180\;\kyr$, a part of gas move toward an upper right direction in the $\nH\operatorname{-}\Tgas$ plane along the line that corresponds to the temperature floor equation (\ref{eq:temperature_floor}). This migration occurs because the mass of the self-gravitationally unstable gas increases with time and therefore a higher pressure is progressively needed to support the cloud. All of the gas with $\nH>10^{8}\;\mathrm{cm^{-3}}$ concentrates into a single gas clump, which is located at $(51.76\;\pc,0,0)$ and has a disk-like shape whose major axes lie in the $yz$ plane. The total mass and the diameter of this clump are $211\;\Msolar$ and $0.084\;\pc$, respectively. The hydrogen column density for the clump is $N^{\mathrm{H}}_{\mathrm{col}}>5.5\times 10^{24}\;\mathrm{cm^{-2}}$. Because of extremely high column densities, X-ray heating by the AGN, whose effect is not included in our simulations, will have little influence on the clump even if we take it into account. If we assume $\Tgas =20\;\mathrm{K}$ and $\mu =2$ as in a normal prestellar core, the Jeans mass is $\approx 0.26\;\Msolar$ for $\nH=10^{8}\;\mathrm{cm^{-3}}$, which is much smaller than the total mass of the clump. Therefore, this clump will collapse and fragment into small sub-clumps if we remove the temperature floor.

The chemical structure and the velocity fields of the model H20 are shown in Fig.~\ref{fig:Time_Evolution_H20_Xi} and \ref{fig:Time_Evolution_H20_Velocity}, respectively. The stripped flow is fully ionized and the velocity of the flow increases to $\approx 200\;\kms$ by it reaches a distance of $\sim \rcl$ from the cloud surface. In order to show contributions of $\Hatm$-photo-ionization and dust absorption in accelerating the flow, we plot in the right panel of Fig.~\ref{fig:Acceleration_H20} the spacial distribution of $|\bmath{a}^{\mathrm{gr}}_{\mathrm{rad}}|/|\bmath{a}^{\Hatm}_{\mathrm{rad}}|$ of the model H20 at $t=90\;\kyr$. It is clear from the figure that the stripped flow is dominantly accelerated by the dust absorption. As in Low-$\mathcal{U}$ models, a part of the stripped flow comes around the shocked layer and becomes neutral. The velocity of the neutral flow is $\lesssim 100\;\kms$.

\begin{figure*}
\centering
\includegraphics[clip,width=\linewidth]{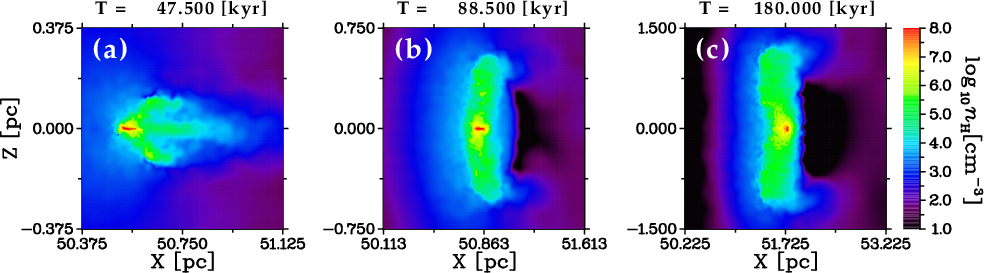}
\caption{Enlarged views of Fig.~\ref{fig:Time_Evolution_High_U_models}d,h,l. Note that the upper limit of the color bar is different from Fig.~\ref{fig:Time_Evolution_High_U_models}.}
\label{fig:High_U_models_Density_enlarged}
\end{figure*}

\begin{figure*}
\centering
\begin{tabular}{cc}
\begin{minipage}{0.5\linewidth}
\includegraphics[clip,width=\linewidth]{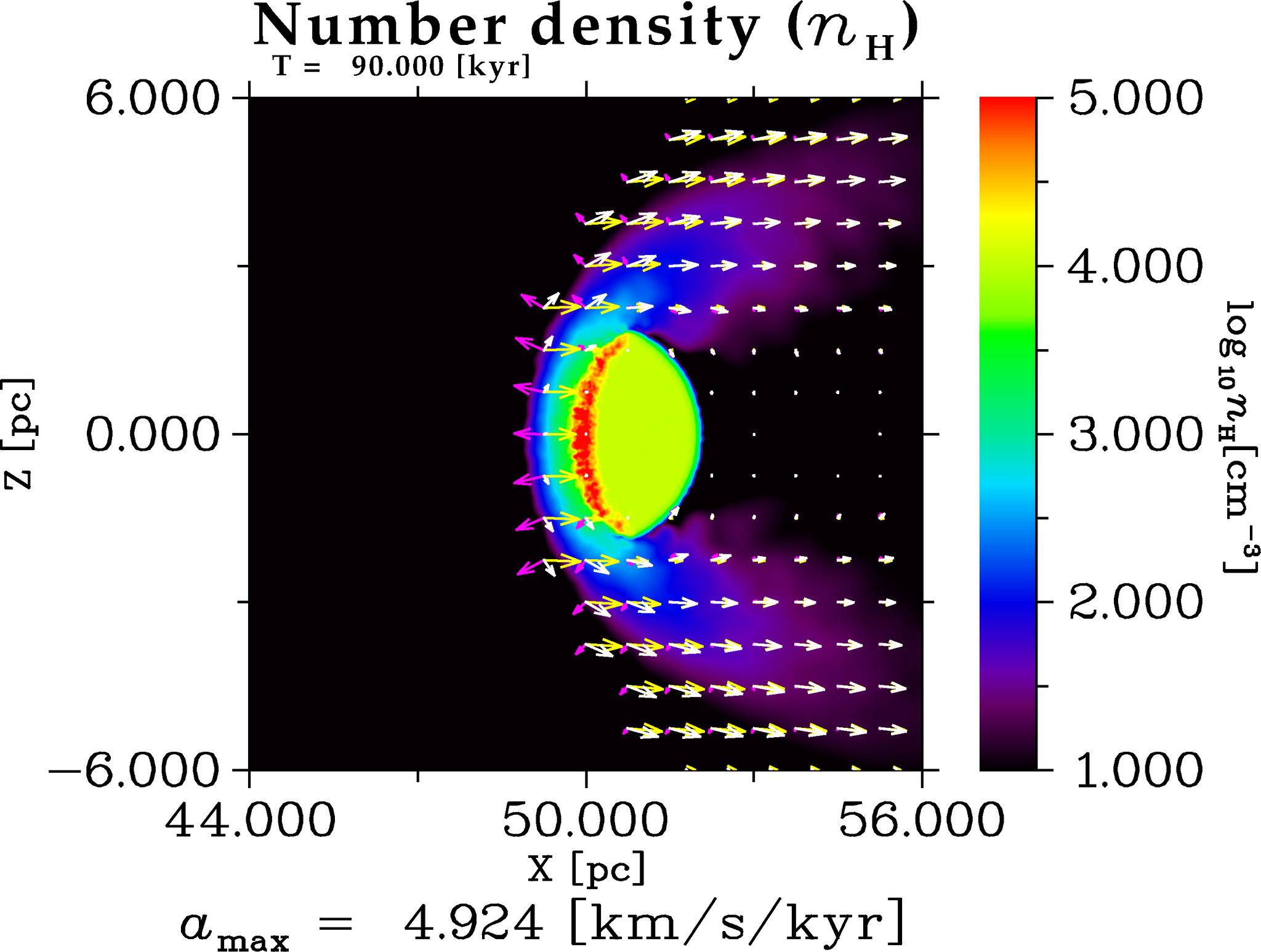}
\end{minipage}
\begin{minipage}{0.5\linewidth}
\includegraphics[clip,width=\linewidth]{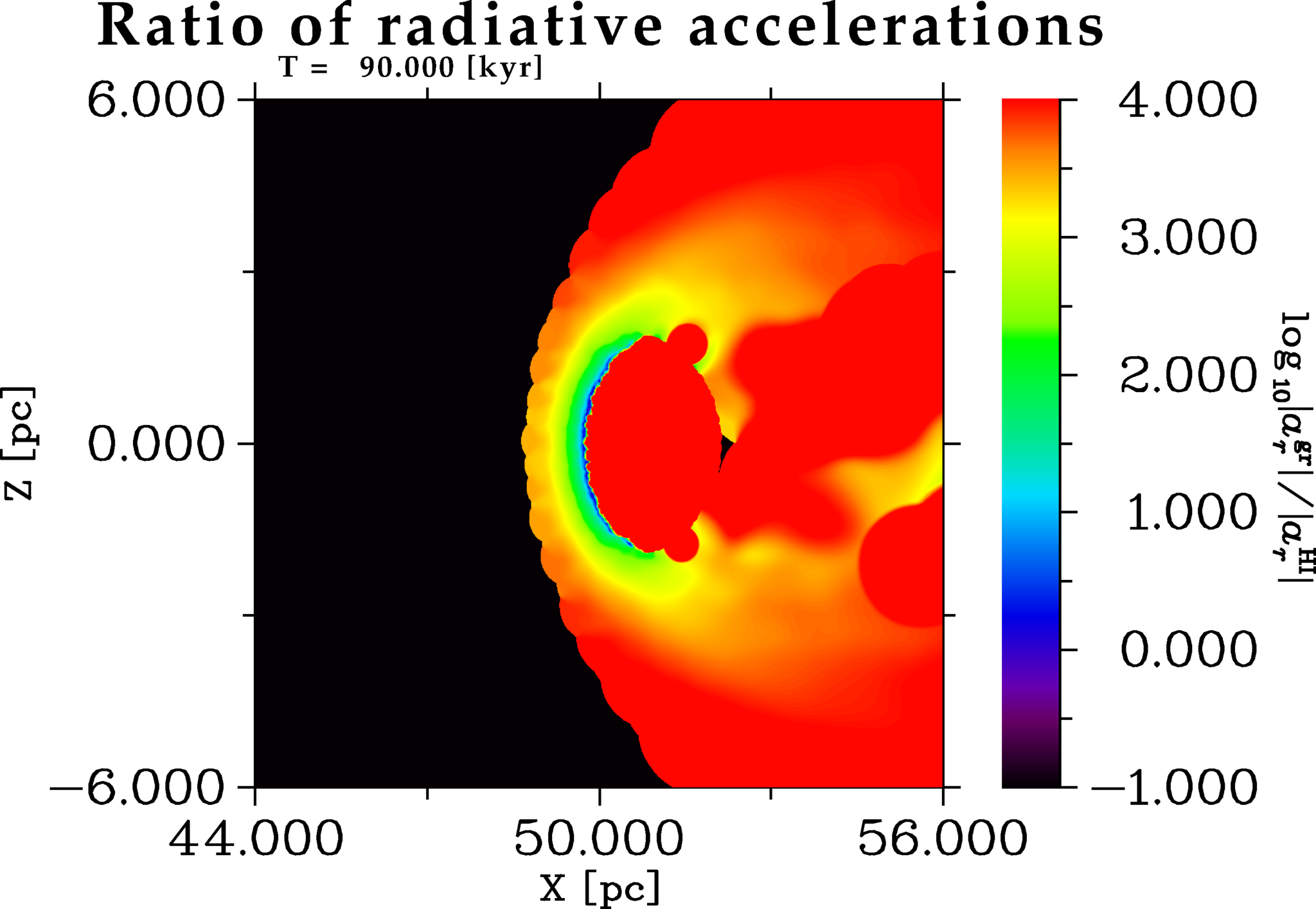}
\end{minipage}
\end{tabular}
\caption{The acceleration fields (\textit{left}) and $|\bmath{a}^{\mathrm{gr}}_{\mathrm{rad}}|/|\bmath{a}^{\Hatm}_{\mathrm{rad}}|$ (\textit{right}) at $t=90\;\kyr$ in the model H20. In the left panel, the arrows show the total acceleration (\textit{white}), the pressure-gradient acceleration (\textit{magenta}), and the radiative acceleration (\textit{yellow}). The maximum arrow length corresponds to the value described on the bottom of the panel. }
\label{fig:Acceleration_H20}
\end{figure*}

\begin{figure}
\begin{tabular}{cc}
\begin{minipage}{0.5\linewidth}
\includegraphics[clip,width=\linewidth]{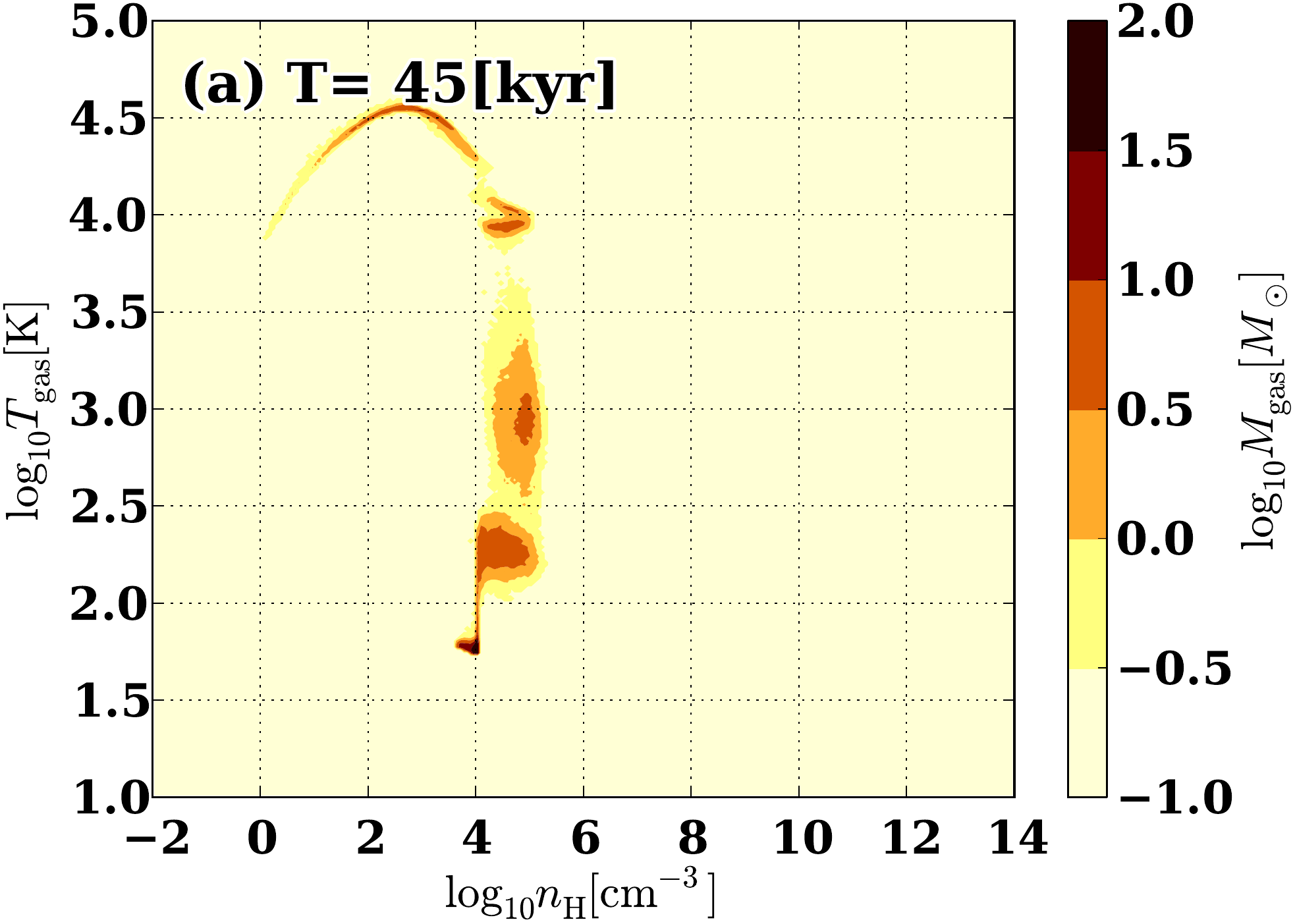}
\includegraphics[clip,width=\linewidth]{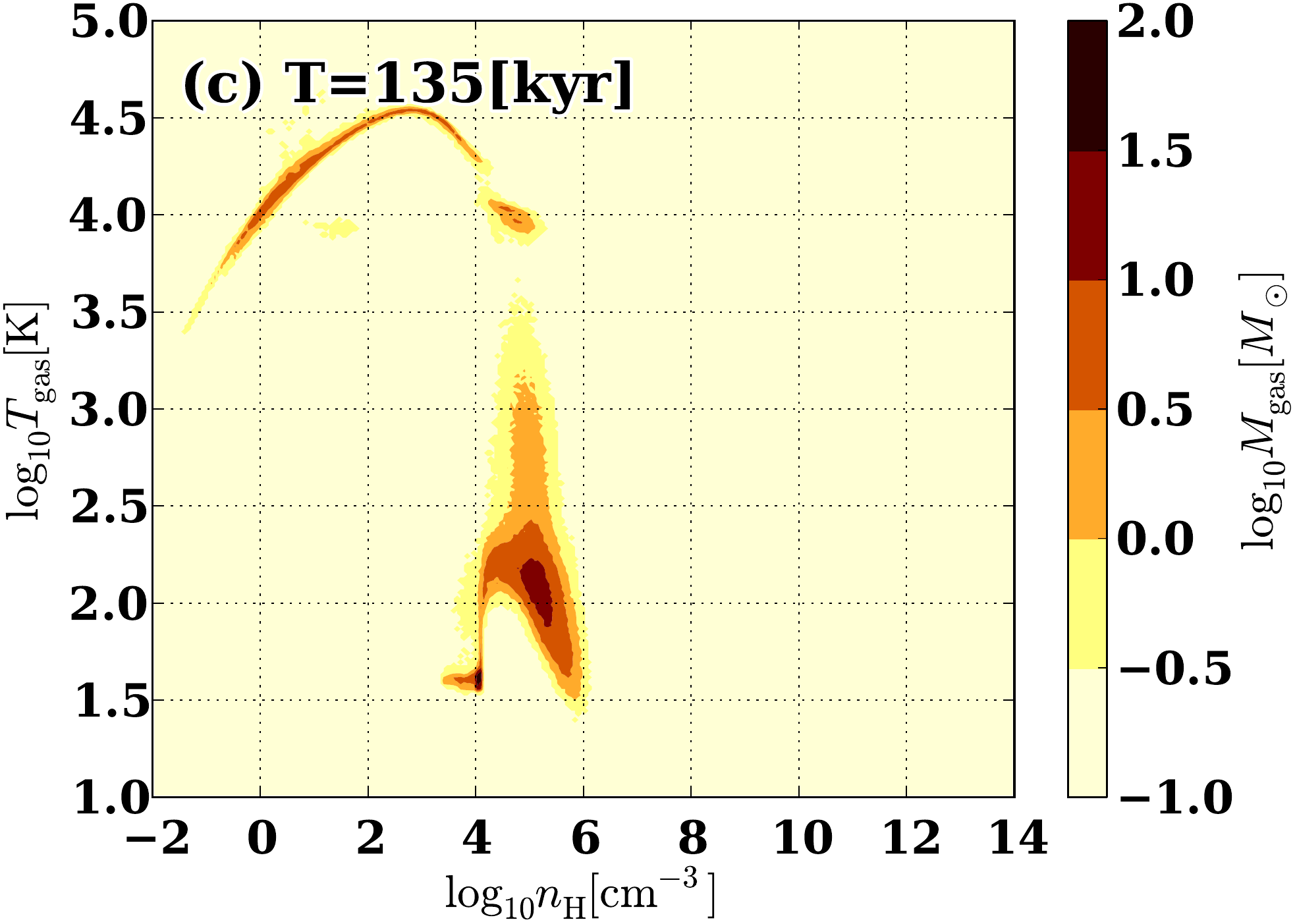}
\end{minipage}
\begin{minipage}{0.5\linewidth}
\includegraphics[clip,width=\linewidth]{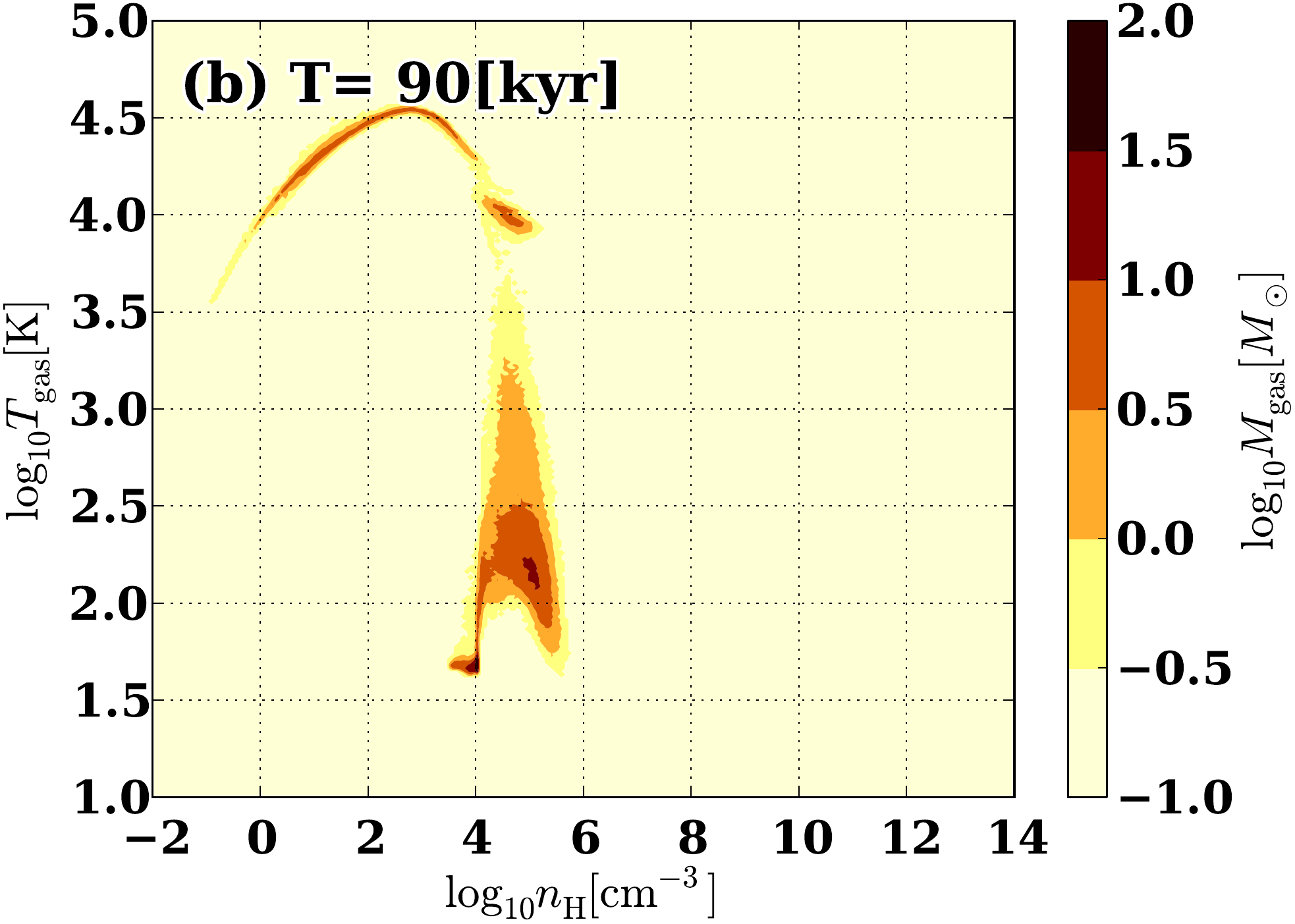}
\includegraphics[clip,width=\linewidth]{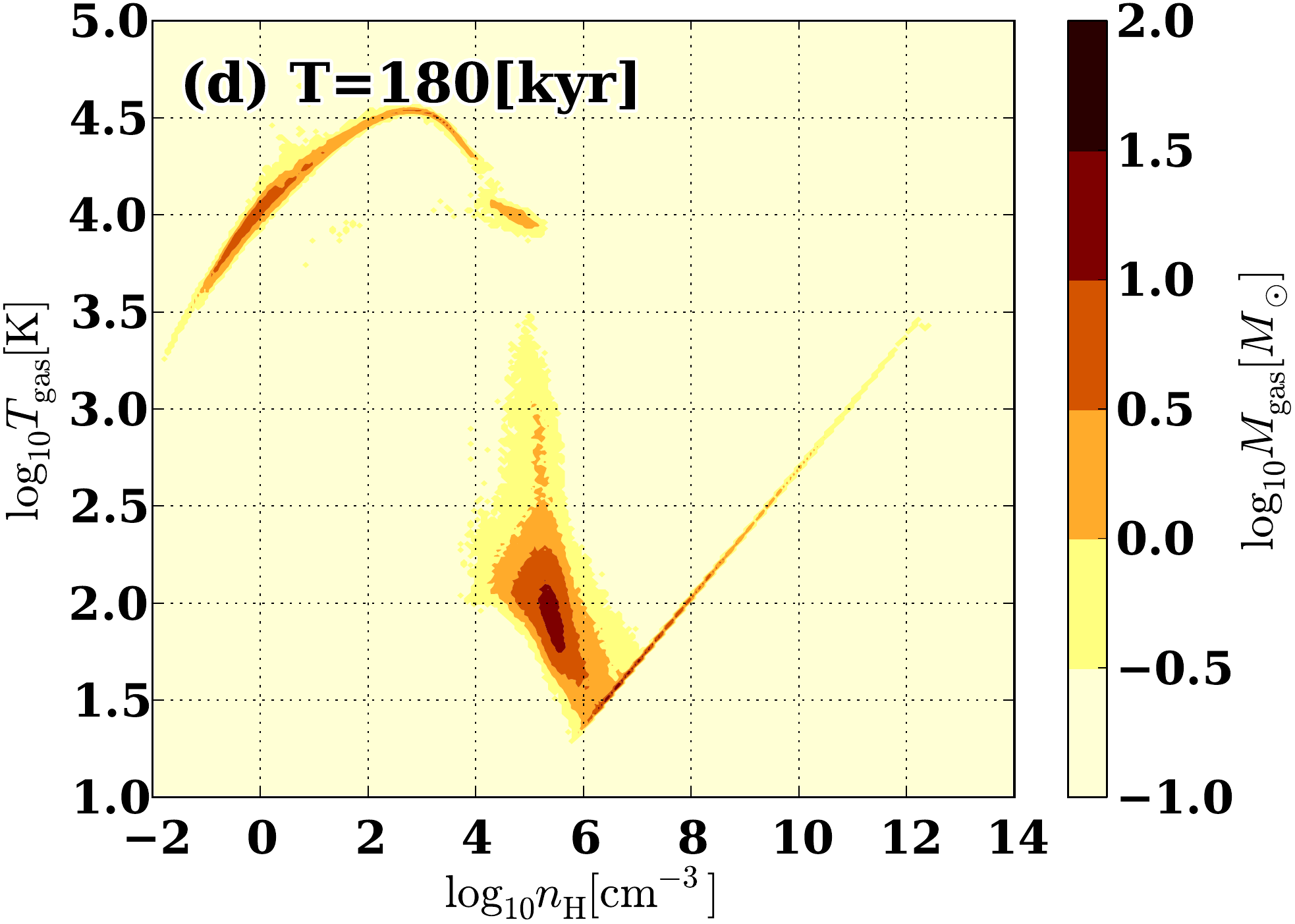}
\end{minipage}
\end{tabular}
\caption{Time evolution of the $\nH$-$\Tgas$ diagram of the model H20. Each panel has $256^{2}$ pixels and the color of the pixel shows the gas mass contained in the pixel.}
\label{fig:Time_Evolution_H20_rhoT}
\end{figure}

\begin{figure*}
\includegraphics[clip,width=\linewidth]{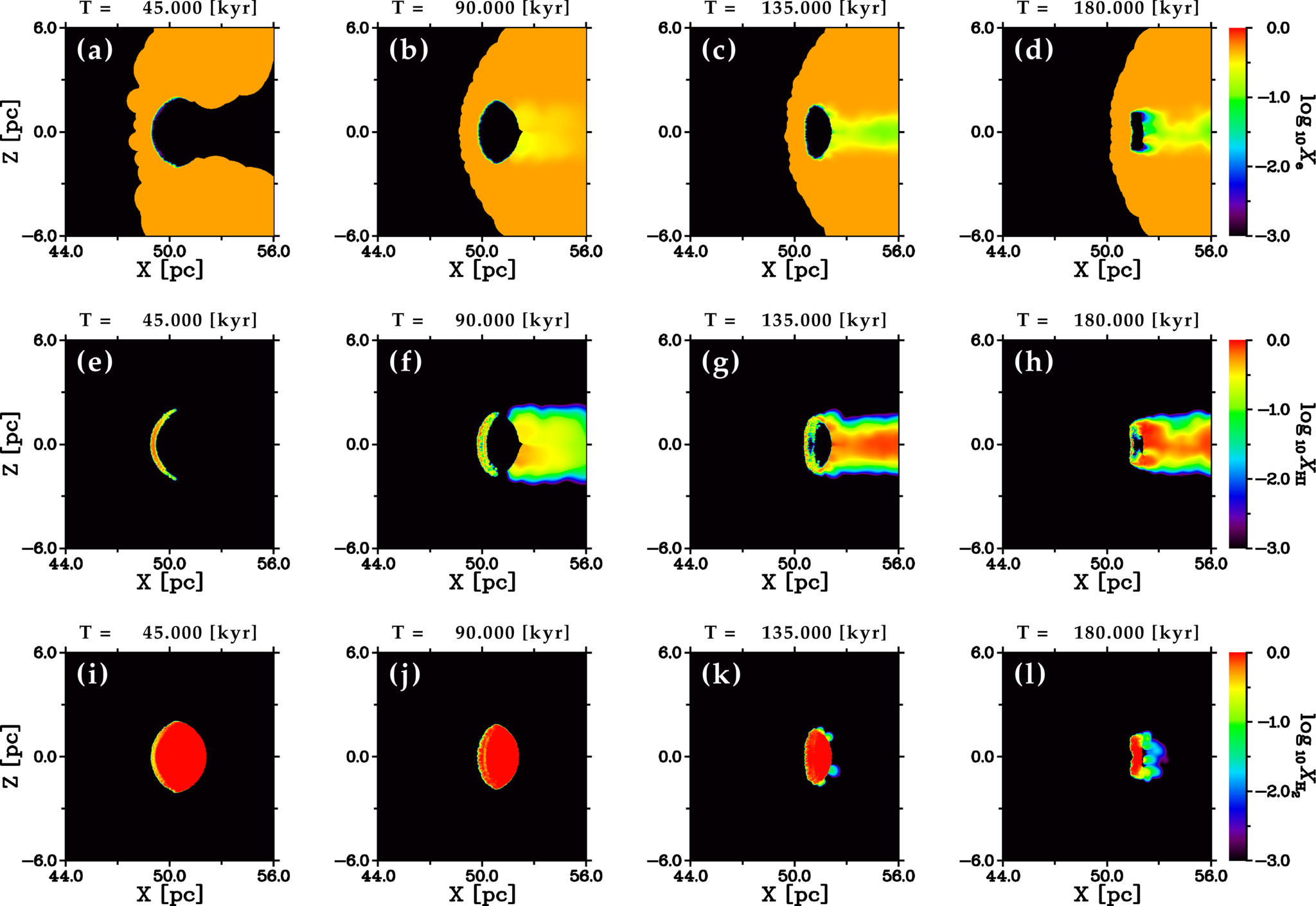}
\caption{The same as Fig.~\ref{fig:Time_Evolution_L20_Xi}, but for the model H20.}
\label{fig:Time_Evolution_H20_Xi}
\end{figure*}

\begin{figure*}
\includegraphics[clip,width=\linewidth]{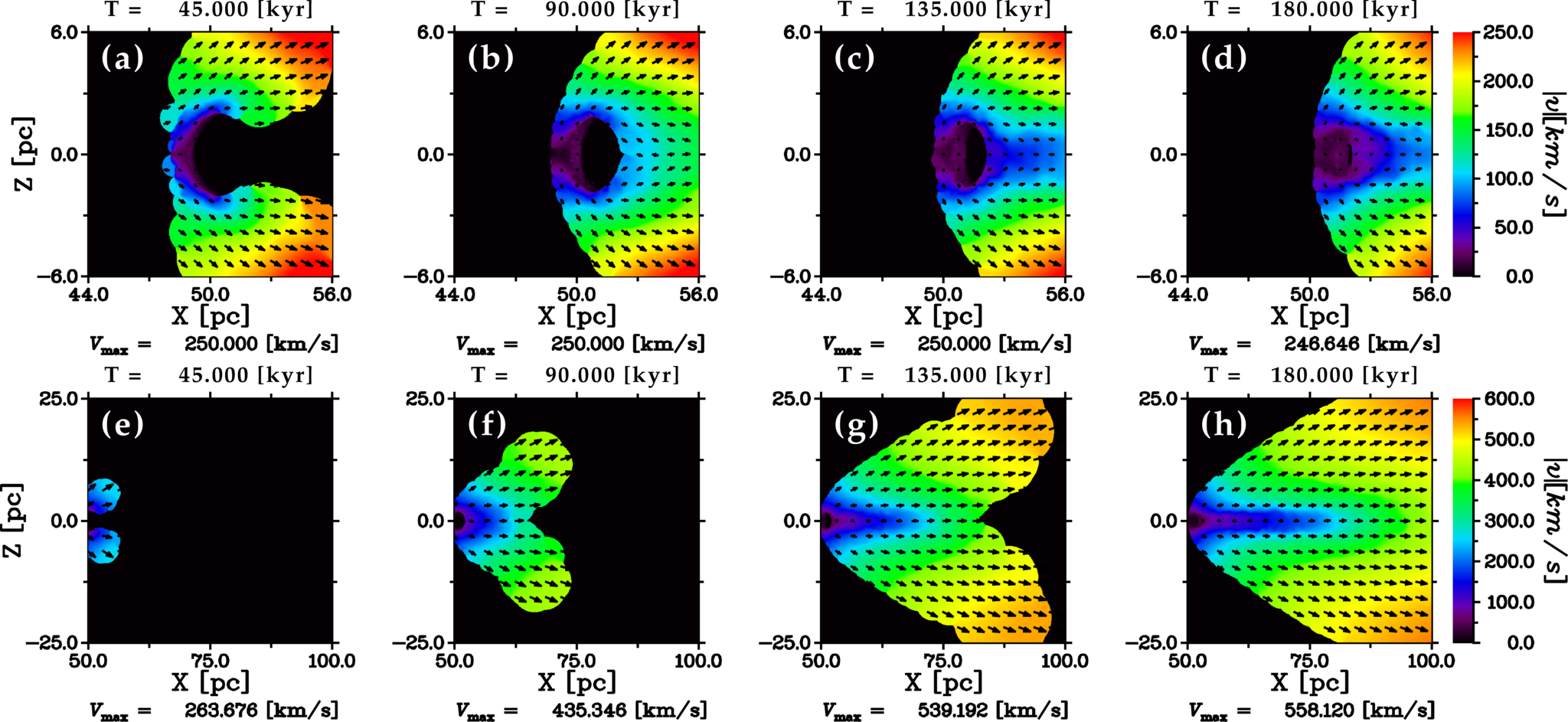}
\caption{The same as Fig.~\ref{fig:Time_Evolution_L20_Velocity}, but for the model H20.}
\label{fig:Time_Evolution_H20_Velocity}
\end{figure*}

\subsection{SC models} \label{subsec:SC_models}
Here, we first explain the simulation conditions of the model \SCmodel.
The cloud has an uniform number density of $6\times 10^{4}\;\nden$ and is located at a distance of $5\;\pc$ from an AGN. Its radius and mass are $1\;\pc$ and $\approx 6\times 10^{3}\;\Msolar$. We assume the same initial gas and dust temperatures, namely, $\Tgas = 100\;\mathrm{K}$ and $\Tgr = 20\;\mathrm{K}$. Given this temperature, the Jeans length $\lambda_{J}$ is $\approx 0.55\;\pc$. Therefore, the cloud is initially self-gravitationally unstable. The cloud is initially at rest. After starting of the simulation, the cloud starts to fall toward a galactic center (GC) in a fixed external gravitational potential of a SMBH and a nuclear star cluster (NSC). In this study, both components are modelled by the Plummer models as follows:
\begin{equation}
\Phi_{\mathrm{ext}}(r)=-\dfrac{G\MBH}{\sqrt{r^{2}+\bBH^{2}}} - \dfrac{G\MNSC}{\sqrt{r^{2}+\bNSC^{2}}},
\end{equation}
where $G$ is the gravitational constant, $\MBH = 8\times 10^{6}\;\Msolar$, $\bBH = 0.1\;\pc$, $\MNSC = 2.2\times 10^{8}\;\Msolar$, and $\bNSC = 25\;\pc$. The simulation conditions described above are almost the same as those of SC00 in \citet{schartmann11:_radiat_seyfer}\footnote{The exception is the treatment of the gravitational potential of the SMBH. In their study, the SMBH is modeled by the Newtonian potential. The reason why we replace this by the Plummer potential is simply to avoid diverging the external gravity to infinity at the origin.}. In order to obtain a higher column density resolution, the number of SPH particles of $2^{21}$ is used in this model.

We first perform the numerical simulation of the model \SCmodel \textit{without} the external gravity to make clear the effect of it and to compare the result with those of Low-$\mathcal{U}$ and High-$\mathcal{U}$ models. Next, we perform the simulation with the external gravity. Hereafter, we call these runs \SCs and \SCff, respectively. We note that in \SCs, the cloud does not move toward the GC because of no external gravity.

\begin{figure*}
\includegraphics[clip,width=\linewidth]{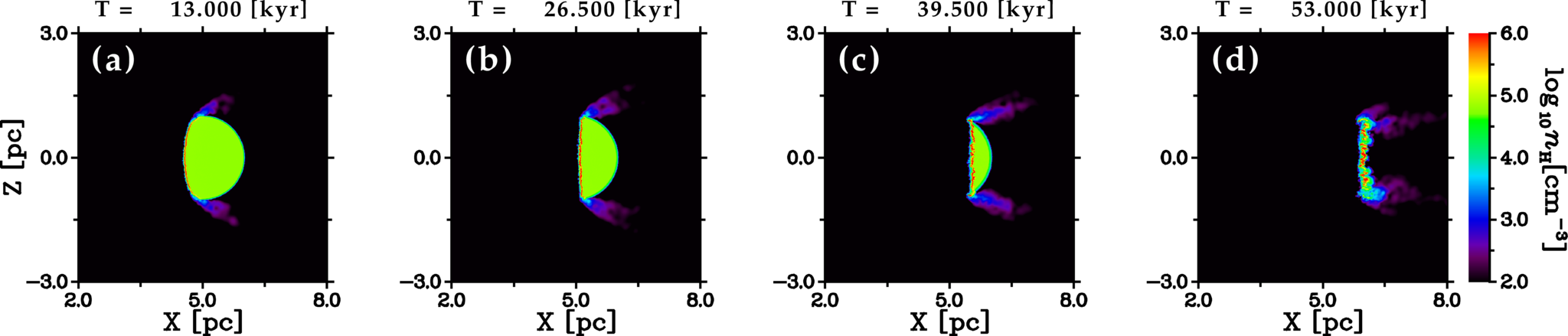}
\caption{The same as Fig.~\ref{fig:Time_Evolution_Low_U_models}, but for the model \SCs.}
\label{fig:Time_Evolution_SC00_3D_static}
\end{figure*}

Figure~\ref{fig:Time_Evolution_SC00_3D_static} and \ref{fig:Time_Evolution_SC00_3D_free_fall} show the time evolution of the number density distribution of \SCs and \SCff, respectively. From these figures, it is obvious that the radiation pressure plays a dominant role in the evolution of the cloud. Because of very intense radiation pressure, no photo-evaporation occurs in both runs, although the gas stripped by the radiation pressure from the cloud edge is promptly photo-ionized. 

\begin{figure}
\includegraphics[clip,width=\linewidth]{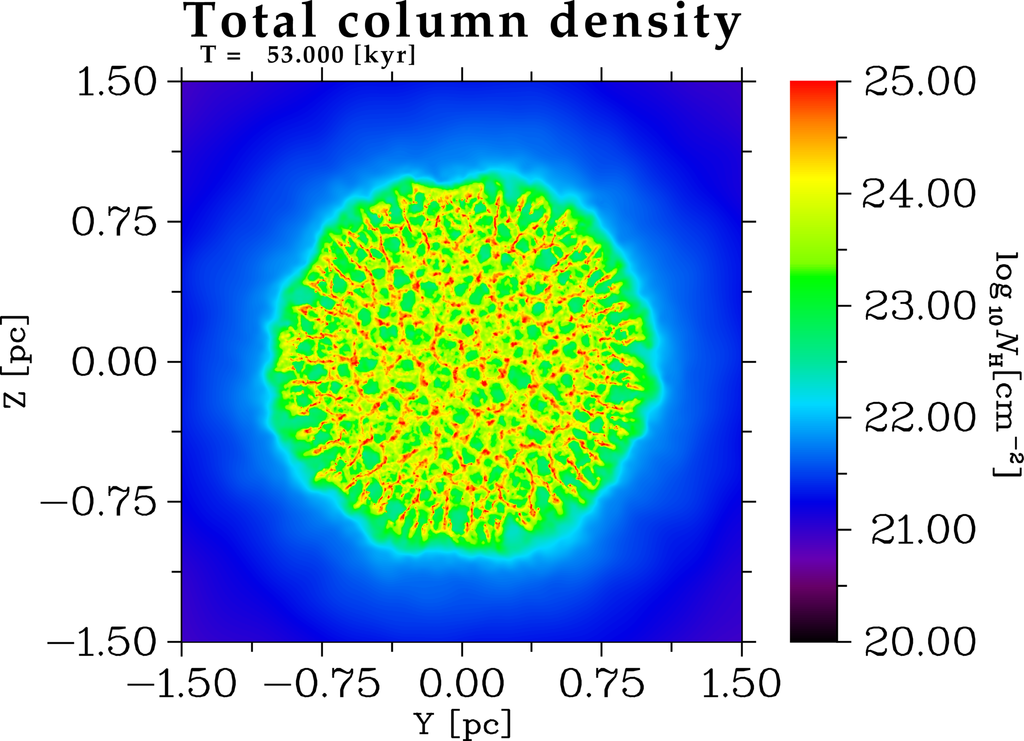}
\caption{The column density distribution of the model \SCs in the $yz$ plane at $t=53\;\kyr$.}
\label{fig:SC00_3D_static_Sigma}
\end{figure}

\begin{figure}
\includegraphics[clip,width=\linewidth]{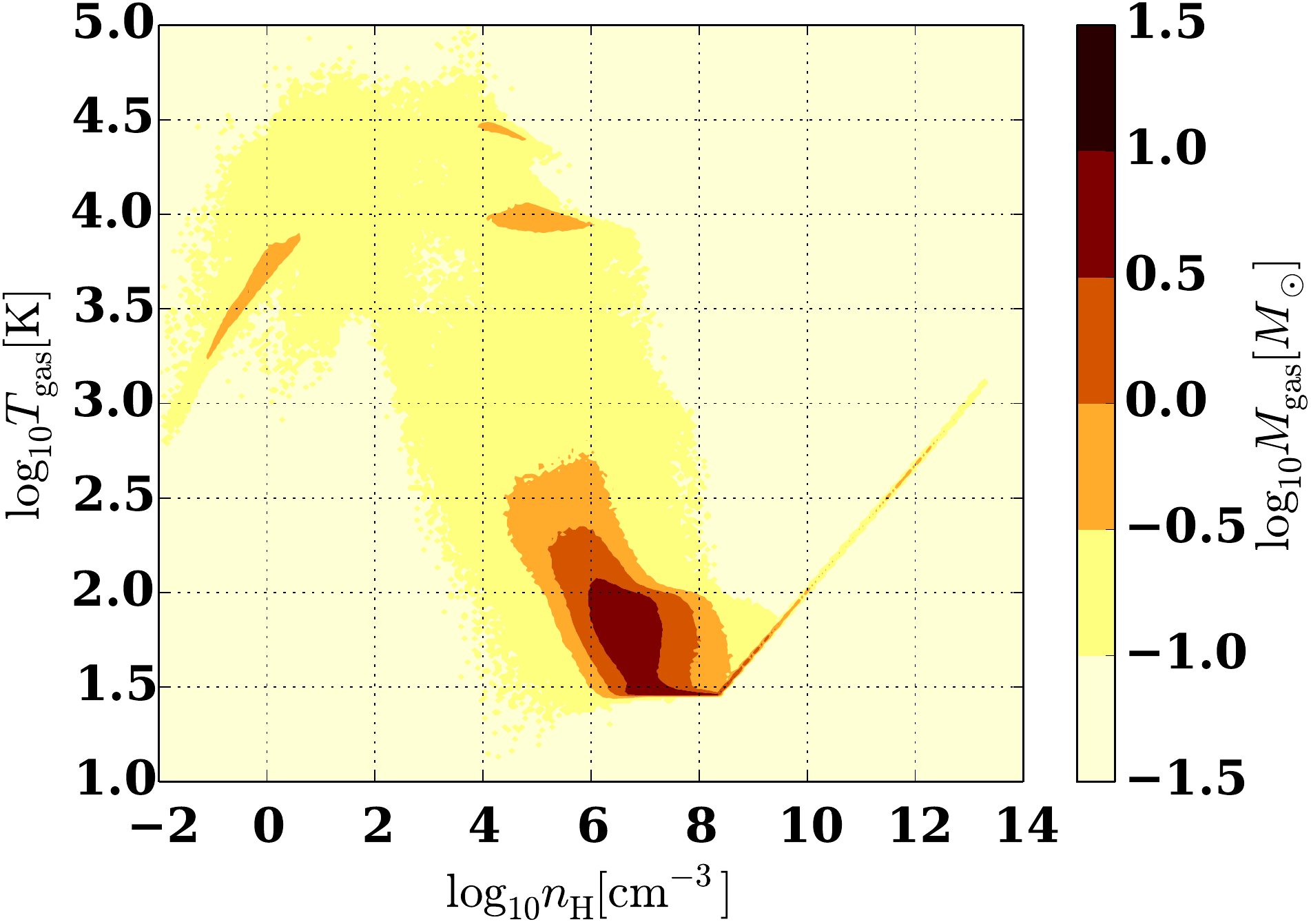}
\caption{The $\nH$-$\Tgas$ diagram of the model \SCs at $t=53\;\kyr$.}
\label{fig:SC00_3D_static_nH_Tgas_diagram}
\end{figure}

\begin{figure*}
\includegraphics[clip,width=\linewidth]{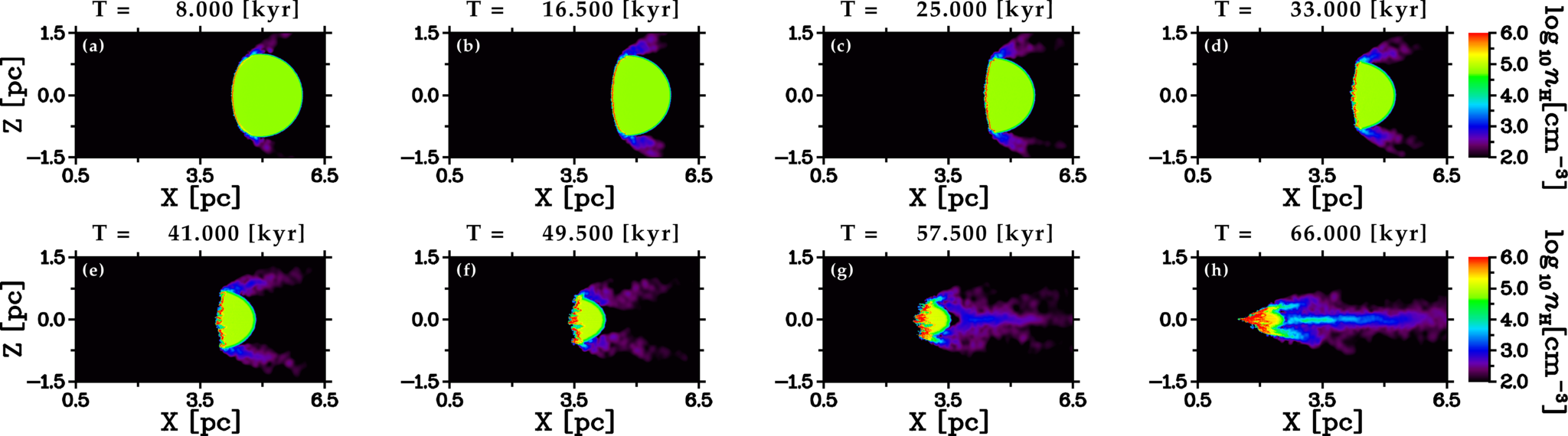}
\caption{The same as Fig.~\ref{fig:Time_Evolution_Low_U_models}, but for the model \SCff.}
\label{fig:Time_Evolution_SC00_3D_free_fall}
\end{figure*}

In \SCs, the cloud is simply destroyed by the passage of the shock driven by the radiation pressure. The shocked layer keeps its shape flat until it reaches the end of the cloud. The time averaged velocity of the shock is $\overline{v}_{\mathrm{sh}}\approx 37.7\;\kms$, which is in good agreement with the estimated value by the razor-thin approximation, $35\;\kms$ (Eq.~(\ref{eq:vshapp})). As the shock sweeps the gas, the mass of the post-shocked layer grows with time. Consequently, the shocked layer fragments into gas clumps and gas filaments by the self-gravitational instability (Fig.~\ref{fig:Time_Evolution_SC00_3D_static}c,d). This is more clearly seen in the column density distribution in the $yz$ plane at $t=53\;\kyr$ (Fig.~\ref{fig:SC00_3D_static_Sigma}). The distinct filamentary structure is developed over the entire cloud. The points of junction of the filaments are beginning to collapse self-gravitationally at $t=53\;\kyr$ as shown by $\nH$-$\Tgas$ diagram at $t=53\;\kyr$ (Fig.~\ref{fig:SC00_3D_static_nH_Tgas_diagram}). It is interesting to examine further evolution of this filamentary structure, since the stellar IMF becomes potentially top-heavy via the selective destruction of low column density regions due to the radiation pressure.

By contrast with \SCs, \SCff shows a different evolution. During the initial $33\;\kyr$, the evolution of the cloud is similar to that of \SCs, but the propagation of the shock in the cloud is slower than that of \SCs. This slowdown is more prominent in the later stage of the simulation (e.g., Fig.~\ref{fig:Time_Evolution_SC00_3D_free_fall}e,f,g). Two mechanisms cause this slowdown. One is the radial component of the gravity and the other is the gravity in the traverse direction. Because the external gravity is stronger at smaller radius, the region between the shock (the near-side edge) and the far-side edge of the cloud is stretched as the cloud approaches the GC. In other words, the velocity of the shock is smaller in a comoving frame of the pre-shock gas than that in \SCs. In addition to this, the gradual increase of the averaged density of the cloud, which is caused by the transverse components of the total gravity, also becomes a factor for the slowness of the shock (see Eq.~\ref{eq:vshapp}). As a result, there exists the pre-shock gas at $t=66\;\kyr$ in \SCff, while all the gas is swept by the shock till $t=53\;\kyr$ in \SCs. In this sense, the slowdown of the shock propagation results in a slight increase of the longevity of the cloud ($10$-$20$\% in this particular model). However, the surviving pre-shock gas will be collapsed along the transverse direction eventually in a short time and star formation is expected to be unavoidable. What fraction of the gas is supplied to the AGN accretion disk depends on the star formation rate in the cloud.

Finally, we compare \SCff with SC00 in \citet{schartmann11:_radiat_seyfer}. The overall evolution of \SCff is similar to SC00, but there are several differences between them. First, the stripped gas is more diffuse in \SCff than SC00, because it is rapidly photo-ionized. By the same reason, long radial filaments found in SC00 are not formed in \SCff. Second, because the stripped gas expands into a region behind the cloud, a tail-like gas structure is formed there in \SCff (Fig.~\ref{fig:Time_Evolution_SC00_3D_free_fall}h), which is not seen in SC00. Third, which is most important, owing to the self-gravity and three-dimensionality, a great deal of gas of the cloud concentrates into the central region of the cloud in \SCff compared to SC00. This alters the density distribution of the cloud and enhances the slowdown of the shock. As a result, there are undamaged gas at $t=66\;\kyr$. In contrast, in SC00, all the gas is affected by the AGN radiation by $t=60\;\kyr$. Thus, the self-gravity and three-dimensionality will enhance the gas supply rate to the galactic center.

\subsection{Cloud evaporation and cloud destruction timescale} \label{subsec:cloud_evaporation}
\begin{figure}
\includegraphics[width=\hsize,clip]{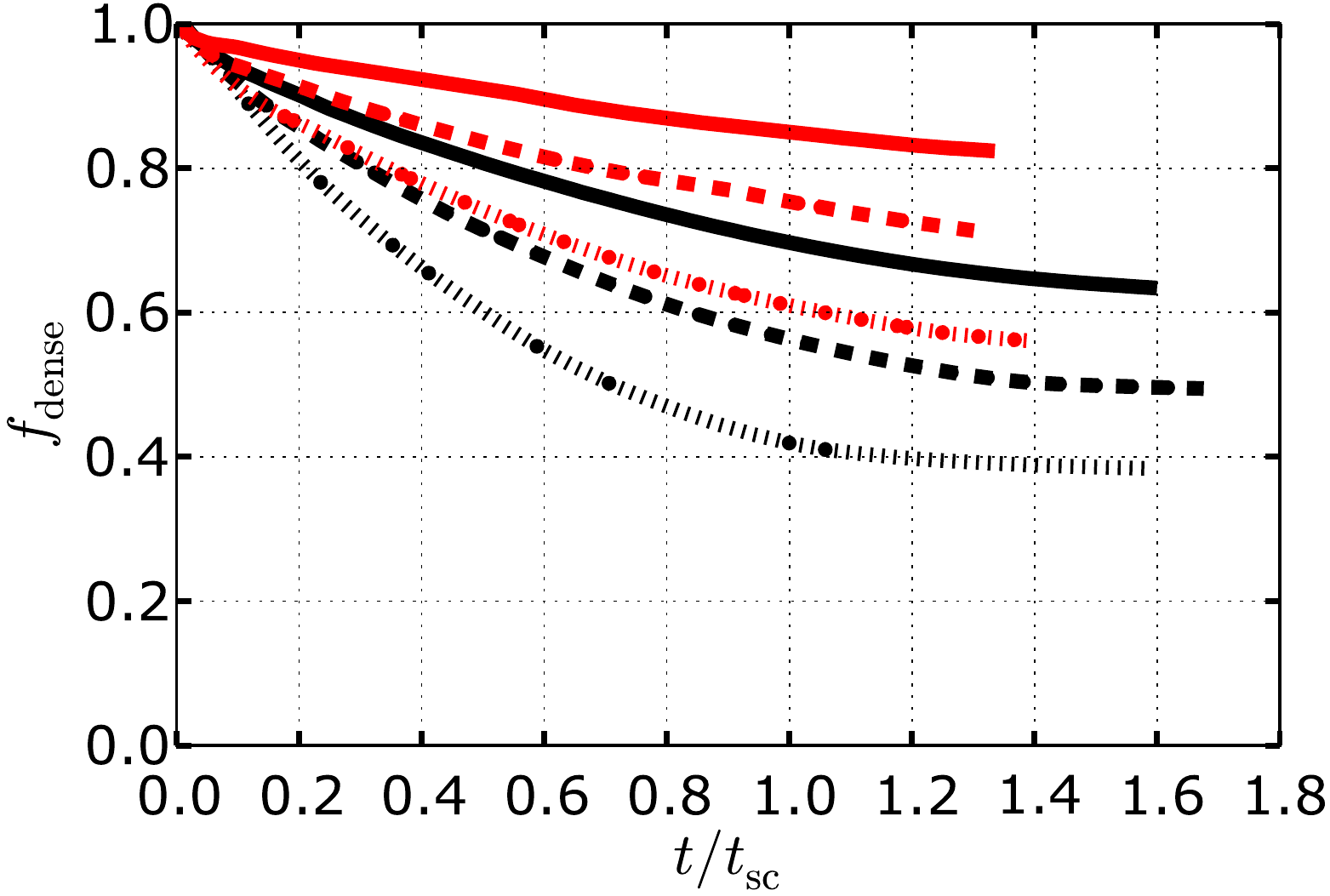}
\caption{The time evolution of the mass fraction of dense gas (\textit{black dotted}:L05, \textit{black dahsed}:L10, \textit{black solid}:L20, \textit{red dotted}:H05, \textit{red dashed}:H10, \textit{red solid}:H20). The calculation time is normalized by $\tsc$. }
\label{fig:f_dense}
\end{figure}

In order to see the dependencies of mass loss rates on $\mathcal{U}$ and $\mathcal{N}_{S}$ quantitatively, we plot in Fig.~\ref{fig:f_dense} the time evolution of dense gas fraction $f_{\mathrm{dense}}\equiv M_{\mathrm{gas}}(>\nH/2)/\Mcl$, where $M_{\mathrm{gas}}(>\nH/2)$ is the total mass of the gas that has a larger density than half of the initial density $\nH$. $f_{\mathrm{dense}}$ decreases with time in all the models and the decreasing rate is smaller in larger $\mathcal{N}_{S}$ model if $\mathcal{U}$ are the same. In Low-$\mathcal{U}$ models, the deceasing rates become small gradually with time, since the counteraction of the photo-evaporation compresses the gas cloud three-dimensionally and the geometric surface area that irradiated directly by the AGN becomes small with time. Because of this, $f_{\mathrm{dense}}$ in the models L05 and L10 saturate since $t=1.2\tsc$ and $1.4\tsc$. At the ends of the simulations, $f_{\mathrm{dense}}\approx 0.4$, $0.5$, $0.6$ for $\mathcal{N}_{S}=5$, $10$, $20$, respectively. In High-$\mathcal{U}$ models, $f_{\mathrm{dense}}$ also decrease but almost linearly with time except for $\mathcal{N}_{S}=5$ case. Main reason of the difference in the dependency of $f_{\mathrm{dense}}$ on $\mathcal{U}$ is the confinement effect by the radiation pressure as pointed out by \citet{pier95:_photoev}. At the ends of the simulations, $f_{\mathrm{dense}}\approx 0.6$, $0.7$, $0.8$ for $\mathcal{N}_{S}=5$, $10$, $20$, respectively. Thus, the fractional mass loss rates are smaller for High-$\mathcal{U}$ at least until $t\lesssim \tsc$. However, $f_{\mathrm{dense}}$ in the models H10 and H20 will decrease further for a while if we continue the simulations and then they will be saturated finally, because the clouds should be compressed by the counteraction of the stripped flow and their geometric cross sections that absorb the AGN radiation should become small (compare, e.g., Fig.~\ref{fig:Time_Evolution_High_U_models}g and Fig.~\ref{fig:Time_Evolution_High_U_models}h).

Cloud destruction timescale $t_{\mathrm{dest}}$ is determined by $2\rcl/\overline{v}_{\mathrm{sh}}$. In higher $\mathcal{U}$ cases (High-$\mathcal{U}$ and \SCmodel models), $t_{\mathrm{dest}}\approx \tsweep$ because of $\overline{v}_{\mathrm{sh}}\approx \vshapp$. On the other hand, in lower $\mathcal{U}$ cases (Low-$\mathcal{U}$ models), $t_{\mathrm{dest}}$ is smaller than $\tsweep$ because $\overline{v}_{\mathrm{sh}}$ is larger than $\vshapp$ due to the rocket effect. In terms of $\tsc$, $t_{\mathrm{dest}}/\tsc$ is in the range of $1\operatorname{-}2$ (Fig.~\ref{fig:f_dense}) for $(\mathcal{U},\mathcal{N}_{S})$ that is examined in this study.

\section{Discussions} \label{sec:discussions}
In this section, we first discuss uncertainties in our numerical results in \S~\ref{subsec:uncertainties}. Then, we compare our numerical results with previous studies in terms of cloud destruction process in \S~\ref{subsec:comparison}. After that, in \S~\ref{subsec:star_formation}, we discuss the possibility of star formation in AGN-irradiated gas clouds. Finally, we give some implications for gas clumps in AGN tori in \S~\ref{subsec:AGN_gas_clumps}.

\subsection{Uncertainties} \label{subsec:uncertainties}
 In this paper, we used a highly-simplified initial cloud model. In reality, gas clouds are not spherical and must have internal density structure and (turbulent) velocity fields. In addition, they should have orbital and spin angular momenta and be surrounded by a low-density gas if the clouds are initially in the pressure equilibrium. Furthermore, we neglected helium and metals for simplicity. Here, we discuss the possible effects on the evolution of the clouds, if these factors are taken into account. 

\subsubsection{Effects of chemical abundance and diffuse photons} \label{subsubsec:chemical_abundance_diffuse_photons}
In this study, we assumed that ISM consists of hydrogen and dust grains only and neglected helium and metals. However, actual ISM contains helium and metals. These elements enhance the cooling rate of gas significantly in a wide range of gas temperature. Therefore, the gas temperatures in the photo-evaporation flow and in the shocked layer will be dropped to lower values compared to the cases presented in this paper. This may result in a decrease of photo-evaporation rate and facilitate self-gravitational fragmentation of the shocked layer. Also, helium and metals increase the opacity of ISM by photon absorption by an enormous number of bound-bound/bound-free transitions. This enhancement may promote the confinement of the photo-evaporation flow, which is seen in higher $\mathcal{U}$ cases. In addition, we did not take into account size distribution of dust grains and propagation of diffuse photons (scattered and re-emitted photons). These will also affect the evolution of the clouds to some extent. Especially, analytic and numerical studies pointed out that photon scattering is potentially capable of increasing radiative acceleration (e.g., \citealt{lamers99:_introd_stell_winds,roth12:_three_radiat_trans_calcul_radiat}). In the following, we examine the possibilities discussed above by using the photo-ionization calculation code \textsc{Cloudy} (version C13.03, last described by \citealt{ferland13:_releas_cloud}). In all the photo-ionization calculations described below, we assume (1) that the gas is at the rest, (2) that the hydrogen number density distribution is the same as that along the $x$-axis of the model L20 at $T=54\;\kyr$, and (3) that the incident radiation field is the same as that used for the model L20.

Before doing the photo-ionization calculations, we first check if the dust absorption and scattering coefficients used in the \textsc{Cloudy} is similar to those used in this study, since a difference affects the interpretation of the results. With this aim, we make the data of absorption and scattering coefficients for a silicate grain with $a_{\mathrm{gr}}=0.05\;\micron$ by the \texttt{compile grains} command of the \textsc{Cloudy}. The coefficients are shown in Fig.~\ref{fig:alpha_dust} by the cyan lines. This figure shows that the dust properties used in the \textsc{Cloudy} is very similar to those in our study.

\begin{figure*}
\begin{tabular}{cc}
\begin{minipage}{0.5\linewidth}
\includegraphics[clip,width=\linewidth]{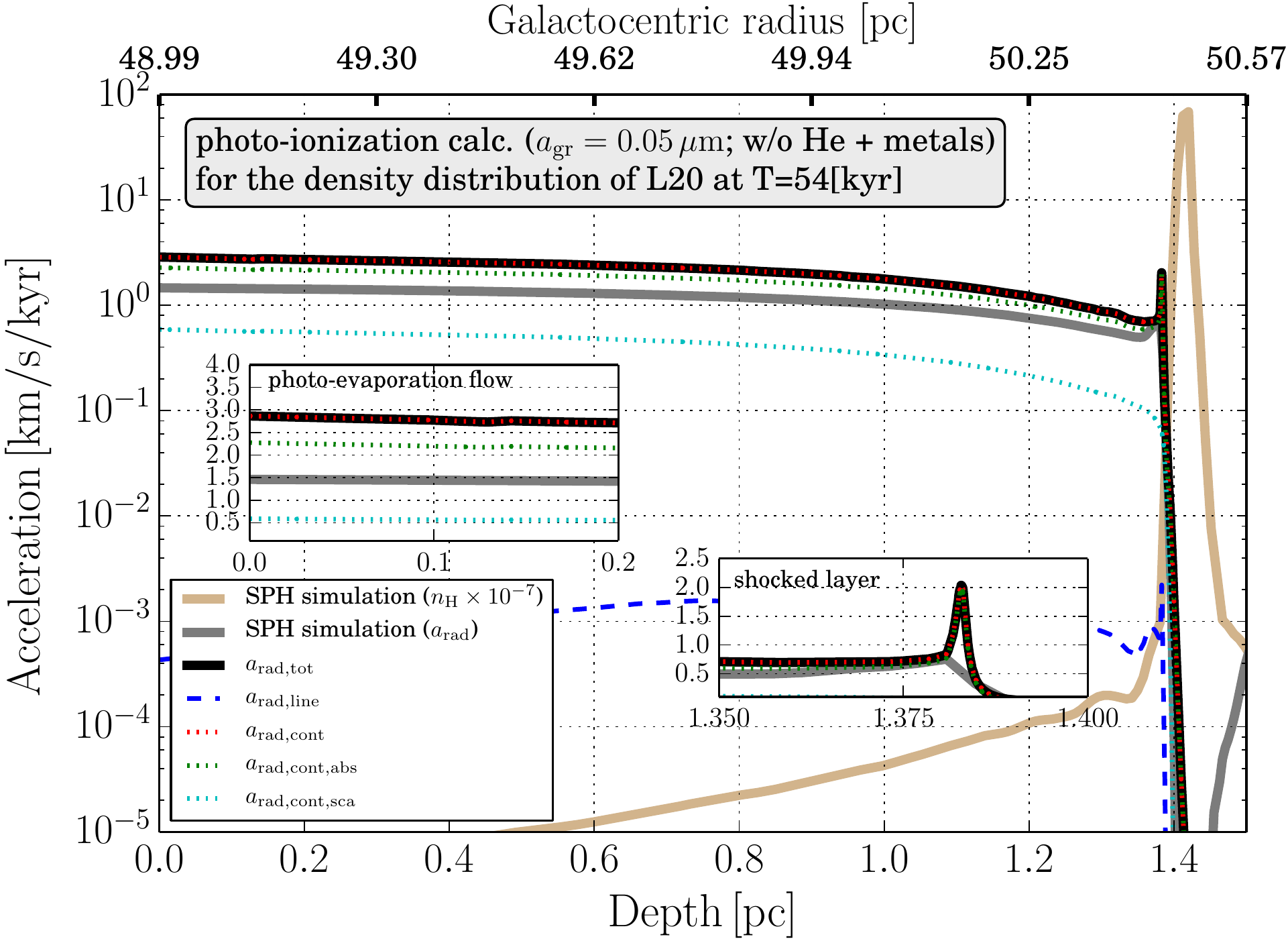}
\end{minipage}
\begin{minipage}{0.5\linewidth}
\includegraphics[clip,width=\linewidth]{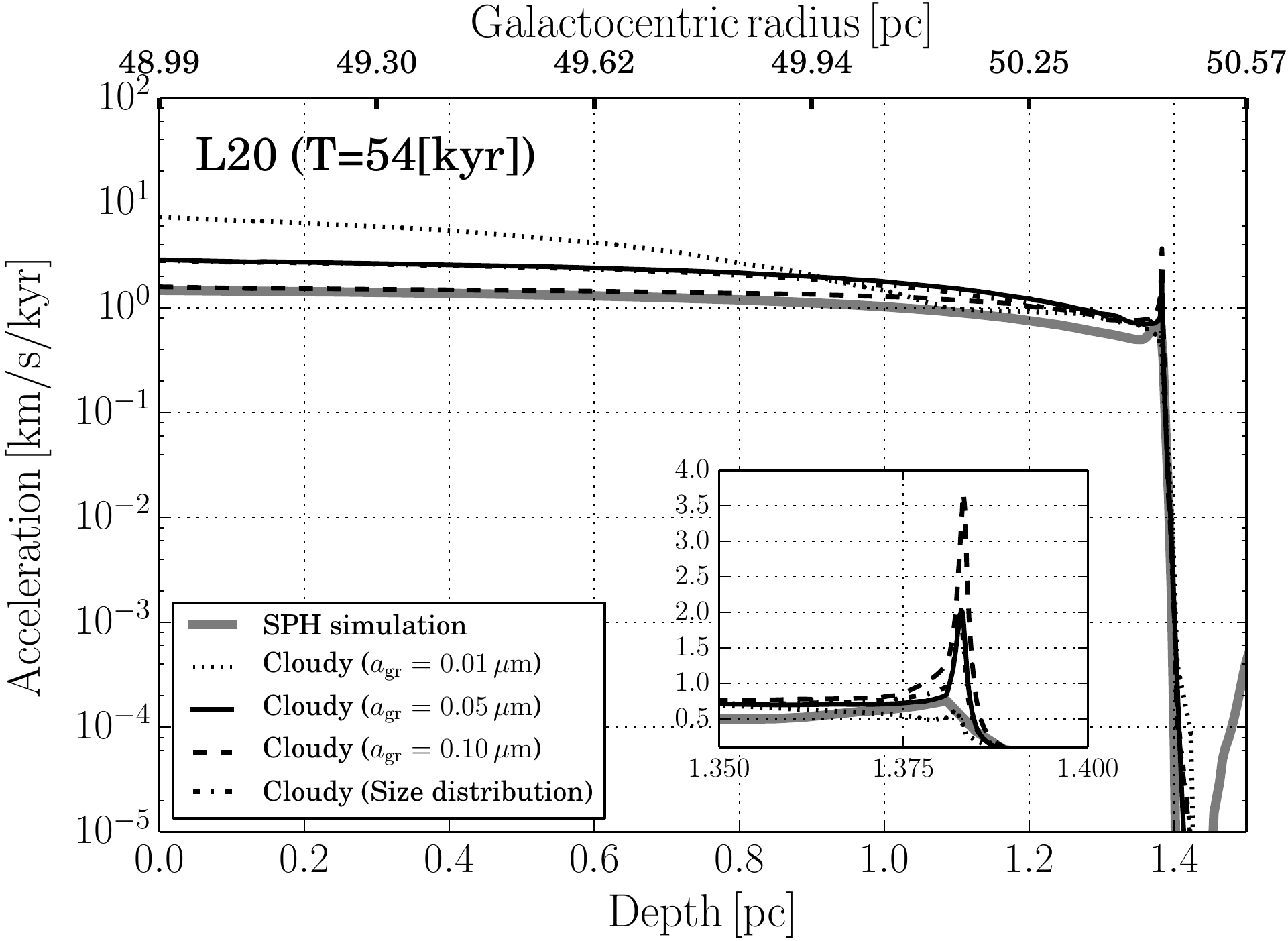}
\end{minipage}
\end{tabular}
\caption{\textit{Left}: A comparison of the radiative acceleration along the $x$-axis of the model L20 at $T=54\;\kyr$ with that obtained by the photo-ionization calculation with the \textsc{Cloudy}. Each component of the radiative acceleration are shown by different lines (\textit{thick black solid line}: total radiative acceleration, \textit{blue dashed line}: line radiative acceleration, \textit{red dotted line}: continuum radiative acceleration, \textit{green dotted line}: contribution of absorption for continuum radiative acceleration, \textit{cyan dotted line}: contribution of scattering for continuum radiative acceleration). The hydrogen number density distribution used in the photo-ionization calculation is plotted by the \textit{brown solid line}. \textit{Right}: Total radiative accelerations for different dust grain sizes. Four different dust size distributions are plotted (three are single-sized dust grains of $a_{\mathrm{gr}}=0.01, 0.05, 0.1\;\mathrm{\mu m}$ and the other is the standard size distribution used in the \textsc{Cloudy}). In both panels, the accelerations are plotted as a function of distance from the illuminated face (\textit{lower horizontal axis}) or from the AGN (\textit{upper horizontal axis}). The small insets in the figures are enlarged views and their vertical axes are linear scale. In the photo-ionization calculations, we assume that an ISM consists of hydrogen and silicate dust grains only and the dust abundance is adjusted so that $f_{\mathrm{gr}}=0.01$.}
\label{fig:L20_rad_accl}
\end{figure*}

To discriminate the effects of the diffuse photons, we perform a photo-ionization calculation assuming that ISM consists of hydrogen and dust grains only. The spacial distribution of the radiative acceleration obtained by the calculation is shown, together with that in the SPH simulation, in the left panel of Fig.~\ref{fig:L20_rad_accl}. This figure shows that in the photo-evaporation flow, the total radiative acceleration in the SPH simulation ($\approx 1.5\;\mathrm{km\;s^{-1}\;kyr^{-1}}$) is a factor of $\approx 2$ lower than that in the photo-ionization calculation ($\approx 3\;\mathrm{km\;s^{-1}\;kyr^{-1}}$). A part of this difference ($\approx 0.5\;\mathrm{km\;s^{-1}\;kyr^{-1}}$) is explained by a scattering component of the continuum radiative acceleration\footnote{In order to obtain absorption and scattering components of the continuum radiative acceleration, we customized \texttt{wind.h}, \texttt{pressure\_total.cpp}, \texttt{parse\_save.cpp}, and \texttt{save\_do.cpp} in the \textsc{Cloudy} code.}. The remaining difference may be explained by absorption of diffuse photons. At the shocked layer, the SPH simulation underestimates the total radiative acceleration by a factor of $3\operatorname{-}4$. This is partly attributed to the deficiency of numerical resolution and the fact that our numerical scheme is not photon-conservative. Another reason may be that we did not take into account the absorption of diffuse photons in the SPH simulations. These analyses suggest that $\overline{v}_{\mathrm{sh}}$ becomes larger and the photo-evaporation rate becomes smaller than those in the SPH simulations presented in \S~\ref{sec:numerical_results} if the diffuse photons is taken into account.

Next, we perform the photo-ionization calculations for different grain size to check the effects of grain size. Three single-sized grains ($a_{\mathrm{gr}}=0.01$, $0.05$, and $0.1\;\mathrm{\mu m}$) and the standard size distribution used in the \textsc{Cloudy} are assumed. The spacial distributions of the total radiative acceleration for different grain sizes are shown in the right panel of Fig.~\ref{fig:L20_rad_accl}. This figure shows that the total radiative acceleration for the $a_{\mathrm{gr}}=0.05\;\mathrm{\mu m}$ case is almost identical to that for the standard size distribution. This is probably because a mean grain size $\langle \agr \rangle$ of $0.05\;\mathrm{\mu m}$ approximately satisfies the following relation:
\begin{eqnarray}
n(\langle\agr\rangle)\int \frac{L_{\nu}}{4\pi r^{2}}\mathcal{Q}_{\mathrm{abs}}(\nu,\langle\agr\rangle)\pi\langle\agr\rangle^{2}d\nu & & \nonumber \\
\quad = \iint \frac{L_{\nu}}{4\pi r^{2}}\mathcal{Q}_{\mathrm{abs}}(\nu,\agr)\pi \agr^{2}n(\agr)d\nu d\agr, & & \label{eq:rad_accl_approx_rel}
\end{eqnarray}
where $n(\agr)d\agr$ is the number density of dust grain between $\agr$ and $\agr+d\agr$ and $n(\langle\agr\rangle)\equiv \int (\frac{4}{3}\pi\agr^{3}\rho_{\mathrm{gr}}n(\agr))/(\frac{4}{3}\pi\langle\agr\rangle^{3}\rho_{\mathrm{gr}})d\agr$. Assuming that $\int \frac{L_{\nu}}{4\pi r^{2}}\mathcal{Q}_{\mathrm{abs}}(\nu,\agr)d\nu$ depends weakly on $\agr$ and introducing $\overline{\agr^{n}}=\int \agr^{n}n(\agr)d\agr/\int n(\agr)d\agr$, Eq.(\ref{eq:rad_accl_approx_rel}) can be rewritten as
\begin{eqnarray}
\langle\agr\rangle \approx \overline{\agr^{3}}/\overline{\agr^{2}}.
\end{eqnarray}
For the standard size distribution used in the \textsc{Cloudy}, $\overline{\agr^{3}}/\overline{\agr^{2}}=0.0354\;\mathrm{\mu m}$, which is close to $0.05\;\mathrm{\mu m}$. Thus, the $\agr=0.05\;\mathrm{\mu m}$ model, which is assumed in the SPH simulations in this paper, is a good approximation at least in respect of radiative acceleration and the evolution of the clouds will not be altered largely if we adopt a standard size distribution.

\begin{figure*}
\begin{tabular}{cc}
\begin{minipage}{0.5\hsize}
\includegraphics[clip,width=\hsize]{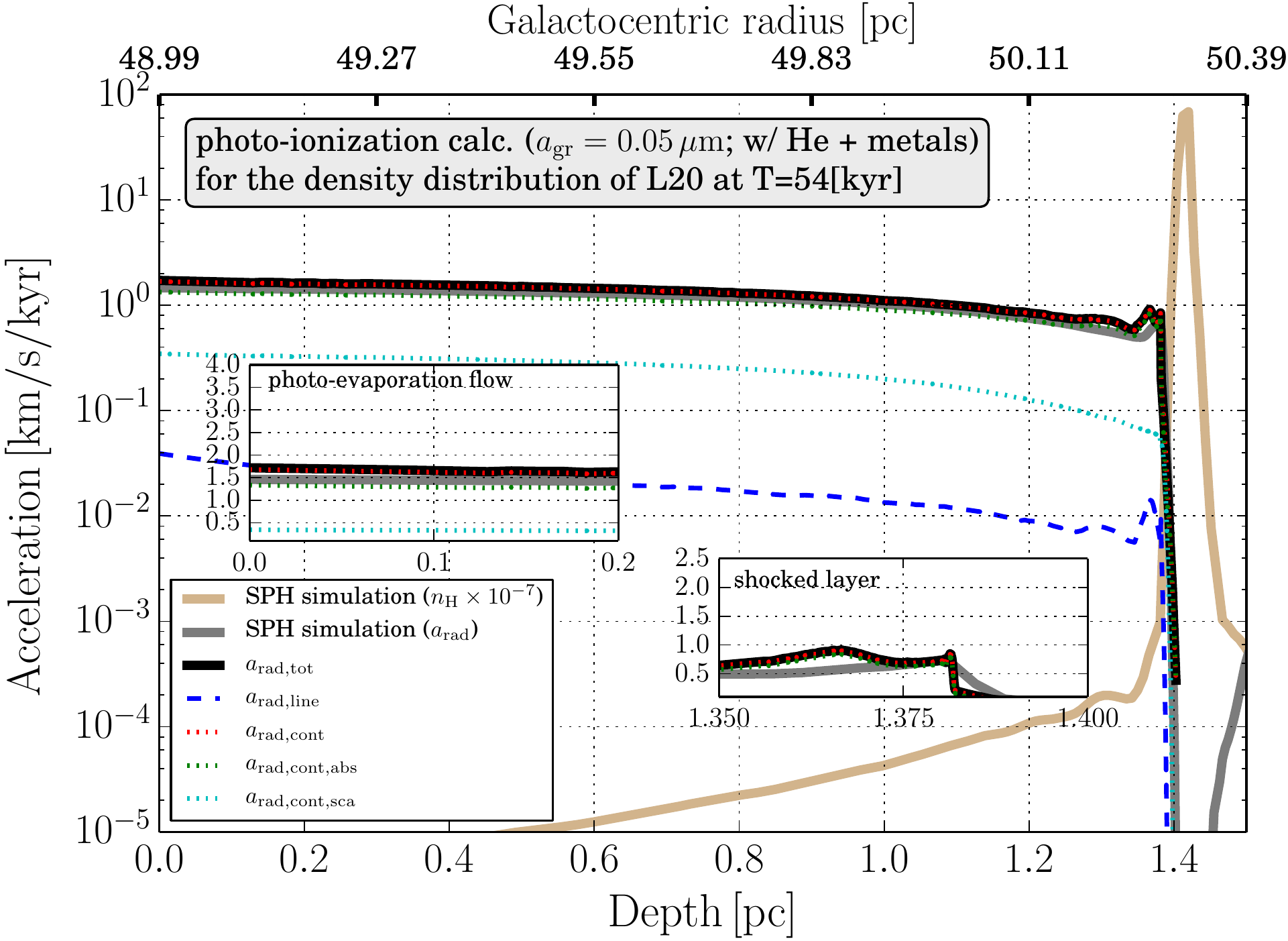}
\end{minipage}
\begin{minipage}{0.5\hsize}
\includegraphics[clip,width=\hsize]{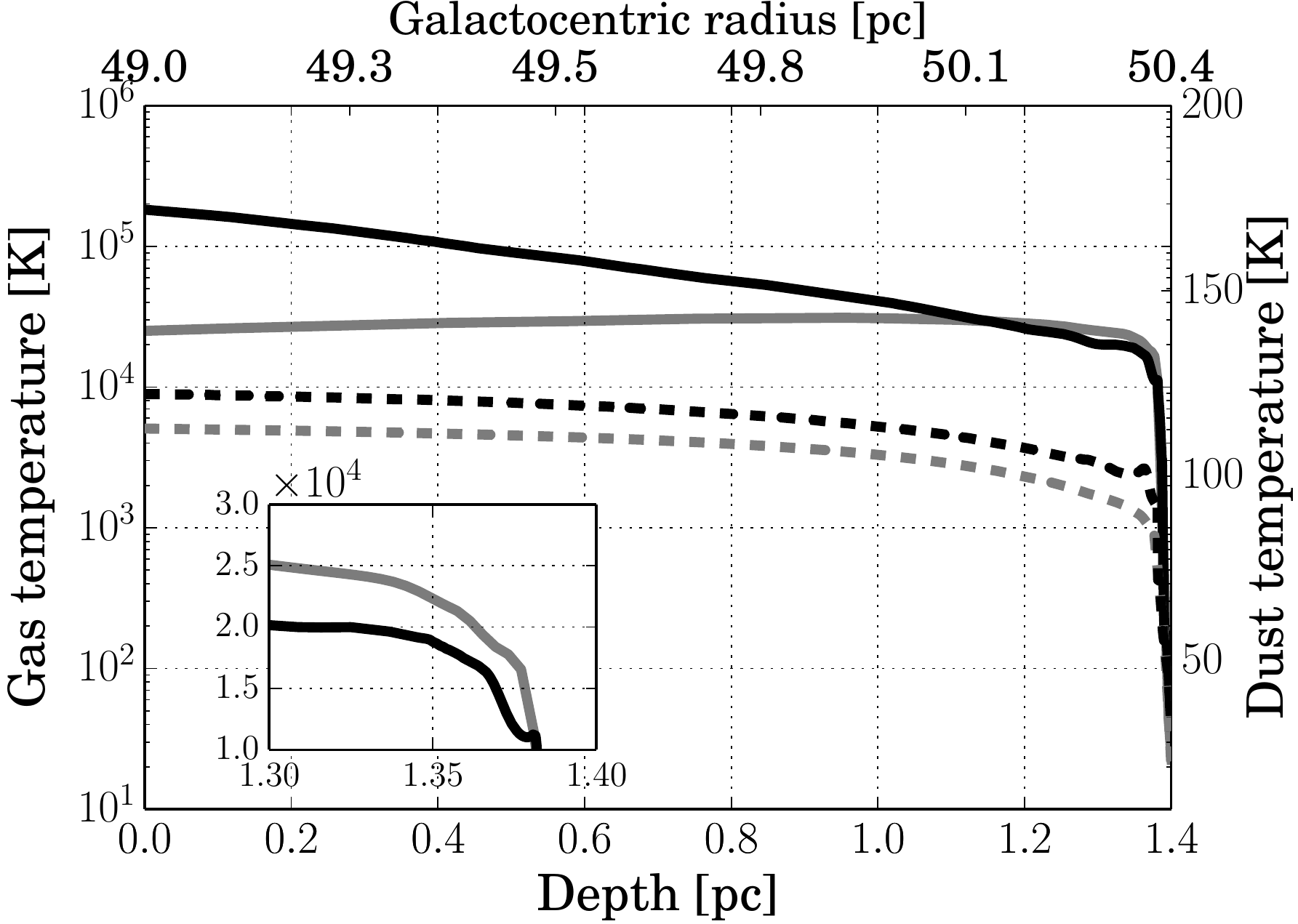}
\end{minipage}
\end{tabular}
\caption{\textit{Left}: The same as the left panel of Fig.~\ref{fig:L20_rad_accl}, but helium and metals is taken into account in this case. \textit{Right}: A comparison of gas and dust temperature distributions between the SPH simulation (\textit{gray}) and the photo-ionization calculation (\textit{black}). The gas and dust temperatures are plotted by the \textit{solid lines} and the \textit{dashed lines}, respectively. }
\label{fig:L20_arad_temp_metals}
\end{figure*}

\begin{figure}
\centering
\includegraphics[clip,width=\hsize]{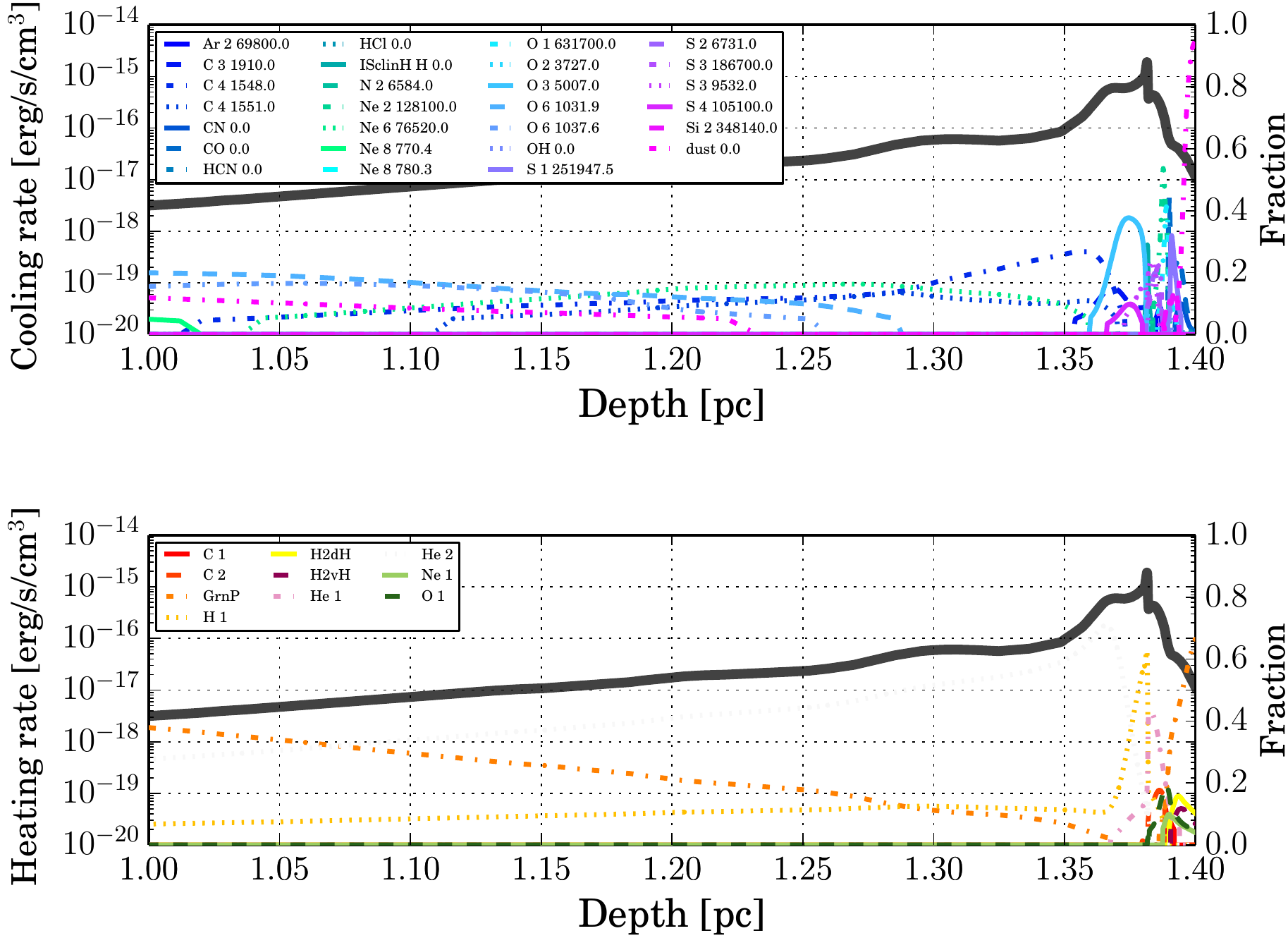}
\caption{Spacial profiles of cooling rate and heating rate (\textit{thick dark gray lines} in both panels) in the photo-ionization calculation shown in Fig.~\ref{fig:L20_arad_temp_metals}. The proportion of cooling (heating) rate of each process to the total cooling (heating) rate is shown by a colored line. In the upper panel, if a label that starts with an element name, an integral number and a real number following the element name indicate the ionization stage and the wavelength of the emission, respectively. The labels that start with a molecular name indicates cooling by emissions from that molecule. The meaning of the other labels in the upper panel is as follows: \texttt{dust 0.0} is the thermal emission from dust grains; \texttt{ISclinH H 0.0} is a cooling due to the H iso-electronic sequence; In the lower panel, photo-electric heating by an element is indicated by a label beginning with the element name and followed by an integral number that indicates its ionization stage. The meaning of the other labels in the lower panel is as follows: \texttt{GrnP} is the photo-electric heating by dust grains; \texttt{H2dH} is the heating due to $\mathrm{H_{2}}$ dissociation; \texttt{H2vH} is the heating due to collisions with $\mathrm{H_{2}}$.}
\label{fig:L20_coolheat_metals}
\end{figure}

Finally, we examine the effects of helium and metals. To this end, we perform a photo-ionization calculation taking into account these elements, but assuming a single-sized silicate grain of $a_{\mathrm{gr}}=0.05\;\mathrm{\mu m}$. The gas-phase abundance assumed in the calculation is shown in Table.~\ref{tbl:elemental_abundance}. This abundance pattern is similar to the \textsc{Cloudy}'s abundance set \texttt{ISM} except that (1) the abundance is increased by a factor of $1.75$ so that $f_{\mathrm{gr}}=0.01$ and (2) carbon locked up in graphite grains is returned to the gas phase (no graphite grains is assumed). The spacial distribution of the radiative acceleration and gas and dust temperatures are shown in Fig.~\ref{fig:L20_arad_temp_metals}. By comparing the left panel of Fig.~\ref{fig:L20_arad_temp_metals} with that of Fig.~\ref{fig:L20_rad_accl}, we can see that the radiative acceleration is smaller than that in the zero-metallicity case. This is mainly because (1) the mean molecular weight of ISM is increased by introducing helium and metals and (2) line absorption by metals saturates easily without a large velocity gradient. The right panel of Fig.~\ref{fig:L20_arad_temp_metals} shows a comparison of the gas and dust temperatures between the SPH simulation and the photo-ionization calculation. Near the shocked layer, the gas temperature in the SPH simulation is a factor of 1.3 larger than that in the photo-ionization calculation. In the photo-evaporation flow, the SPH simulation underestimates the gas temperature. To examine the source of these differences, we plot in Fig.~\ref{fig:L20_coolheat_metals} the spacial profile of the fraction of cooling (heating) rate of important processes to the total cooling (heating) rate. Near the shocked layer, emission lines by oxygen, neon, and sulfur contributes largely to the total cooling rate. Helium plays a role in heating the ISM (Depth $\approx 1.38\;\pc$). On the other hand, the photo-evaporation flow is predominantly heated by the photo-electric heating by dust grains. Thus, metals decreases the pressure gradient in the ionized region at the irradiated face of the cloud and the photo-evaporation rate will be smaller than those in the SPH simulations. For further details, radiation hydrodynamic simulations taking into account helium and metals are necessary and we will address this in the future.

\begin{table}
\centering
\begin{minipage}{\hsize}
\caption{Elemental abundance in gas-phase assumed in the photo-ionization calculation.}
\label{tbl:elemental_abundance}
\begin{tabular}{@{}ll|ll@{}}
\hline
Element & Abundance$^{\sharp}$ & Element & Abundance$^{\sharp}$\\
\hline
H  &  0.0      & S  & -4.246462 \\
He & -0.765762 & Cl & -6.756962 \\
Li & -10.02456 & Ar & -5.306762 \\
B  & -9.807562 & K  & -7.715562 \\
C  & -3.033152 & Ca & -9.144162 \\
N  & -3.857162 & Ti & -8.993562 \\
O  & -3.256613 & V  & -9.756962 \\
F  & -7.455962 & Cr & -7.756962 \\
Ne & -3.667062 & Mn & -7.395262 \\
Na & -6.257262 & Fe & -9.614866 \\
Mg & -4.678865 & Co & -7.986062 \\
Al & -6.857162 & Ni & -7.496862 \\
Si & -5.353955 & Cu & -8.580862 \\
P  & -6.552862 & Zn & -7.455962 \\
\hline
\end{tabular}
\begin{flushleft}
$^{\sharp}$ $\log_{10}(n_{X}/n_{\mathrm{H}})$ where $n_{X}$ is the number density of element $X$.
\end{flushleft}
\end{minipage}
\end{table}

\subsubsection{Effects of surrounding medium}\label{subsubsec:surrounding_medium}
In this study, we neglected the surrounding medium of the gas clouds. However, the gas clouds are actually confined by the surrounding medium if they are in the pressure equilibrium. The presence of the surrounding medium may affect the photo-evaporation rates from the gas clouds. In order to check this possibility, we perform one-dimensional (1D) spherically-symmetric RHD simulations for different densities of the surrounding medium. The details of the 1D RHD code will be presented elsewhere (Namekata et al., in prep.). Here, the following two cases are examined.
\begin{description}
\item[(Case I)] $\nH=10^{-4}\;\mathrm{cm^{-3}}$. This case mimics the vacuum condition assumed in the SPH simulations.
\item[(Case II)] $\nH\approx 8.33\;\mathrm{cm^{-3}}$. In this case, the gas cloud is initially in the pressure equilibrium with the ambient gas.
\end{description}
In both cases, the surrounding medium is assumed to be fully-ionized gas. Its initial gas temperature and $f_{\mathrm{gr}}$ are set to be $3\times 10^{4}\;\mathrm{K}$ and $10^{-10}$, respectively. The physical conditions of the gas cloud is the same as those of the model L20.

\begin{figure}
\centering
\includegraphics[clip,width=\linewidth]{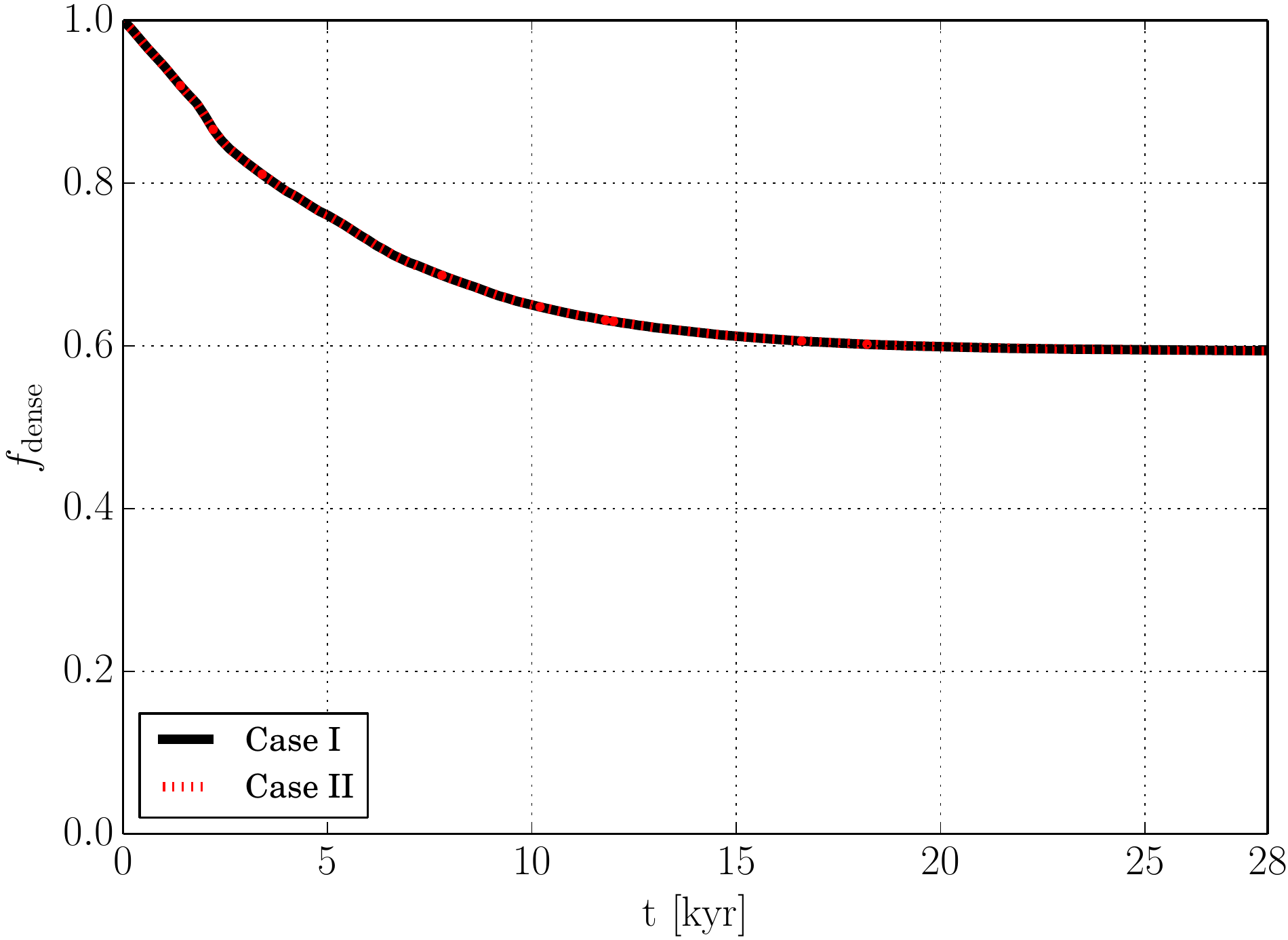}
\caption{Time evolution of $f_{\mathrm{dense}}$ in the 1D spherically-symmetric RHD simulations described in \S~\ref{subsubsec:surrounding_medium}. The \textit{black solid line} and the \textit{red dotted line} indicate Case I and II, respectively. The 1D RHD simulations are stopped if the shock reaches the center of the gas cloud.}
\label{fig:f_dense_comp_L20}
\end{figure}

Figure~\ref{fig:f_dense_comp_L20} shows time evolution of $f_{\mathrm{dense}}$, which is defined in \S~\ref{subsec:cloud_evaporation}, in the 1D RHD simulations. $f_{\mathrm{dense}}$ in both cases are very similar to each other. Thus, the surrounding medium has little impact on the evolution of the gas clouds. Similarly, the results of the 1D RHD simulations for the other models listed in Table.\ref{tbl:simulation_runs} do not depend on the density of the surrounding medium. However, if the surrounding medium contains dust grains with a normal abundance, a dusty gaseous shell, which will be formed at the front of the photo-evaporation flow, would stop the photo-evaporation when it becomes optically-thick. A time for the photo-evaporation to stop depends on the physical conditions of the surrounding medium, which should be determined by more realistic simulations.

\subsubsection{Effects of orbital and spin motions}\label{subsubsec:orbital_spin_motions}
As we have shown in \S~\ref{sec:numerical_results}, the clouds are compressed in a timescale $\tsc$ for Low-$\mathcal{U}$ case or $\tsweep$ for High-$\mathcal{U}$ case. We can neglect the orbital and spin motion of the clouds, if these timescales are much shorter than the orbital and spin periods. In a typical galaxy, circular velocity $\vorb$ is roughly $200\;\kms$, if the cloud is far from the SMBH. Then, the orbital period is $\torb=2\pi r/\vorb\approx 1.57\times 10^{6}\;\yr (r/50\;\pc)$. For the clouds located at $r=50\;\pc$, $\torb$ is much larger than both $\tsc$ and $\tsweep$ (see Eq.(\ref{eq:tsc}) and (\ref{eq:tsweep})). The spin period $\tspin$ has possibly the same order as the orbital period, if the cloud's spin is originated in the galactic shear\footnote{According to \citet{kim02:_three_simul_parker_magnet_swing}, the galactic shear velocity can be written as $\bmath{v}_{0}=-q\Omega x\hat{y}$ in a local Cartesian reference frame, where $\Omega$ is the orbital frequency and $q=(4-\kappa^{2}/\Omega^{2})/2$ and $\kappa\equiv \{r\frac{d\Omega^{2}}{dr}+4\Omega^{2}\}^{1/2}$ is the epicyclic frequency. Because of $q\sim \mathcal{O}(1)$, $\tspin \approx 2\pi\rcl/|-q\Omega\rcl|\approx 2\pi/\Omega=\torb$.}, although it is possible that the cloud spins up in the process of formation. Near a SMBH, these timescales can be estimated as Eq.(\ref{eq:app:tsweep}), (\ref{eq:app:torb}), and (\ref{eq:app:tsc}) for a particular AGN case (see \S~\ref{subsec:AGN_gas_clumps}) and $\torb$ is larger than $\tsc$ and $\tsweep$ at $r=1\;\pc$ (see also Fig.~\ref{fig:t_orb}). Therefore, we conclude that the orbital motion and spin of the clouds do not affect largely the global evolution of the clouds. But, we note that the infall motion toward a galactic center can affect the evolution of the cloud, if the cloud is initially located near the SMBH, because the tidal force increases significantly as the cloud approaches the galactic center (see \S~\ref{subsec:SC_models}).

\subsubsection{Effects of density structure and velocity fields} \label{subsubsec:density_velocity}
The shape of the cloud as well as the internal density structure and velocity fields are likely to have a great impact on how the cloud deformation occurs. Since we have assumed the spherical uniform cloud in this study, the counteraction of the photo-evaporation flow direct toward the center of the cloud in Low-$\mathcal{U}$ case, leading to formation of a dense filament along the symmetric axis of the cloud ($x$-axis in the simulation coordinate; see Fig.~\ref{fig:Time_Evolution_Low_U_models}d,h). This result likely changes if the cloud has a different shape and/or a density structure. The internal velocity fields also affect spin angular momenta of sub-clumps which will be formed by the self-gravitational instability in a later phase of the cloud evolution, which we cannot follow in this study because of very small timestep. The effects of these factors on the evolution of the clouds should be explored in the future.

\subsubsection{Other uncertainties} \label{subsubsec:other_uncertainties}
In Low- and High-$\mathcal{U}$ models, we used the same SED assumed in \citet{nenkova08a:_agn_dusty_tori} (see Eq.(\ref{eq:AGN_SED_Nenkova08})). The extreme-ultraviolet part of this spectrum is steeper than average spectrum of radio-quiet quasi-stellar objects derived by \citet{telfer02}. In order to check the effects of SED, we performed 1D spherically-symmetric RHD simulations assuming the SED described by Eq.(\ref{eq:AGN_SED_Schartmann05}), which is more similar to the average spectrum. We found that the numerical results are hardly different from those in which the Nenkova's SED is assumed. Therefore, we conclude that the shape of the SED does not affect the evolution of the gas clouds at least for the range of $(\mathcal{U},\mathcal{N}_{S})$ that are investigated in this paper.

\subsection{Comparison to previous studies}\label{subsec:comparison}
Evolution of an irradiated gas cloud have been studied extensively by many authors in various fields other than the field of AGN. For early phase of galaxy formation, photo-evaporation of a minihalo or a dwarf galaxy by the reionization photons have been discussed (e.g., \citealt{barkana99,shapiro04:_photoev,susa04:_format,susa04:_the_effec_of_early_cosmic,iliev05:_minih}) In the context of star formation, photo-evaporation of a gas clump or a protoplanetary disk that is exposed to a nearby massive young star is a subject of research (e.g., \citealt{oort55:_accel_inter_cloud_o_stars,kahn69,zeldovich69,bertoldi89,bertoldi90,lefloch94:_comet_i,draine96:_struc_of_station_photod_front,johnstone98:_photoev_disks_clump_nearb_masiv_stars,stoerzer99:_photod_orion,richling00:_photoev_protos_disks,gorti02:_photoev,susa06:_secon_iii,motoyama07:_hii,susa07:_photod_iii,susa09:_format_iii,hasegawa09:_radiat_popul_iii}). Here, we briefly compare the results of our study with those of some of previous studies in terms of cloud destruction timescale.

\begin{figure}
\centering
\includegraphics[clip,width=\linewidth]{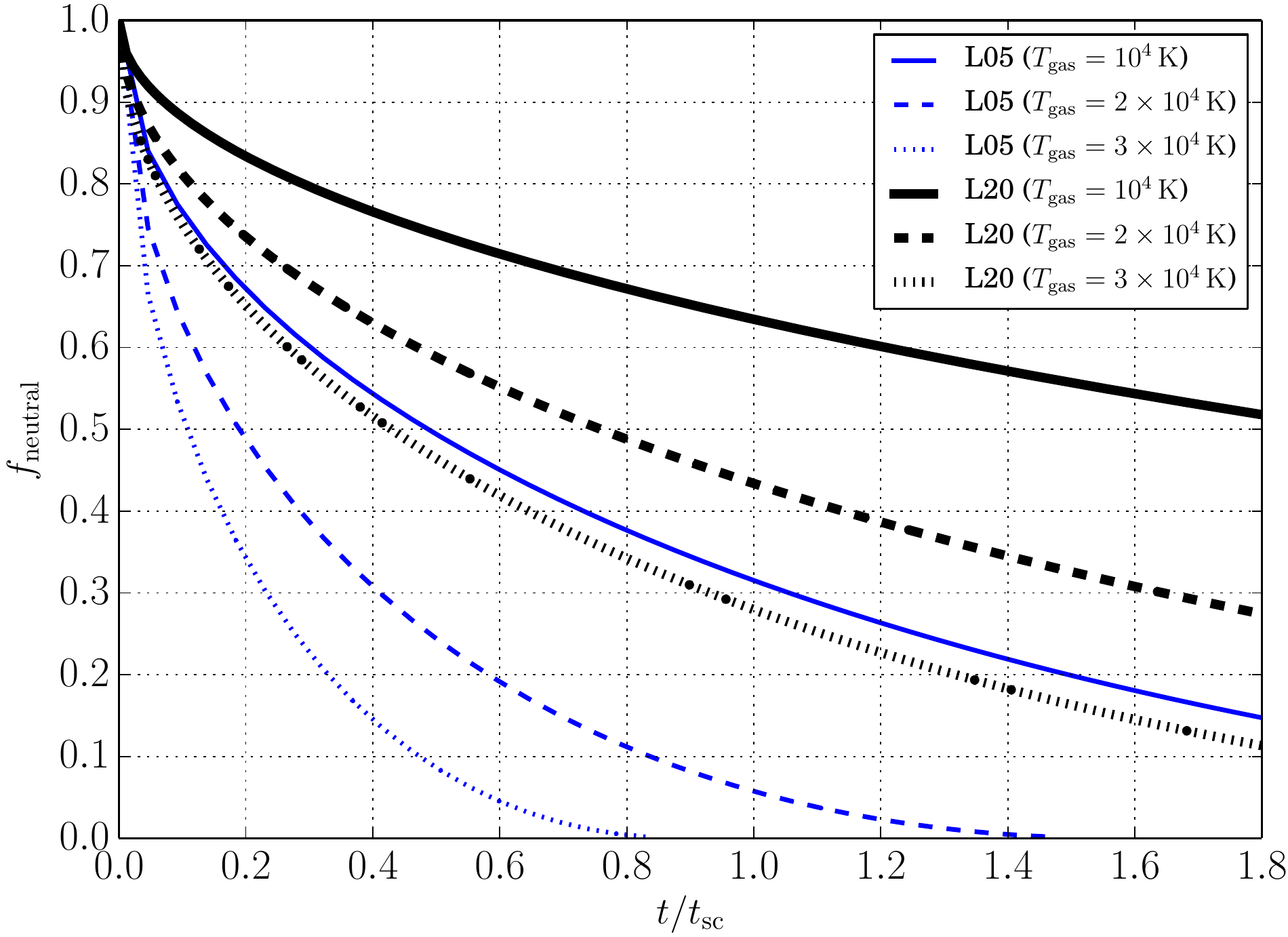}
\caption{Time evolution of mass fraction of the neutral gas predicted by Eq.(\ref{eq:outflow_rate_ZS69}). Time is normalized by the sound crossing time, Eq.(\ref{eq:tsc}). The \textit{blue} and \textit{black} lines denote the predictions for the models L05 and L20, respectively. The different line styles indicate the gas temperature of the ionized gas (see the legend in the figure for details). The ranges of both axis are adjusted to coincide with those of Fig.~\ref{fig:f_dense}}
\label{fig:frem_ZS69}
\end{figure}

\citet{zeldovich69} derived a self-similar solution of outflow of ionized gas from an ionization front for the case when the optical thickness of the outflowing gas becomes important. The self-similar solution is written as
\begin{eqnarray}
v & = & \frac{c_{\mathrm{T}}}{\sqrt{2}}+\frac{x}{t}, \\
n & = & \sqrt{\frac{\sqrt{2}F_{\mathrm{ion}}}{\alpha_{\mathrm{B}}c_{\mathrm{T}}}} t^{-1/2}\exp\left(-\frac{x}{\sqrt{2}c_{\mathrm{T}}t}\right),
\label{eq:solution_ZS69}
\end{eqnarray}
where $v$ is the velocity of the ionized gas, $n$ is the number density of proton, $c_{\mathrm{T}}$ is the isothermal sound speed, $F_{\mathrm{ion}}$ is the number flux of ionizing photon, $\alpha_{\mathrm{B}}$ is the case B recombination coefficient, $x$ is the distance from the ionization front, and $t$ is the time. The mass outflow rate at the front is given by
\begin{eqnarray}
\rho v|_{\mathrm{IF}} & = & m_{\mathrm{H}} \frac{c_{\mathrm{T}}}{\sqrt{2}}\left(\frac{\sqrt{2}F_{\mathrm{ion}}}{\alpha_{\mathrm{B}}c_{\mathrm{T}}}\right)^{1/2} t^{-1/2}.
\label{eq:outflow_rate_ZS69}
\end{eqnarray}
Using Eq.(\ref{eq:outflow_rate_ZS69}) and assuming that the outflowing gas streams along lines that connect between the radiation source and the front, we can estimate the photo-evaporation rate for the cloud models described in Table~\ref{tbl:simulation_runs}. Figure~\ref{fig:frem_ZS69} shows the time evolution of the mass fraction of the \textit{neutral} gas, $f_{\mathrm{neutral}}$. For a given gas cloud, $f_{\mathrm{neutral}}$ monotonically decreases with time and its decreasing rate is larger for higher gas temperature. Also, a smaller gas cloud loses neutral gas more rapidly as expected. The comparison of Fig.~\ref{fig:frem_ZS69} with Fig.~\ref{fig:f_dense} shows that the time evolution of $f_{\mathrm{neutral}}$ is in disagreement with that of $f_{\mathrm{dense}}$. Especially, $f_{\mathrm{neutral}}$ does not saturate unlike $f_{\mathrm{dense}}$ and the mass of the neutral gas is underestimated. Thus, we should only use the self-similar solution by \citet{zeldovich69} to obtain a rough estimate of photo-evaporation timescale from an spherical cloud that is irradiated from one direction.

\citet{bertoldi89} investigated the structure and evolution of a gas clump that is exposed to a nearby early-type star for a wide range of conditions by solving the steady state hydrodynamic equations coupled with the ionization equations for hydrogen and helium. The clouds in Low- and High-$\mathcal{U}$ cases belong to the region III in his study (the clouds in our study have $\log\Gamma=-1.886\operatorname{-}-1.284$ and $\delta'=0.2\operatorname{-}0.05$ and see Fig.1 in his study) and have the photo-evaporation parameter $\psi\equiv \alpha_{\mathrm{B}}F_{\mathrm{ion}}\rcl/c^{2}_{s,i}>10^{6}$, where $c_{s,i}$ is the gas temperature of ionized gas. Although a result for the case of $\psi>10^{6}$ was not shown in his study, a simple extrapolation of the result for $\psi=10^{3}$ and $\delta'=10^{-0.5}$ (his Fig.8c) suggests that for such clouds a strong ionization-shock (IS) front forms at the irradiated face and it sweeps the neutral gas into the symmetry axis of the cloud, resulting in a cylindrical accretion shock. Such evolution is exactly seen in the numerical results for Low-$\mathcal{U}$ cases (Fig.~\ref{fig:Low_U_models_Density_enlarged}a,b). He also estimated the IS-front crossing time for the case of $\psi>10$,
\begin{eqnarray}
t_{\mathrm{IS}} & = & 2\times 10^{5}\;\yr\;\psi^{-1/4}\left(\frac{\Mcl}{\Msolar}\right)^{1/2} \label{eq:tIS}.
\end{eqnarray}
Table \ref{tbl:tIS_tend} lists $t_{\mathrm{IS}}$ for our cloud models (Table \ref{tbl:simulation_runs}), together with $t_{\mathrm{end}}$, which is the time when the SPH simulation is stopped and corresponds to the actual shock crossing time. $t_{\mathrm{IS}}$ is in very good agreement with $t_{\mathrm{end}}$ for small $\mathcal{N}_{S}$ or small $\mathcal{U}$ models (L05, L10, and H05). For larger $\mathcal{N}_{S}$ models (H20 and \SCmodel), $t_{\mathrm{IS}}$ is poorly matched with $t_{\mathrm{end}}$. A large difference between $t_{\mathrm{IS}}$ and $t_{\mathrm{end}}$ for the model \SCmodel is certainly due to that radiation pressure is not taken into account in the Bertoldi's formula $t_{\mathrm{IS}}$. Thus, $t_{\mathrm{IS}}$ is a good estimator for shock crossing time only for smaller $\mathcal{N}_{S}$ and $\mathcal{U}$ cases.

Finally, we comment on the relation between the radiation pressure stripping described in \S~\ref{subsec:Time_Evolution_High_U_models} and the derimming process described in \citet{mathews86:_struc}, in which he studied the structure and stability of a broad emission line cloud nearby a quasar semi-analytically and estimated that the periphery (rim region) of a flatten cloud is preferentially stripped off in a timescale of $< 10 \tsc$ in a typical environment nearby a quasar because column density-averaged radiative acceleration, $\overline{|\bmath{a}_{\mathrm{rad}}|}$, is larger in the periphery than the central part of the cloud. A difference of $\overline{|\bmath{a}_{\mathrm{rad}}|}$ also exists in our simulations and this derimming process plays a role in stripping the gas of the cloud. But, unlike the pure derimming process, a part of the stripped gas comes from the central part of the cloud in the radiation pressure stripping by the effect of the photo-evaporation (Fig.~\ref{fig:Acceleration_H20}).

\begin{table}
\centering
\begin{minipage}{\hsize}
\caption{Comparison of $t_{\mathrm{IS}}$ with $t_{\mathrm{end}}$.}
\label{tbl:tIS_tend}
\begin{tabular}{@{}lll@{}}
\hline
Model name & $t_{\mathrm{IS}}$ $^{\sharp}$ & $t_{\mathrm{end}}$ $^{\dagger}$ \\
& [kyr] & [kyr] \\
\hline
L05      & 12.1 & 12   \\
L10      & 28.9 & 28.5 \\
L20      & 68.7 & 54   \\
H05      & 48.6 & 47.5 \\
H10      & 115  & 88.5 \\
H20      & 274  & 180  \\
\SCmodel & 132  & 53   \\
\hline
\end{tabular}
\begin{flushleft}
$^{\sharp}$ The IS-front crossing timescale defined by Eq.(\ref{eq:tIS}) where we assumed $c_{s,i}=11.4\;\kms$ as with \citet{bertoldi89}. A higher $c_{s,i}$ results in a lower $t_{\mathrm{IS}}$ because of $t_{\mathrm{IS}}\propto c_{s,i}^{-0.5}$. \\
$^{\dagger}$ The times when the SPH simulations are stopped.
\end{flushleft}
\end{minipage}
\end{table}

\subsection{Star formation} \label{subsec:star_formation}
Star formation in AGN-irradiated gas clouds plays important roles in determining the gas supply rate for the following reasons. First, once gas is converted into stars, it is very difficult to supply the gas locked in the stars into a SMBH unless the stars are located at a distance enough close to the SMBH that the tidal disruption mechanism works. Then, what fraction of the gas in the cloud is converted into stars is quite important. Second, it could affect self-gravitational stability of the shocked layer via stellar feedback processes such as molecular outflow and stellar radiation. If the stellar feedback is strong enough that part of the shocked layer is brown off into the direction to the AGN, the photo-evaporation might be enhanced, changing longevity of the cloud.

\begin{figure}
\centering
\includegraphics[clip,width=\linewidth]{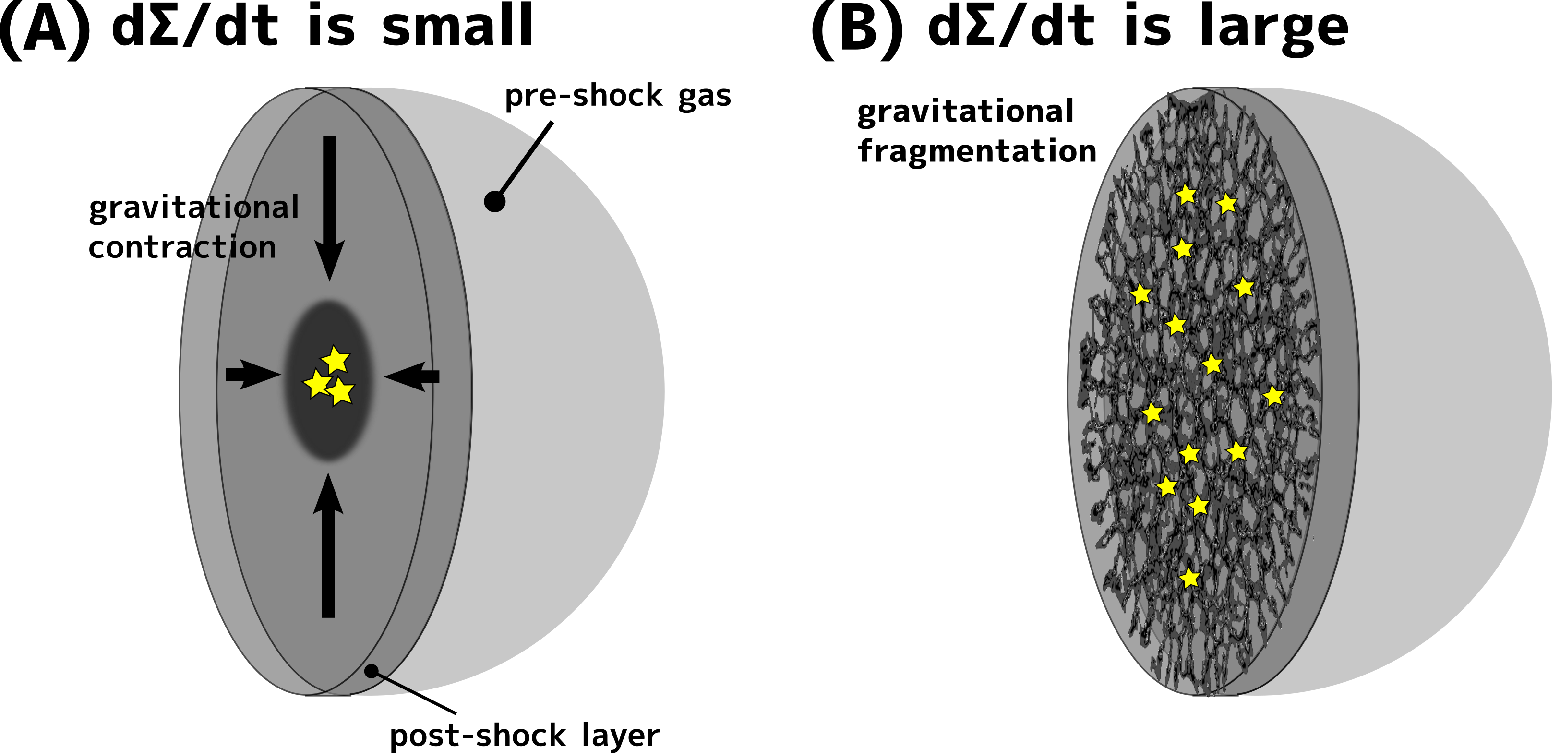}
\caption{A schematic illustration of two types of star formation in a gas cloud exposed to AGN radiation.
If an increasing rate of surface density of a post-shock layer, $\dot{\Sigma}_{\mathrm{gas}}$, is small,
star formation occurs in a dense gas core at the center of the post-shock layer,
which is formed by global gravitational collapse of the post-shock layer (\textit{left}).
On the other hand, when $\dot{\Sigma}_{\mathrm{gas}}$ is sufficiently large, star formation will begin
in dense gas filaments in the post-shock layer,
which are formed by (two-dimensional) self-gravitational instability of the post-shock layer (\textit{right}).
\label{fig:SFtype}}
\end{figure}

Star formation probably occurs in a gas cloud having the same $(\mathcal{U},\mathcal{N}_{S})$ that is investigated in this study, if the initial cloud is sufficiently massive, since density of a dense filament or a shocked layer is quite high. For example, a self-gravitating dense clump is formed in the model H20 as already discussed in \S~\ref{subsec:Time_Evolution_High_U_models}. The density in the center of the clump increases as large as $10^{12}\;\mathrm{cm^{-3}}$ (Fig.~\ref{fig:Time_Evolution_H20_rhoT}), although we do not include metal line cooling and the gas is artificially heated by the temperature floor. The thermal pressure in the center of the clump is comparable to that of first core in low-mass star formation simulations (e.g., see $\rho$-$\Tgas$ planes in \citealt{masunaga98:_radiat_hydrod_model_protos_collap,masunaga00:_radiat_hydrod_model_protos_collap,tomida12:_radiat_magnet_simul_protos_collap}). This suggests that the dense clump is eligible to form stars. If we take into account line cooling (e.g., CO, C{\scriptsize II}, O{\scriptsize I}) and remove the temperature floor, this clump will fragment into sub-clumps rapidly and collapse dynamically. Such process is reminiscent of current massive star formation scenario (e.g., \citealt{zinnecker07:_towar_under_massiv_star_format,bergin07:_cold_dark_cloud}). Unlike the model H20, a different type of star formation will be expected in the model \SCs, although the initial cloud properties other than $\mathcal{U}$ are similar each other (see Table~\ref{tbl:simulation_runs}). In the model \SCs, the shock velocity is higher than the model H20. Consequently, the surface density of the post-shock layer increases rapidly compared to the model H20, resulting in the self-gravitational fragmentation in the course of the passage of the shock. On the other hand, in the model H20, the gas in the post-shock layer falls into the center of the layer. Thus, there can be two types of star formation depending on the increasing rate of the surface density (Fig.~\ref{fig:SFtype}). 

In connection with the current discussion, \citet{hocuk10:_x,hocuk11:_star_agn_variat} recently investigate the influence of X-ray emitted by an AGN on the initial mass function (IMF) in a gas cloud by performing three-dimensional hydrodynamic simulations with the X-ray transfer calculation. They assume that the cloud is in a location where most of the optical/UV lights are blocked by many intervening clouds and the X-ray only illuminates the cloud. Their cloud shows similar evolution with that of the lower $\mathcal{U}$ cases in this study. They show that the X-ray heating makes gaseous fragments more massive and the resulting IMF becomes top-heavy compared to the Salpeter IMF. In contrast, the gas clouds that are investigated in this study are irradiated directly by an AGN and in this sense our study is complementary with their studies, although we do not take into account star formation at this time. It is very interesting to examine the star formation properties in a directly-irradiated gas clouds and we will address this issue in a future study.

\subsection{Implications for gas clumps in AGN tori} \label{subsec:AGN_gas_clumps}
Here, we discuss possible effects of the AGN radiation on the evolution of gas clumps in AGN tori and mass supply process based on our results. To this end, we first derive the physical properties of the gas clumps in the AGN tori. They can be estimated by considering stabilities against both the tidal shearing by the SMBH and the internal pressure (see appendix in \citealt{kawaguchi11:_near_infrar_rever_by_dusty}). For the sake of completeness, we re-derive a part of the results relevant to our study employing the following assumptions: (1) any external pressures are ignored, and (2) we consider the gravity of the SMBH alone.

From the tidal stability condition, we can determine uniquely the mass density of the clump that located at a distance of $r$ from the SMBH as,
\begin{eqnarray}
\rho_{\mathrm{cl}} & = & 3.54\times 10^{-16}\;\mathrm{g\;cm^{-3}} \left(\dfrac{\MBH}{10^{7}\;\Msolar}\right)\left(\dfrac{r}{1\;\pc}\right)^{-3}. \label{eq:rhocl_AGN}
\end{eqnarray}
The corresponding number density of the hydrogen nuclei is
\begin{eqnarray}
n_{\mathrm{H,cl}} & = & 2.12\times 10^{8}\;\mathrm{cm^{-3}}\left(\dfrac{\MBH}{10^{7}\;\Msolar}\right)\left(\dfrac{r}{1\;\pc}\right)^{-3}. \label{eq:nHcl_AGN}
\end{eqnarray}
The radius of the clump can be determined by the balance between its internal pressure and self-gravity, namely by the Jeans length, as
\begin{eqnarray}
\rcl & = & 0.0182\;\pc \left(\dfrac{\MBH}{10^{7}\;\Msolar}\right)^{-0.5} \nonumber \\
&& \quad \times\left(\dfrac{r}{1\;\pc}\right)^{1.5}\left(\dfrac{c_{s}}{3\;\kms}\right), \label{eq:rcl_AGN}
\end{eqnarray}
where the normalization of the sound speed is based on the result of \citet{krolik89:_physic_state_obscur_torus_seyfer_galax}, in which they showed that the gas temperature of the clump is mainly determined by the X-ray radiation and is $\approx 10^{3}\;\mathrm{K}$, leading to $c_{s}\approx 3\;\kms$. We can take into account the magnetic pressure by replacing $c_{s}$ by $c_{\mathrm{eff}}\equiv (c_{s}^{2}+c_{A}^{2})^{0.5}$, where $c_{A}$ is the Alfv{\' e}n velocity, although the strength of the magnetic fields inside the clump is currently unknown.
The mass and hydrogen column density of the clump are
\begin{eqnarray}
\Mcl & = & 120\;\Msolar \left(\dfrac{\MBH}{10^{7}\;\Msolar}\right)^{-0.5}\nonumber \\
&& \quad \times \left(\dfrac{r}{1\;\pc}\right)^{1.5}\left(\dfrac{c_{s}}{3\;\kms}\right)^{3}, \label{eq:Mcl_AGN} \\
\NHcl & = & 1.2\times 10^{25}\;\mathrm{cm^{-2}} \left(\dfrac{\MBH}{10^{7}\;\Msolar}\right)^{0.5} \nonumber\\
&& \quad \times \left(\dfrac{r}{1\;\pc}\right)^{-1.5}\left(\dfrac{c_{s}}{3\;\kms}\right), \label{eq:NHcl_AGN}
\end{eqnarray}
The size, mass, and column density of the clump can be smaller values if the clump is confined by the external pressure.
Note that the values above are just rough estimates, because we ignore dynamical effects.
For example, if the creation and the destruction of the clumps are repeatedly occurred in the AGN torus and the clumps are merely transient objects, we cannot necessarily consider the clumps are stable against the tidal shear.

\begin{figure*}
\centering
\includegraphics[clip,width=\linewidth]{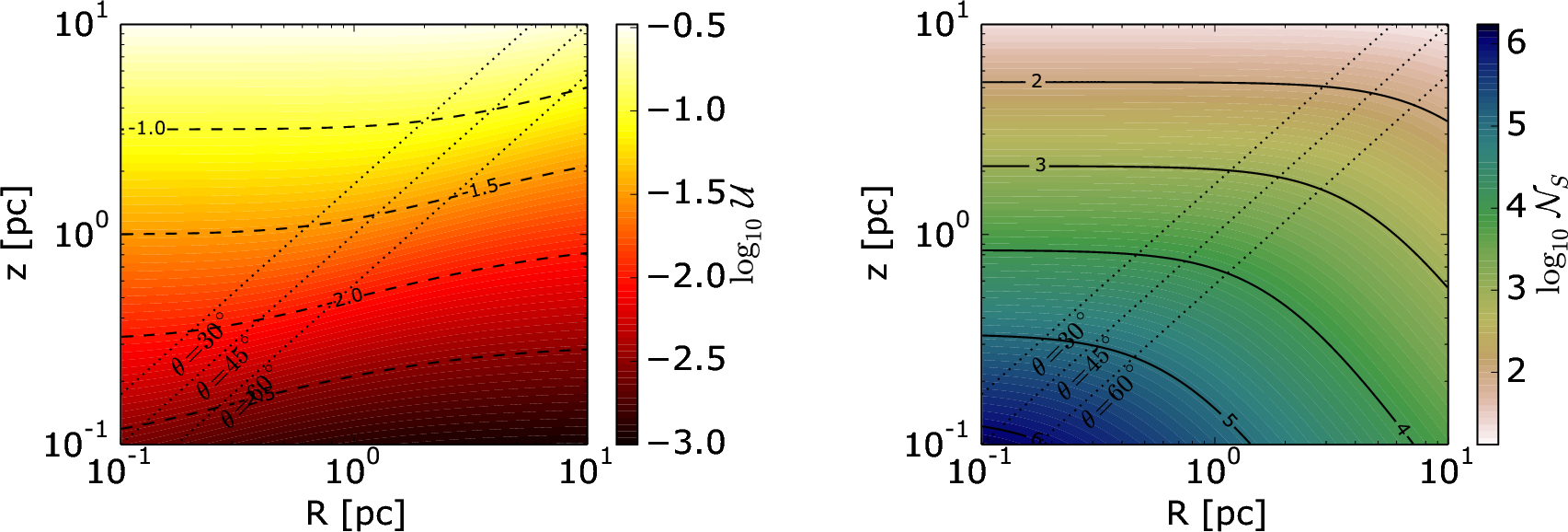}
\caption{The spacial distributions of $\mathcal{U}$ and $\mathcal{N}_{S}$ for unshielded gas clumps in an AGN torus in the case of $\Lbol=10^{45}\;\mathrm{erg\;s^{-1}}$ and $\MBH = 10^{7}\;\Msolar$ (This value of $\Lbol$ is almost the same as the classical Eddington luminosity). The $z$-axis is the symmetric axis of the accretion disk and $R$ is the distance from the SMBH along the equatorial plane of the AGN torus. We assume the symmetric axis of the accretion disk is aligned with the symmetric axis of the AGN torus. We also assume that the radiation from the accretion disk is anisotropic (\citealt{netzer87:_quasar}). For simplicity, we assume that the X-ray, which is isotropically radiated from the accretion disk corona, has the same flux distribution as \citet{netzer87:_quasar}. The dotted lines show $z=\tan(\pi/2-\theta)R$, where $\theta$ is the angle measured from the symmetric axis of the accretion disk. In the calculations of $\mathcal{U}$ and $\mathcal{N}_{S}$, we used the equations (\ref{eq:nHcl_AGN}) and (\ref{eq:rcl_AGN}).}
\label{fig:U_Ns_AGN_tori}
\end{figure*}

Using these values and assuming the isotropic AGN radiation, we can obtain the ionization parameter $\mathcal{U}$ and the Str{\" o}mgren number $\mathcal{N}_{S}$ for the clump that is directly irradiated by the AGN as
\begin{eqnarray}
\mathcal{U} & = & 1.3\times 10^{-2} \left(\dfrac{\Lbol}{10^{45}\;\mathrm{erg\;s^{-1}}}\right) \nonumber\\
&& \quad \times \left(\frac{\MBH}{10^{7}\;\Msolar}\right)^{-1}\left(\dfrac{r}{1\;\pc}\right), \nonumber\\
& \approx & 1.3\times 10^{-2} \left(\dfrac{r}{1\;\pc}\right)\left(\dfrac{\lambda_{\mathrm{Edd}}}{1}\right), \\
\mathcal{N}_{S} & = & 1.67\times 10^{4} \left(\dfrac{\Lbol}{10^{45}\;\mathrm{erg\; s^{-1}}}\right)^{-1}\left(\dfrac{\MBH}{10^{7}\;\Msolar}\right)^{-0.5} \nonumber\\
&& \quad \times  \left(\dfrac{r}{1\;\pc}\right)^{-2.5}\left(\dfrac{c_{s}}{3\;\kms}\right), \label{eq:app:Ns}
\end{eqnarray}
where we assume the SED shown in Eq.(\ref{eq:AGN_SED_Nenkova08}) and $\lambda_{\mathrm{Edd}}$ is the Eddington ratio. The Str{\" o}mgren length of the clumps $l_{s}$ is
\begin{eqnarray}
l_{S} & = & 0.45\;\mathrm{AU} \left(\frac{\Lbol}{10^{45}\;\mathrm{erg\;s^{-1}}}\right) \nonumber\\
&& \quad \times \left(\dfrac{\MBH}{10^{7}\;\Msolar}\right)^{-2}\left(\dfrac{r}{1\;\pc}\right)^{4},
\end{eqnarray}
where we $\alpha_{\mathrm{B}}=2.59\times 10^{-13}\;\mathrm{cm^{3}\;s^{-1}}$ (\citealt{hui97:_equat}).
Thus, the clumps in the AGN torus are extremely optically-thick if the clumps are stable for the tidal shear and the self-gravity. Figure \ref{fig:U_Ns_AGN_tori} shows the spacial distributions of $\mathcal{U}$ and $\mathcal{N}_{S}$ of \textit{unshielded} gas clumps in the AGN torus in the case of $\Lbol=10^{45}\;\mathrm{erg\;s^{-1}}$ and $\MBH=10^{7}\;\Msolar$. In the calculations, we assume the SED described in \S\ref{subsec:AGN_SED} and take into account the anisotropy of the radiation from the accretion disk. The range of $\mathcal{U}$ that is investigated in this study is $(1.3\operatorname{-}5.2) \times 10^{-2}$, which corresponds to the surface layer of the AGN torus at $r=1\;\pc$. Based on our numerical results, the photo-evaporation flow may be launched from the surface layer of the AGN torus and then the flow are blown off by the radiation force on the dust. If the AGN torus is sufficiently clumpy, the AGN radiation might drive the photo-evaporation of the clumps that lie deep in the torus. Such evaporation flow may contribute the obscuration of the AGN and affect the mass supply rate to the galactic center as with the simulations by \citet{wada12:_radiat}.  

\begin{figure}
\centering
\includegraphics[clip,width=\linewidth]{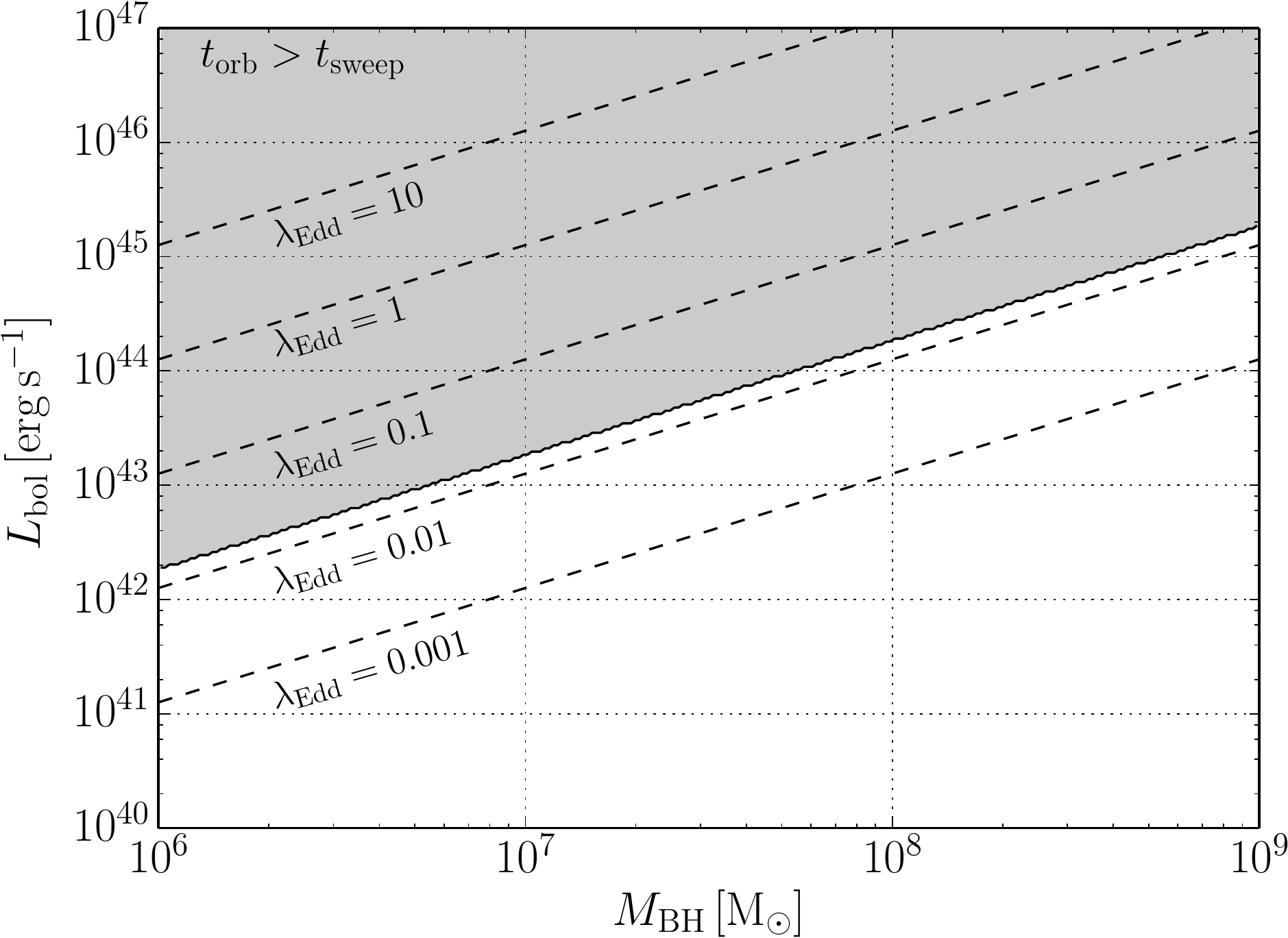}
\caption{A parameter space of ($\MBH$, $\Lbol$) where $\torb > \tsweep$ at $r=1\;\pc$ (the filled region). The \textit{dashed lines} show bolometric luminosities for $\lambda_{\mathrm{Edd}}=10^{-3}\operatorname{-}10$.}
\label{fig:t_orb}
\end{figure}

For the clump located at $r$, we can estimate the velocity of the shocked layer and the sweep time as 
\begin{eqnarray}
\vshapp & = & 9.04\;\kms \left(\dfrac{\Lbol}{10^{45}\;\mathrm{erg\;s^{-1}}}\right)^{0.5} \nonumber\\
&& \quad \times \left(\dfrac{\MBH}{10^{7}\;\Msolar}\right)^{-0.5}\left(\dfrac{r}{1\;\pc}\right)^{0.5}, \\
\tsweep & = & 4016\;\yr \left(\dfrac{\Lbol}{10^{45}\;\mathrm{erg\;s^{-1}}}\right)^{-0.5} \nonumber\\
&& \quad \times \left(\dfrac{r}{1\;\pc}\right)\left(\dfrac{c_{s}}{3\;\kms}\right), \label{eq:app:tsweep}
\end{eqnarray}
where we again assume the AGN radiates isotropically. This razor-thin approximation breaks down if $\vshapp < c_{s}$, because the shock does not form. If $c_{s}\approx 3\;\kms$ throughout the torus, the critical radius is
\begin{eqnarray}
r_{\mathrm{cr}} & = & 0.1\;\pc \left(\dfrac{\Lbol}{10^{45}\;\mathrm{erg\;s^{-1}}}\right)^{-1} \left(\dfrac{\MBH}{10^{7}\;\Msolar}\right), \label{eq:rcr}
\end{eqnarray}
which is coincidentally close to the dust sublimation radius for this AGN model that we're looking at.
At $r<r_{\mathrm{cr}}$, the clumps are simply pushed by the radiation force if $\mathcal{U}$ is high. For small $\mathcal{U}$ cases, $\vshapp$ may not be a good approximation according to the results of Low-$\mathcal{U}$ models and therefore $r_{\mathrm{cr}}$ may be different from Eq.(\ref{eq:rcr}). In order to investigate the evolution of clouds with small $\mathcal{U}$ and high $\mathcal{N}_{S}(>10^{3})$, Namekata et al. (in prep.) have performed one-dimensional spherically-symmetric RHD simulation for the cloud having $\mathcal{U}\approx 1.3\times 10^{-2}$ and $\mathcal{N}_{S}\approx 1300$, and have found that the shock velocity averaged over the cloud evolution is roughly close to $\vshapp$ because the photo-evaporation flow is finally confined by the radiation pressure and the rocket effect becomes inefficient. Therefore, $\tsweep$ is a good approximation for the cloud destruction time if $\mathcal{N}_{S}$ is extremely high. For this reason, we expect that the clumps are not destroyed at $r<\mathrm{r}_{\mathrm{cr}}$ if $\mathcal{U}$ is small. It is useful to compare $\tsweep$ with other various timescales. The orbital time and the sound crossing time are 
\begin{eqnarray}
t_{\mathrm{orb}} & = & 2.98\times 10^{4}\;\yr \left(\dfrac{\MBH}{10^{7}\;\Msolar}\right)^{-0.5}\left(\dfrac{r}{1\;\pc}\right)^{1.5}, \label{eq:app:torb} \\
\tsc & = & \dfrac{2\rcl}{c^{\mathrm{irr}}_{s}}  =  1.27\times 10^{3}\;\yr \left(\dfrac{\MBH}{10^{7}\;\Msolar}\right)^{-0.5} \nonumber\\
&& \qquad\quad \times \left(\dfrac{r}{1\;\pc}\right)^{1.5}\left(\dfrac{c_{s}}{3\;\kms}\right), \label{eq:app:tsc}
\end{eqnarray}
where $c^{\mathrm{irr}}_{s}$ is the sound speed of fully-ionized pure hydrogen gas of $\Tgas=30000\;\mathrm{K}$. Thus, both $\tsweep$ and $\tsc$ are much smaller than $t_{\mathrm{orb}}$ and the gas clumps should be compressed in a short time if they are directly irradiated. This may suggest that the AGN tori with $\lambda_{\mathrm{Edd}}\approx 1$ can be short-lived compared to the orbital timescale. Because $\tsweep \propto \mathcal{U}^{-0.5}$ (see Eqs.(\ref{eq:vshapp2}) and (\ref{eq:tsweep})), $\tsweep$ remains smaller than $\torb$ even for the AGN with $\lambda_{\mathrm{Edd}}\approx 0.1$, which is a typical Eddington ratio (e.g., \citealt{lusso12:_bolom_eddin_x_xmm_cosmos}). Figure~\ref{fig:t_orb} shows combinations of ($\MBH$, $\Lbol$) where $\torb>\tsweep$ at $r=1\;\pc$.

On the other hand, if the actual AGN torus maintains its clumpy structure for long period, there are the following possibilities:
\begin{description}
\item[I)] The most of the clumps in the torus are exposed to sufficiently weak radiation field (small $\mathcal{U}$) and therefore the shock does not occur in the clumps. The causes to reduce $\mathcal{U}$ include (i) the anisotropy of the AGN radiation, (ii) the higher averaged density of the clumps than the critical density with respect to the tidal shear, and (iii) extinction by the inter-clump medium. Another interesting possibility is that a dusty wind that launched from the inner edge of the torus shields the main body of the torus as if the hitch-hiking gas does in the accretion disk wind model (\citealt{murray95:_accret_disk_winds_activ_galac_nuclei}). This possibility seems to be compatible with the recent numerical study (\citealt{wada12:_radiat}) and observational studies (\citealt{czerny11,tristram12:_agn,onig12:_parsec_scale_dust_emiss_from,onig13:_dust_polar_region_majour_contr}).
\item[II)] The clumps are destroyed by the shock in short timescales, but, the creation of the clumps are continuously occurred to offset the destruction of the clumps by some mechanism. The stellar feedback-driven turbulence might be a candidate mechanism.
\end{description}
Obviously, further studies are needed to examine these possibilities.

\section{Summary} \label{sec:summary}
In this paper, we have performed 3D RHD simulations of the gas clouds exposed to the AGN radiation varying the ionization parameter $\mathcal{U}$ and the Str{\"o}mgren number $\mathcal{N}_{S}$ to investigate combined effects of the photo-evaporation and the radiation pressure force on the evolution of the cloud. We have found that the evolution of the clouds can be classified into two cases depending on $\mathcal{U}$:
\begin{enumerate} 
\item In Low-$\mathcal{U}$ case ($\mathcal{U}\approx 1.3\times 10^{-2}$), the photo-evaporation determines the evolution of the cloud independent of $\mathcal{N}_{S}$. The photo-evaporation flow is launched from the irradiated face. The flow is almost spherical and its velocity is $\approx 100\;\kms$. The mass loss is realized as this photo-evaporation flow and the gas clouds are compressed by the counteraction of the flow. The compression is completed in a timescale comparable with $\tsc$.

\item In High-$\mathcal{U}$ case ($\mathcal{U}\gtrsim 5.2\times 10^{-2}$), the radiation pressure turn the photo-evaporation flow that launched from the outskirt of the irradiated face into the direction opposite to the AGN, outward from the the galactic center, while it confines the flow that launched from the central part of the irradiated face. Because of this, fractional mass loss rate is smaller than Low-$\mathcal{U}$ case, at least until $t \lesssim \tsc$. The shocked layer that is formed by both the radiation pressure and the counteraction of the photo-evaporation flow sweep the main body of the cloud in a timescale of $\tsweep$.
\end{enumerate}

In both cases of $\mathcal{U}$, mass fraction of dense gas is larger for larger $\mathcal{N}_{S}$. Star formation will occurs in the compressed part of the cloud, if the cloud has initially sufficient mass to form stars. We have analyzed a self-gravitating dense clump formed at the center of the shocked layer in the model H20 and have found that the clump has physical properties similar to those in high-mass star-forming region in the Galaxy. We speculate high-mass star formation occurs in this clump.

We also have performed simulations of more higher-$\mathcal{U}$ case ($\mathcal{U}\approx 0.17$; the models \SCs and \SCff) in order to investigate the evolution of the cloud under a more stronger radiation field as well as the dependency on the initial condition. In the model \SCs, the cloud is simply destroyed by the passage of the shock driven by the intense radiation pressure and no photo-evaporation occurs. On the other hand, in the model \SCff, which models a cloud that infalls in an external gravitational potential of a SMBH and a nuclear star cluster, the propagation speed of the shock is decreased by the effects of the tidal force. As a result, the pre-shock gas survives longer than that in \SCs. However, it is expected that the cloud collapses by the transverse component of the total gravity before it reaches the galactic center. In order to elucidate the gas supply process to the galactic center by a low-angular momentum cloud such as the cloud in \SCff, we have to investigate how the star formation proceeds in the cloud.

Based on the numerical results, we have discussed the properties of the gas clumps in the AGN torus with high Eddington ratio in \S~\ref{subsec:AGN_gas_clumps}. A simple estimate suggests that the clumps are destroyed in timescales that are shorter than the orbital period. For the clumpy structure to be maintained over long period, the incident radiation field needs to be sufficiently weaken $\mathcal{U}$ for most of the clumps, or, some mechanism that creates the clumps continuously is needed.

\section*{Acknowledgments}
We thank Ataru Tanikawa for his advice about the implementation of the Phantom-GRAPE library, and Toshihiro Kawaguchi for helpful discussions about properties of AGN tori. We would like to thank Nitadori Keigo for publication of the Phantom-GRAPE library on the Web. We also thank the anonymous referee for his/her helpful suggestions that improved our manuscript. Finally, we thank Akitoshi Oshima who greatly improved the computing environments of the medium-scale PC clusters installed at Center for Computational Astrophysics (CfCA) of National Astronomical Observatory of Japan (NAOJ) to meet our difficult requests, and the developers of the softwares \texttt{DISLIN}, \texttt{matplotlib}, and \texttt{Asymptote}, with which we visualized our numerical results. The numerical simulations were carried out on Cray XT4, Cray XC30, and the medium-scale PC clusters at CfCA of NAOJ, and FIRST and T2K Tsukuba at Center for Computational Sciences (CCS) in University of Tsukuba. This work was supported by the Ministry of Education, Culture, Sports, Science and Technology (MEXT) Grant-in-Aid for Scientific Research (S)20224002 (M.U.), MEXT Grant-in-Aid for Young Scientists (B)25800100 (D.N.), and by MEXT SPIRE Field 5 and JICFuS.

\bibliographystyle{mn2e}
\bibliography{ms}

\appendix

\section{Doppler-shift and dust-gas coupling} \label{appendix:sec:doppler_and_dust}
In this study, we ignored the effects of the Doppler shift and assumed that dust is tightly coupled with gas. Here, we discuss these assumptions briefly. The inclusion of the Doppler shift does not change our results largely as far as the motion of photo-evaporation flow is mainly determined by the absorption of continuum photons by neutral hydrogen and dust grains, because the shape of incident spectrum keeps almost unchanged even if we take into account the Doppler shift (the wavelength shift is at most $\approx 1\ring{\mathrm{A}}$ at $\lambda=10^{3}\ring{\mathrm{A}}$ for $v=300\;\kms$). However, in a situation that the motion of photo-evaporation flow is controlled by the absorption of line photons, the Doppler shift must play important roles to determine the velocity structure of the flow, as in the case of accretion disk wind (e.g., \citealt{murray95:_accret_disk_winds_activ_galac_nuclei}). Such situation is realized if dust is completely destroyed by thermal sputtering process and metal opacities become important. The dust destruction by the thermal sputtering was studied by \citet{draine79}. Based on their results, the dust destruction timescale $\tau^{\mathrm{dest}}_{\mathrm{gr}}$ is given by
\begin{eqnarray}
\tau^{\mathrm{dest}}_{\mathrm{gr}} \sim \left\{
\begin{array}{ll}
100 \;\myr  \left(\dfrac{\Tgas}{56000\;\mathrm{K}}\right)^{-3} & \vspace{0.5em} \\
\quad \times \left(\dfrac{\nH}{1\;\mathrm{cm^{-3}}}\right)^{-1}\left(\dfrac{a_{\mathrm{gr}}}{0.01\;\micron}\right) & \Tgas \leq 10^{6}\;\mathrm{K} \vspace{0.5em}\\ 
20  \;\kyr \left(\dfrac{\nH}{1\;\mathrm{cm^{-3}}}\right)^{-1}\left(\dfrac{a_{\mathrm{gr}}}{0.01\;\micron}\right) & \Tgas > 10^{6}\;\mathrm{K}.
\end{array}
\right.
\end{eqnarray}
The typical gas temperature and number density of the photo-evaporation flow in our simulations are $\Tgas=30000\;\mathrm{K}$ and $\nH=10^{2}$-$10^{3}\;\mathrm{cm^{-3}}$. For $a_{\mathrm{gr}}=0.05\;\micron$, $\tau^{\mathrm{dest}}_{\mathrm{gr}}\gtrsim 3.25\;\myr$. Therefore, the thermal sputtering is not important for our simulations. But, the thermal sputtering can be effective in a denser photo-evaporation flow, which may occur in the inner part of an AGN torus where average gaseous density is expected to be high.

\section{Tree-accelerated long characteristic method}\label{appendix:sec:TreeLONG}
In this section, we explain tree-accelerated long characteristic method in detail. First of all, we introduce some terminology. For a given radiation source $s$ and a target particle $i$ to which we want to evaluate the column densities, we define \textit{upstream particles} as particles that obscure the target particle at least partially. Conversely, the particles that are obscured by a target particle at least partially for a given radiation source are called \textit{downstream particles}. 

\subsection{Column density calculation}\label{appendix:subsec:Ncol_calculation}
\begin{figure}
\centering
\includegraphics[clip,width=\linewidth]{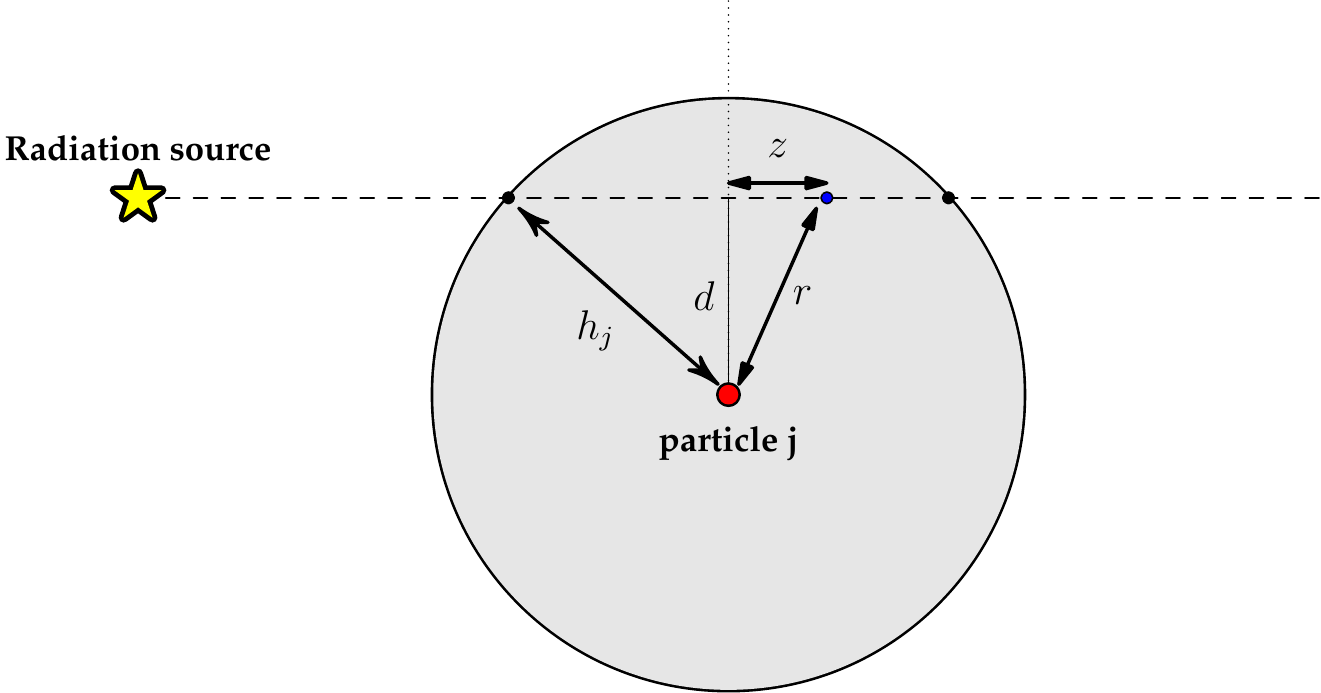}
\caption{A schematic illustration of the calculation of the column density. The \textit{dashed line} shows the line that connects between radiation source $s$ and particle $i$ to which we want to evaluate the column density. The \textit{red point} represents one of the upstream particles of particle $i$. Here, we call it particle $j$. It is located at a distance $d$ from the line $\overline{i\operatorname{-}s}$ and its smoothing length is $h_{j}$. Particle $i$ is located outside the figure.}
\label{fig:dNcol_calc_case1}
\end{figure}

Here, we explain calculation method of obscuring column density due to particle $j$ to particle $i$.
As shown in Fig.~\ref{fig:dNcol_calc_case1}, if the line segment $\linesegment{i}{s}$ that connects between particle $i$ and radiation source $s$ passes through the SPH kernel of particle $j$, obscuring column density due to particle $j$, $\delta N^{(k)}_{\mathrm{col}}(j\rightarrow i)$, can be computed by
\begin{eqnarray}
\delta N^{(k)}_{\mathrm{col}}(j\rightarrow i)  =  2\int^{\sqrt{h^{2}_{j}-d^{2}}}_{0}Y^{(k)}_{j}m_{j}W(r,h_{j})dz,
\end{eqnarray}
where $k$ denotes species, $Y^{(k)}_{j}$ is the mass abundance of species $k$ of particle $j$, $d$ is the distance between particle $j$ and the line $\overline{i\operatorname{-} s}$, $W$ is the SPH kernel function. Using normalized variables $\tilde{d}=d/h_{j}$, $\tilde{r}=r/h_{j}$ and $\tilde{z}=z/h_{j}$ and the relation $W(r,h)=h^{-3}w(\tilde{r})$, we can rewrite the equation above into
\begin{eqnarray}
\delta N^{(k)}_{\mathrm{col}}(j\rightarrow i)  =  Y^{(k)}_{j}m_{j}h_{j}^{-2} F(\tilde{d}),
\end{eqnarray}
where
\begin{eqnarray}
F(\tilde{d}) = 2\int^{\sqrt{1-\tilde{d}^{2}}}_{0}w(\widetilde{r})d\widetilde{z}.
\end{eqnarray}
Thus, we calculate $\delta N^{(k)}_{\mathrm{col}}(j\rightarrow i)$ easily using the physical quantities of particle $j$ and a look-up table for $F(\tilde{d})$.

\begin{figure}
\centering
\includegraphics[clip,width=\linewidth]{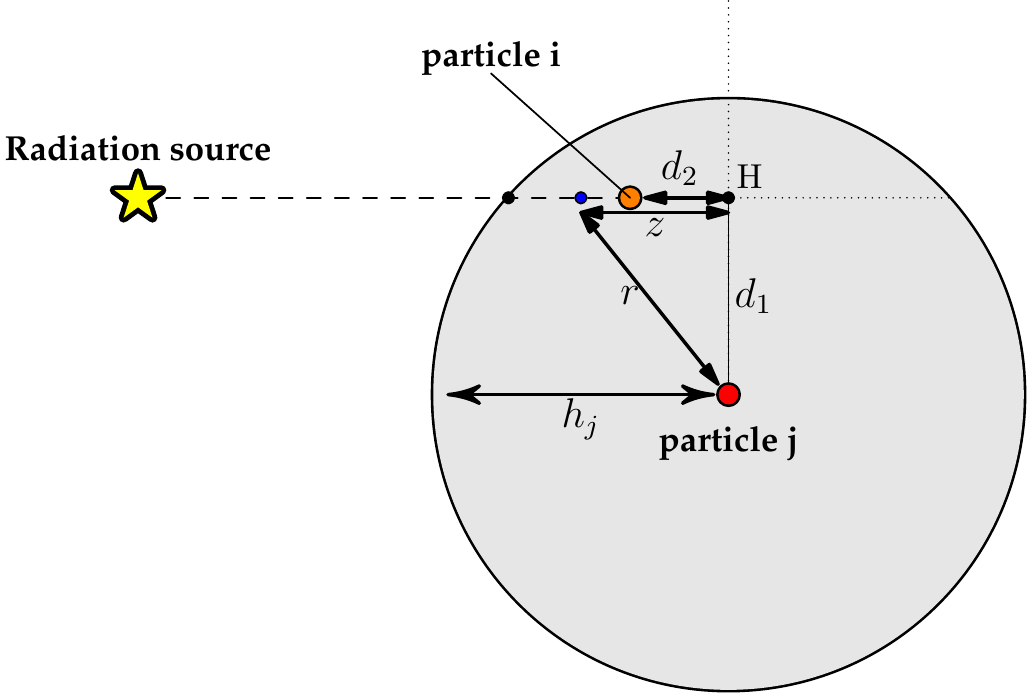}
\caption{The same as Fig.~\ref{fig:dNcol_calc_case1} but for the case that particle $i$ is located within the SPH kernel of particle $j$ and is nearer to the radiation source $s$ than the point H. $d_{1}$ is the same as $d$ in Fig.~\ref{fig:dNcol_calc_case1}, while $d_{2}$ is the distance between the point H and particle $i$.}
\label{fig:dNcol_calc_case2}
\end{figure}

In the cases that either of particle $i$ or radiation source $s$ is located within the SPH kernel of particle $j$, $\delta N^{(k)}_{\mathrm{col}}(j\rightarrow i)$ can be calculated using a two dimensional table. As an example, we consider the situation shown in Fig.~\ref{fig:dNcol_calc_case2}, in which particle $i$ is in the SPH kernel of particle $j$ and its position is nearer to the radiation source $s$ than the point H. In this case, $\delta N^{(k)}_{\mathrm{col}}(j\rightarrow i)$ is computed by
\begin{eqnarray}
\delta N^{(k)}_{\mathrm{col}}(j\rightarrow i) & = & \int^{\sqrt{h^{2}_{j}-d^{2}_{1}}}_{d_{2}}Y^{(k)}_{j}m_{j}W(r,h_{j})dz, \\
& = & Y^{(k)}_{j}m_{j}h^{-2}_{j}G(\tilde{d_{1}},\tilde{d_{2}}), \label{eq:dNcol_ji_2nd}
\end{eqnarray}
where
\begin{eqnarray}
G(\tilde{d_{1}},\tilde{d_{2}}) = \int^{\sqrt{1-\tilde{d}_{1}^{2}}}_{\tilde{d_{2}}}w(\tilde{r})d\tilde{z}.
\end{eqnarray}
When particle $i$ is farther to the radiation source $s$ than the point H, $\delta N^{(k)}_{\mathrm{col}}(j\rightarrow i)$ is computed using the same two-dimensional table but by replacing $G(\tilde{d_{1}},\tilde{d_{2}})$ in the equation (\ref{eq:dNcol_ji_2nd}) by $2G(\tilde{d}_{1},0)-G(\tilde{d}_{1},\tilde{d}_{2})$. The same method is applicable to the cases that radiation source $s$ is located within the SPH kernel of particle $j$. The situations not discussed so far are the cases that \textit{both} the radiation source $s$ and particle $i$ are contained in the SPH kernel of particle $j$ (there are three cases depending on the relative positions of $s$ and $i$ to the point H). In these cases, we calculate $\delta N^{(k)}_{\mathrm{col}}(j\rightarrow i)$ by using a linear combination of $G(\tilde{d}_{1},0)$, $G(\tilde{d}_{1},\tilde{d}_{2}^{s})$, and $G(\tilde{d}_{1},\tilde{d}_{2}^{i})$, where $\tilde{d}_{2}^{s}$ and $\tilde{d}_{2}^{i}$ are normalized distances between the point H and $s$/$i$, respectively.

\subsection{Acceleration of column density calculation}\label{appendix:subsec:acceleration_method}
In the original long characteristic method, we seek all of the upstream particles for a target particle $i$ and compute $N^{(k)}_{\mathrm{col},i}=\sum_{j}\delta N^{(k)}_{\mathrm{col}}(j\rightarrow i)$. This calculation is repeated for all the particles. However, such calculation is numerically expensive and a simulation does not end within a realistic time. Therefore, we need to reduce the computational cost with keeping its accuracy reasonable. To achieve it, we use a tree structure in this paper.

Let's consider a small number ($< n^{\mathrm{crit}}_{i_{G}}$) of the particles that are close each other. We call these particles the target particles or simply group $i_{G}$ and evaluate the column densities at their locations for a given radiation source $s$ which is placed at $\bmath{x}_{s}$. Because each of the target particles has almost the same upstream particles, we calculate column densities for particle $i\in i_{G}$ using the same list of the upstream particles. This approach is the same as the group interaction list technique used in the $N$-body calculation (\citealt{barnes90:_modif_tree_code,makino91:_treec_special_purpos_proces}) and $i_{G}$ is chosen from the group nodes. The list of the upstream particles is constructed by making the node list according to the procedure described in Algorithm.\ref{alg:identify_UPs}\footnotemark, and then expanding it.

In order to speed up the calculation further, we approximate obscuring column densities due to an upstream particle $j$ to all the particles $i\in i_{G}$ by a common value of $\delta N^{(k)}_{\mathrm{col}}(j\rightarrow i_{G})$, if the upstream particle $j$ satisfies the following two conditions, 
\begin{enumerate}
\item The SPH kernel of particle $j$ does not overlap group node $i_{G}$,
\item particle $j$ satisfies the inequality,
\begin{eqnarray}
\theta \equiv \dfrac{\max(l_{x},l_{y},l_{z})}{r} < \theta_{\mathrm{RT}},
\end{eqnarray}
where $l_{i}$ is the physical size of group node $i_{G}$ in each dimension, $r$ is the distance between the geometric center of group $i_{G}$, $\bmath{x}_{\mathrm{cen},i_{G}}$, and $\bmath{x}_{j}$, and $\theta_{\mathrm{RT}}$ is a tolerance parameter which determines the accuracy. In this study, we use $\theta_{\mathrm{RT}}=1$ for all the simulations.
\end{enumerate}
$\delta N^{(k)}_{\mathrm{col}}(j\rightarrow i_{G})$ is calculated using $\bmath{x}_{j}$ and $\bmath{x}_{\mathrm{cen},i_{G}}$ according to the method described in \S~\ref{appendix:subsec:Ncol_calculation}. Otherwise, we compute $\delta N^{(k)}_{\mathrm{col}}(j\rightarrow i)$ individually for all the particle $i\in i_{G}$.

Thus, this method can reduce the calculation cost by a factor of $n_{i_{G}}$ in the best-case. In our computational environments, optimal choice were $n^{\mathrm{crit}}_{i_{G}}=n^{\mathrm{crit}}_{j_{G}}=32$. Further speed-up can be possible if we replace a group of \textit{upstream} particles by a virtual particle or cell based on some criterion. This will reduce the cost by a factor of $n_{j_{G}}$. But, a care must be taken because a density structure of upstream region is smoothed out.

\begin{figure}
\centering
\includegraphics[clip,width=\linewidth]{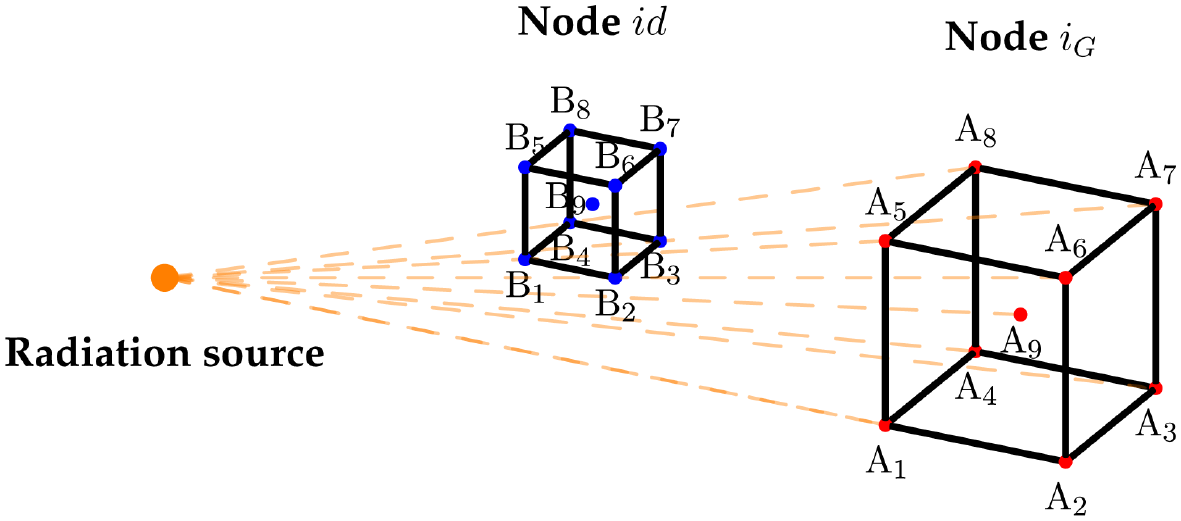}
\caption{A schematic illustration of the intersection test. The two cubes represent tree nodes and node $id$ is the node that exposed to the intersection test. For brevity, we do not plot the SPH particles contained in the nodes. The \textit{red} and \textit{blue} points indicate the vertices of the nodes ($A_{1}\sim A_{8}$, $B_{1}\sim B_{8}$) and the geometric centers of the nodes ($A_{9}$, $B_{9}$). The \textit{orange} dashed line segments connect the radiation source and the points $\{A_{i}\}$.} 
\label{fig:intersection_test}
\end{figure}

\footnotetext{
Here, we explain Algorithm.\ref{alg:identify_UPs} in more detail.
This procedure is a recursive one and makes a list of tree nodes that intersects either of the line segments that connect the radiation source $s$ and particle $i\in i_{G}$. For a given node $id$, we first investigate the number of the particle contained in this node, which is represented by $\mathcal{T}.\nptcl (id)$ in the figure ($\mathcal{T}$ is a data structure for the tree and $\nptcl()$ is a member array of $\mathcal{T}$). If $\mathcal{T}.\nptcl (id)=1$, node $id$ is added to the list unconditionally. 

If $\mathcal{T}.\nptcl(id)>1$, several tests are performed to see if node $id$ has upstream particles of the particles contained in node $i_{G}$: 
\begin{description}
\item[\textbf{(1) Disjoint test}] This test investigates whether node $id$ and group node $i_{G}$ are disjoint each other or not and the result is stored into $f_{\mathrm{disjoint}}$,
\item[\textbf{(2) Enclose test}] This test examines whether node $id$ contains the radiation source $s$ or not and the result is stored into $f_{\mathrm{enclose}}$,
\item[\textbf{(3) Intersection test}] If $f_{\mathrm{disjoint}}$ is \texttt{True} and $f_{\mathrm{enclose}}$ is \texttt{False}, an additional test is performed, in which we check whether some particles in node $id$ can be upstream particles of the particles $i\in i_{G}$. A straightforward way to examine this possibility is to check whether the SPH kernel of particle $j$ in node $id$ intersects the line segment $\linesegment{s}{i}$ for all combinations of ($i$,$j$). However, it is inefficient. Instead we use the geometric information of the nodes. More specifically, as shown in Figure~\ref{fig:intersection_test}, we investigate the positional relations between the vertices and the geometric center of node $id$ and the line segments that connect the radiation source $s$ and the vertices and the geometric center of node $i_{G}$.
We decide that node $id$ is an upstream node of node $i_{G}$ if $0\leq t_{ij,\mathrm{H}}\leq 1$ and $d_{ij}\leq d_{\mathrm{crit}}$ are satisfied by at least one combination of $(i,j)$. $t_{ij,\mathrm{H}}$ and $d_{ij}$ are defined as
\begin{eqnarray}
t_{ij,\mathrm{H}} & = & \dfrac{\bmath{x}_{is}\cdot\bmath{x}_{js}}{|\bmath{x}_{is}|^{2}}, \\
d_{ij} & = & |t_{\mathrm{H}}\bmath{x}_{is} - \bmath{x}_{js}|,
\end{eqnarray}
where $s$ indicates the radiation source, $i$ and $j$ take $A_{1}..A_{9}$ and $B_{1}...B_{9}$, respectively (see Fig.~\ref{fig:intersection_test}). 
$t_{ij,\mathrm{H}}$ represents the position of the foot of the perpendicular of the point $j$ onto the line that passes through $s$ and $i$. 
$0\leq t_{ij,\mathrm{H}}\leq 1$ means that the foot is laid in the line segment $\linesegment{s}{i}$.
$d_{ij}$ is the distance between the point $j$ and $\linesegment{s}{i}$.
$d_{\mathrm{crit}} \equiv \max(l_{i_{G}},l_{j_{G}})$ and $l_{\alpha}$ is defined as
\begin{eqnarray}
l_{\alpha} = \sqrt{1.05(l^{2}_{x,\alpha}+l^{2}_{y,\alpha}+l^{2}_{z,\alpha})}.
\end{eqnarray}
If $f_{\mathrm{intersect}}=\mathtt{False}$, we return the process to the parent node.
\end{description}

If (i) either of $f_{\mathrm{disjoint}}=\mathtt{False}$ or $f_{\mathrm{enclose}}=\mathtt{True}$ is satisfied, and (ii) $\mathcal{T}.\nptcl(id) < n^{\mathrm{crit}}_{j_{G}}$, node $id$ is added to the list. Otherwise, we descend the tree further. 

}

\begin{algorithm}
\caption{\textsc{MakeNodelist}}
\label{alg:identify_UPs}
\SetFuncSty{textsc}%
\SetKwFunction{MakeNodelist}{MakeNodelist}%
\SetKwInOut{Input}{input}
\SetKwInOut{Output}{output}
\Input{$\bmath{x}_{\min,i_{G}}$,$\bmath{x}_{\max,i_{G}}$,$\bmath{x}_{s}$,$\mathcal{T}$,$id$}
\Output{a list of nodes}

\BlankLine
\uIf{$\mathcal{T}.n_{\mathrm{ptcl}}(id)>1$}{
   $f_{\mathrm{disjoint}} \gets \text{Disjoint test}$ \;
   $f_{\mathrm{enclose}} \gets \text{Enclose test}$ \;
   \If{$f_{\mathrm{disjoint}} = \mathtt{True}$ and $f_{\mathrm{enclose}} = \mathtt{False}$}{
      $f_{\mathrm{intersect}} \gets \text{Intersection test}$ \;
      \If{$f_{\mathrm{intersect}} = \mathtt{False}$}{
         Return \;
      }
   }
   \uIf{$\mathcal{T}.n_{\mathrm{ptcl}}(id)<n^{\mathrm{crit}}_{j_{G}}$}{
      Append $id$ to the list \;
   }
   \Else{
      \ForEach{a child node, $cid$, of the node $id$}{
         \MakeNodelist{$\bmath{x}_{\min,i_{G}}$,$\bmath{x}_{\max,i_{G}}$,$\bmath{x}_{s}$,$\mathcal{T}$,$cid$} \;
      }
   }
}
\ElseIf{$\mathcal{T}.n_{\mathrm{ptcl}}(id)=1$}{
   Append $id$ to the list \;
}
\end{algorithm}

\subsection{Tests}\label{appendix:subsec:test}
In order to check the accuracy of our method described in the previous sections, we perform the same tests as Test 1 and Test 2 in \citet{iliev06:_cosmol}, which are the calculations of an expansion of a H{\scriptsize II} region in an uniform static medium. We use the same physical parameters and settings as \citet{iliev06:_cosmol} except for the size of the computational box and the numerical resolution. Here, a $13.2\;\kpc$ cubic computational box is assumed. We place an ionizing radiation source at the center of the box. The uniform medium is realized by a glass-like distribution of $128^{3}$ SPH particles. Therefore, the numerical resolution is two times lower than that used in \citet{iliev06:_cosmol}.

\begin{figure*}
\centering
\includegraphics[clip,width=\linewidth]{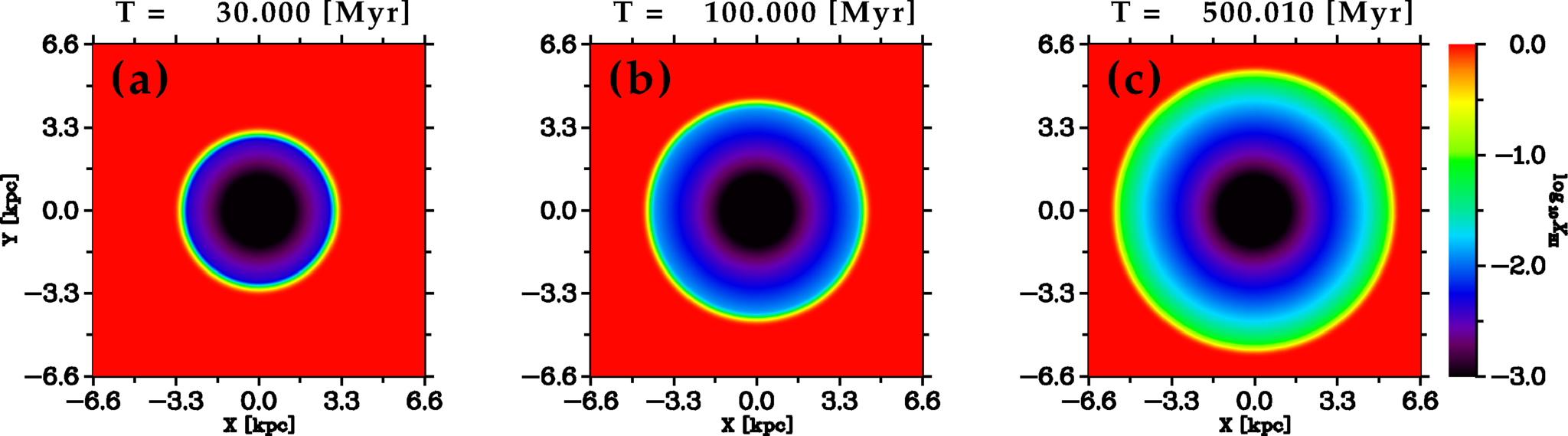}
\caption{H{\scriptsize I} fraction at different times in Test 1 of \citet{iliev06:_cosmol}. All the snapshots are slices at $z=0$.}
\label{fig:Iliev06_Test1}
\end{figure*}

\begin{figure}
\centering
\includegraphics[clip,width=\linewidth]{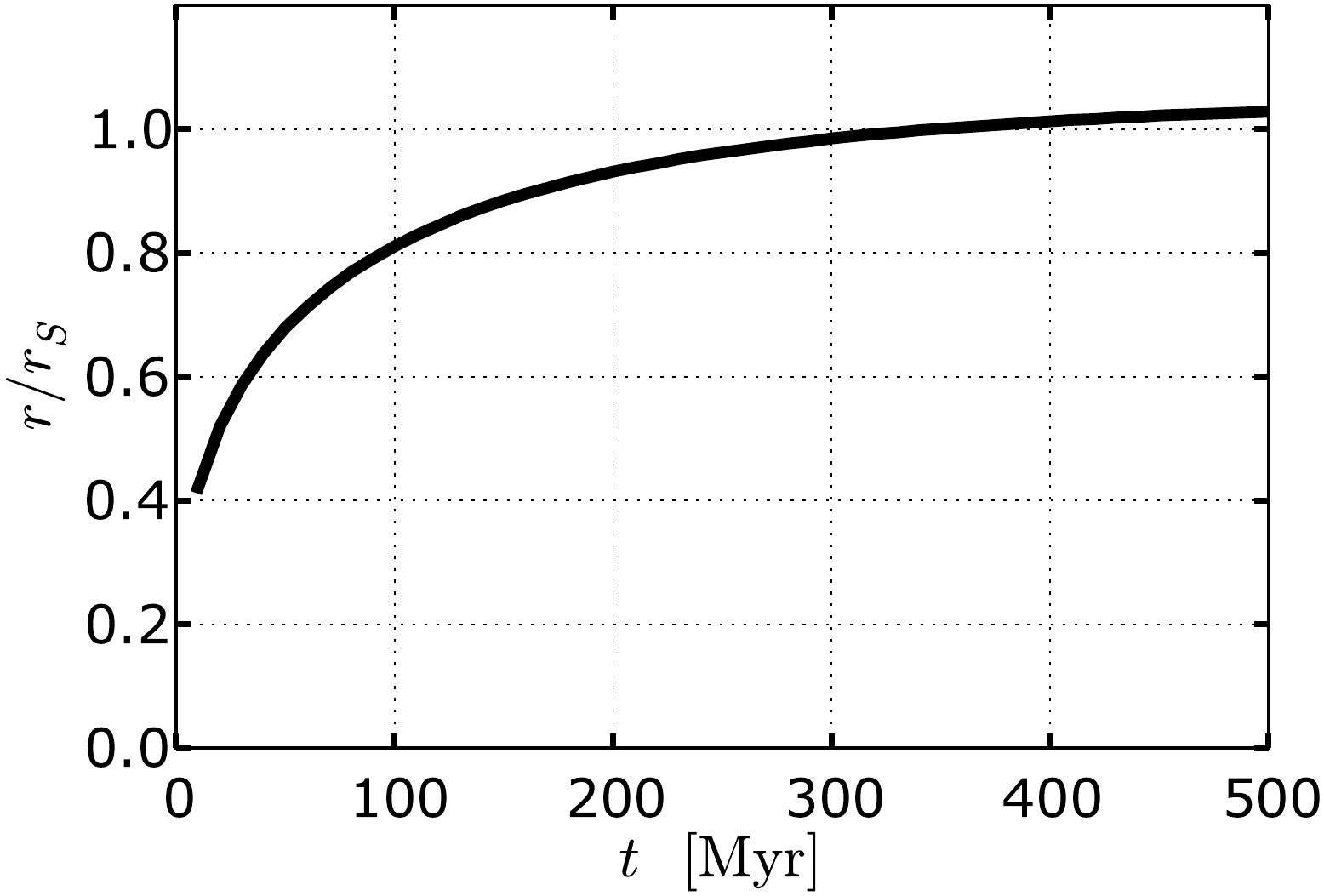}
\caption{The propagation of the ionization front in Test 1 of \citet{iliev06:_cosmol}. The radius of the ionization front is normalized by the initial Str{\"o}mgren radius $r_{S}$, which is $5.4\;\kpc$ in this case.}
\label{fig:IF_Iliev06_Test1}
\end{figure}

\begin{figure*}
\centering
\includegraphics[clip,width=\linewidth]{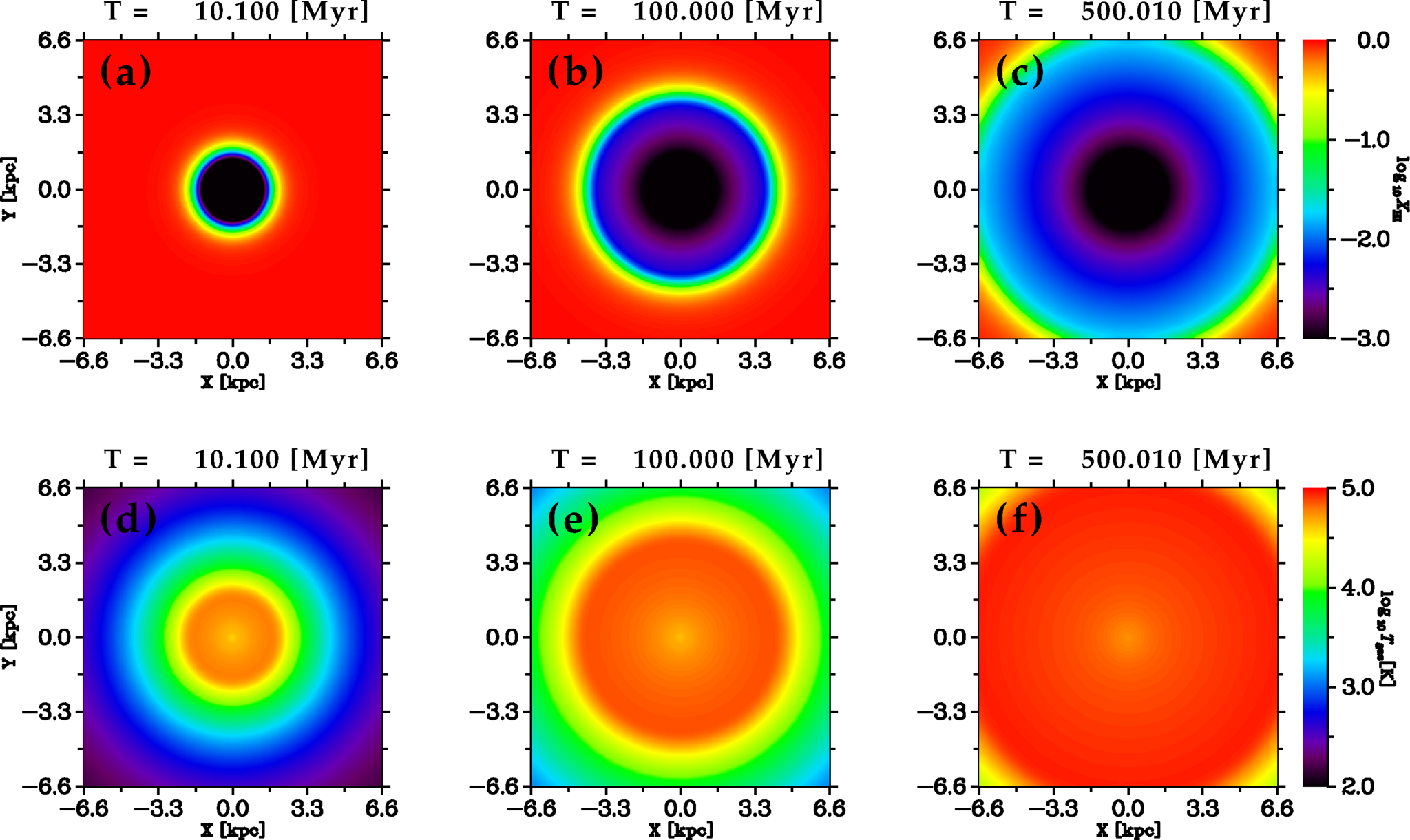}
\caption{H{\scriptsize I} fraction (\textit{upper panels}) and gas temperature (\textit{lower panels}) at different times in Test 2 of \citet{iliev06:_cosmol}. All the snapshots are slices at $z=0$.}
\label{fig:Iliev06_Test2}
\end{figure*}

The time evolution of H{\scriptsize I} fraction, $X_{\mathrm{H_{I}}}$, in Test 1 is shown in Fig.~\ref{fig:Iliev06_Test1}. The result at $t=500\;\myr$ is in good agreement with those shown in Fig.~6 of \citet{iliev06:_cosmol}. In order to see the accuracy more quantitatively, we show in Fig.~\ref{fig:IF_Iliev06_Test1} the time evolution of the radius of the ionization front (IF), which is defined as the radius at which $X_{\mathrm{H_{I}}}=0.5$. The IF saturates at a value of $r\approx 1.03r_{S}$, where $r_{S}$ is the initial Str{\"o}mgren radius. Ideally, the saturation value should be $r=r_{S}$. This small overestimation of the IF position is due to a drawback of our method that the ionization of particle $i$ located at $r_{i}$ causes a decrease of the column densities over a range of $r=r_{i}-h_{i}\sim r_{i}+h_{i}$. Thus, the spatial resolution of the column density calculation in our method is at most of the order of $h$. Finally, we show the results of Test 2 in Fig.~\ref{fig:Iliev06_Test2}, which are also in good agreement with Fig.~11, 12, 13, 14 of \citet{iliev06:_cosmol}.

\newpage

\section{Radiative transfer in optically-thick medium}\label{appendix:sec:RT_optically_thick}
In optically-thick medium, it becomes difficult to follow the propagation of an ionization front properly because the photo-ionization rates at the centers of SPH particles become very small although the incident radiation field is strong enough to ionize the SPH particles. In order to avoid this problem, \citet{susa06:_smoot_partic_hydrod_coupl_radiat_trans} used volume-averaged photo-ionization rate instead of photo-ionization rate evaluated at the center of a SPH particle. Following this, we use volume-averaged photo-ionization rate in the simulations in this study. In this study, the volume-averaged photo-ionization rate of HI, $\overline{k}_{\mathrm{HI}}(r)$, is calculated as follows.
\begin{eqnarray}
\overline{k}_{\mathrm{HI}}(r) & \equiv & \dfrac{\int^{r+\Delta r/2}_{r-\Delta r/2}4\pi r^{2} k_{\mathrm{HI}}(r)\:dr}{\int^{r+\Delta r/2}_{r-\Delta r/2}4\pi r^{2}\:dr}, \\
& = & \dfrac{\int^{r+\Delta r/2}_{r-\Delta r/2}\left[\int^{\infty}_{\nu_{\mathrm{L}}}\dfrac{L_{\nu} e^{-\tau_{\nu}}}{h\nu}n_{\mathrm{HI}}\sigma_{\mathrm{abs}}^{\mathrm{HI}}(\nu)\:d\nu \right]dr}{\int^{r+\Delta r/2}_{r-\Delta r/2}4\pi r^{2}\:dr}, \\
& \approx & \dfrac{1}{1+\dfrac{1}{12}\left(\dfrac{\Delta r}{r}\right)^{2}} \nonumber\\
& & \quad \times \dfrac{1}{2}\left[k_{\mathrm{HI}}(r-\Delta r/4)+k_{\mathrm{HI}}(r+\Delta r/4)\right],
\end{eqnarray}
where we assumed that $n_{\mathrm{HI}}$ is constant over the interval of $[r-\Delta r/2,r+\Delta r/2]$ along a ray and used an approximate relation
\begin{eqnarray}
\int^{r+\Delta r/2}_{r-\Delta r/2}e^{-\tau_{\nu}}dr & \approx & \dfrac{\Delta r}{2}e^{-\tau_{\nu}(r=r-\frac{\Delta r}{4})} \nonumber\\
& + & \dfrac{\Delta r}{2}e^{-\tau_{\nu}(r=r+\frac{\Delta r}{4})}.
\end{eqnarray}
The photo-heating rates and radiative accelerations (Eqs.(\ref{eq:kHI}) $\sim$ (\ref{eq:aDust}), (\ref{eq:kH2I}), and (\ref{eq:aH2I})) are also evaluated in this method and we use $\Delta r_{i}=2h_{i}$, where $h_{i}$ is the smoothing length of SPH particle $i$. Note that this method is not photon-conservation scheme unlike the method in \citet{susa06:_smoot_partic_hydrod_coupl_radiat_trans}\footnote{In order to make the method photon-conservative, we have to calculate the integrals in Eqs.(\ref{eq:kHI}) $\sim$ (\ref{eq:aDust}) at each time step.}.

\bsp

\label{lastpage}

\end{document}